\documentclass{aa}  

\usepackage{graphicx}
\usepackage{amsmath,amsfonts,amssymb}
\usepackage{natbib}

\usepackage{txfonts}
\usepackage{xcolor}

\usepackage{blindtext}
\usepackage{float}
\usepackage{dblfloatfix}
\usepackage{afterpage}
\usepackage{ifthen}
\usepackage[morefloats=12]{morefloats}

\usepackage{placeins}
\usepackage{multicol}
\bibpunct{(}{)}{;}{a}{}{,}
\usepackage[switch]{lineno}
\definecolor{linkcolor}{rgb}{0.6,0,0}
\definecolor{citecolor}{rgb}{0,0,0.75}
\definecolor{urlcolor}{rgb}{0.12,0.46,0.7}
\usepackage[breaklinks, colorlinks, urlcolor=urlcolor,
    linkcolor=linkcolor,citecolor=citecolor,pdfencoding=auto]{hyperref}
\hypersetup{linktocpage}
\usepackage{bold-extra}

\def\setsymbol#1#2{\expandafter\def\csname #1\endcsname{#2}}
\def\getsymbol#1{\csname #1\endcsname}

\def\Planck{\textit{Planck}}

\newbox\tablebox    \newdimen\tablewidth
\def\leaderfil{\leaders\hbox to 5pt{\hss.\hss}\hfil}

\def\endPlancktablewide{\tablewidth=\textwidth 
    $$\hss\copy\tablebox\hss$$
    \vskip-\lastskip\vskip -2pt}
\def\tablenote#1 #2\par{\begingroup \parindent=0.8em
    \abovedisplayshortskip=0pt\belowdisplayshortskip=0pt
    \noindent
    $$\hss\vbox{\hsize\tablewidth \hangindent=\parindent \hangafter=1 \noindent
    \hbox to \parindent{$^#1$\hss}\strut#2\strut\par}\hss$$
    \endgroup}
\def\doubleline{\vskip 3pt\hrule \vskip 1.5pt \hrule \vskip 5pt}

\def\L2{\ifmmode L_2\else $L_2$\fi}

\def\DeltaT{\ifmmode \Delta T\else $\Delta T$\fi}
\def\deltat{\ifmmode \Delta t\else $\Delta t$\fi}
\def\fknee{\ifmmode f_{\rm knee}\else $f_{\rm knee}$\fi}
\def\Fmax{\ifmmode F_{\rm max}\else $F_{\rm max}$\fi}
\def\solar{\ifmmode{\rm M}_{\mathord\odot}\else${\rm M}_{\mathord\odot}$\fi}
\def\Msolar{\ifmmode{\rm M}_{\mathord\odot}\else${\rm M}_{\mathord\odot}$\fi}
\def\Lsolar{\ifmmode{\rm L}_{\mathord\odot}\else${\rm L}_{\mathord\odot}$\fi}
\def\inv{\ifmmode^{-1}\else$^{-1}$\fi}
\def\mo{\ifmmode^{-1}\else$^{-1}$\fi}
\def\sup#1{\ifmmode ^{\rm #1}\else $^{\rm #1}$\fi}
\def\expo#1{\ifmmode \times 10^{#1}\else $\times 10^{#1}$\fi}
\def\,{\thinspace}
\def\lsim{\mathrel{\raise .4ex\hbox{\rlap{$<$}\lower 1.2ex\hbox{$\sim$}}}}
\def\gsim{\mathrel{\raise .4ex\hbox{\rlap{$>$}\lower 1.2ex\hbox{$\sim$}}}}

\def\simprop{\mathrel{\raise .4ex\hbox{\rlap{$\propto$}\lower 1.2ex\hbox{$\sim$}}}}
\def\deg{\ifmmode^\circ\else$^\circ$\fi}
\def\pdeg{\ifmmode $\setbox0=\hbox{$^{\circ}$}\rlap{\hskip.11\wd0 .}$^{\circ}
          \else \setbox0=\hbox{$^{\circ}$}\rlap{\hskip.11\wd0 .}$^{\circ}$\fi}
\def\arcs{\ifmmode {^{\scriptstyle\prime\prime}}
          \else $^{\scriptstyle\prime\prime}$\fi}
\def\arcm{\ifmmode {^{\scriptstyle\prime}}
          \else $^{\scriptstyle\prime}$\fi}
\newdimen\sa  \newdimen\sb
\def\parcs{\sa=.07em \sb=.03em
     \ifmmode \hbox{\rlap{.}}^{\scriptstyle\prime\kern -\sb\prime}\hbox{\kern -\sa}
     \else \rlap{.}$^{\scriptstyle\prime\kern -\sb\prime}$\kern -\sa\fi}
\def\parcm{\sa=.08em \sb=.03em
     \ifmmode \hbox{\rlap{.}\kern\sa}^{\scriptstyle\prime}\hbox{\kern-\sb}
     \else \rlap{.}\kern\sa$^{\scriptstyle\prime}$\kern-\sb\fi}
\def\ra[#1 #2 #3.#4]{#1\sup{h}#2\sup{m}#3\sup{s}\llap.#4}
\def\dec[#1 #2 #3.#4]{#1\deg#2\arcm#3\arcs\llap.#4}
\def\deco[#1 #2 #3]{#1\deg#2\arcm#3\arcs}
\def\rra[#1 #2]{#1\sup{h}#2\sup{m}}

\def\dots{\relax\ifmmode \ldots\else $\ldots$\fi}
\def\WHzsr{\ifmmode $W\,Hz\mo\,sr\mo$\else W\,Hz\mo\,sr\mo\fi}
\def\mHz{\ifmmode $\,mHz$\else \,mHz\fi}
\def\GHz{\ifmmode $\,GHz$\else \,GHz\fi}
\def\mKs{\ifmmode $\,mK\,s$^{1/2}\else \,mK\,s$^{1/2}$\fi}
\def\muKs{\ifmmode \,\mu$K\,s$^{1/2}\else \,$\mu$K\,s$^{1/2}$\fi}
\def\muKRJs{\ifmmode \,\mu$K$_{\rm RJ}$\,s$^{1/2}\else \,$\mu$K$_{\rm RJ}$\,s$^{1/2}$\fi}
\def\muKHz{\ifmmode \,\mu$K\,Hz$^{-1/2}\else \,$\mu$K\,Hz$^{-1/2}$\fi}
\def\MJysr{\ifmmode \,$MJy\,sr\mo$\else \,MJy\,sr\mo\fi}
\def\MJysrmK{\ifmmode \,$MJy\,sr\mo$\,mK$_{\rm CMB}\mo\else \,MJy\,sr\mo\,mK$_{\rm CMB}\mo$\fi}
\def\microns{\ifmmode \,\mu$m$\else \,$\mu$m\fi}

\def\muK{\ifmmode \,\mu$K$\else \,$\mu$\hbox{K}\fi}
\def\microK{\ifmmode \,\mu$K$\else \,$\mu$\hbox{K}\fi}
\def\muW{\ifmmode \,\mu$W$\else \,$\mu$\hbox{W}\fi}
\def\kms{\ifmmode $\,km\,s$^{-1}\else \,km\,s$^{-1}$\fi}
\def\kmsMpc{\ifmmode $\,\kms\,Mpc\mo$\else \,\kms\,Mpc\mo\fi}

\providecommand{\sorthelp}[1]{}

\def\Cosmoglobe{\textsc{Cosmoglobe}}
\def\commanderthree{\texttt{Commander3}}
\def\commander{\texttt{Commander}}
\def\Planck{\textit{Planck}}
\def\WMAP{\textit{WMAP}}
\def\COBE{\textit{COBE}}
\def\GAIA{\textit{Gaia}}
\def\gaia{\textit{Gaia}}
\def\Gaia{\textit{Gaia}}
\def\WISE{WISE}
\def\AKARI{\textrm{{AKARI}}}
\def\IRAS{\textrm{{IRAS}}}

\newcommand{\CII}{\ensuremath{\textsc{C\,ii}}}

\newcommand{\dv}[0]{\vec{d}}
\renewcommand{\t}[0]{\vec{t}}

\newcommand{\B}[0]{\tens{B}}

\newcommand{\G}[0]{\tens{G}}
\newcommand{\n}[0]{\vec{n}}

\newcommand{\s}[0]{\vec{s}}
\renewcommand{\a}[0]{\vec{a}}

\renewcommand{\L}[0]{\tens{L}}

\newcommand{\N}[0]{\tens{N}}
\newcommand{\M}[0]{\tens{M}}

\renewcommand{\r}[0]{\vec{r}}

\renewcommand{\P}[0]{\tens{P}}

\newcommand{\Te}[0]{T_{\rm e}}

\newcommand{\BP}{\textsc{BeyondPlanck}}
\newcommand{\bp}{\textsc{BeyondPlanck}}
\newcommand{\cosmoglobe}{\textsc{Cosmoglobe}}

\newcommand{\npipe}[0]{\texttt{NPIPE}}
\newcommand{\sroll}[0]{\texttt{SRoll}}

\newcommand{\e}{\mathrm e}

\def\Tcmb{\ifmmode T_\mathrm{CMB}\else $T_{\mathrm{CMB}}$\fi}
\def\Tcold{\ifmmode T_\mathrm{c}\else $T_{\mathrm{c}}$\fi}
\def\Thot{\ifmmode T_\mathrm{h}\else $T_{\mathrm{h}}$\fi}
\def\Tnear{\ifmmode T_\mathrm{n}\else $T_{\mathrm{n}}$\fi}
\def\scmb{\ifmmode s_\mathrm{CMB}\else $s_{\mathrm{CMB}}$\fi}
\def\squad{\ifmmode s_\mathrm{quad}\else $s_{\mathrm{quad}}$\fi}
\def\ssynch{\ifmmode s_\mathrm{s}\else $s_\mathrm{s}$\fi}
\def\sdust{\ifmmode s_\mathrm{d}\else $s_{\mathrm{d}}$\fi}
\def\ssdust{\ifmmode s_\mathrm{sd}\else $s_{\mathrm{sd}}$\fi}
\def\same{\ifmmode s_\mathrm{AME}\else $s_{\mathrm{AME}}$\fi}
\def\ssrc{\ifmmode s_\mathrm{src}\else $s_{\mathrm{src}}$\fi}
\def\sco{\ifmmode s_\mathrm{CO}\else $s_{\mathrm{CO}}$\fi}
\def\sff{\ifmmode s_\mathrm{ff}\else $s_{\mathrm{ff}}$\fi}
\def\gff{\ifmmode g_\mathrm{ff}\else $g_{\mathrm{ff}}$\fi}
\def\fsynch{\ifmmode f_\mathrm{s}\else $f_{\mathrm{s}}$\fi}
\def\fsd{\ifmmode f_\mathrm{sd}\else $f_{\mathrm{sd}}$\fi}
\def\fame{\ifmmode f_\mathrm{AME}\else $f_{\mathrm{AME}}$\fi}
\def\alphasrc{\ifmmode \alpha_\mathrm{src}\else $\alpha_{\mathrm{src}}$\fi}
\def\bcold{\ifmmode \beta_\mathrm{c}\else $\beta_{\mathrm{c}}$\fi}
\def\bhot{\ifmmode \beta_\mathrm{h}\else $\beta_{\mathrm{h}}$\fi}
\def\bnear{\ifmmode \beta_\mathrm{n}\else $\beta_{\mathrm{n}}$\fi}
\def\bsynch{\ifmmode \beta_\mathrm{s}\else $\beta_{\mathrm{s}}$\fi} 
\def\bsun{\ifmmode \beta_\mathrm{sun}\else $\beta_{\mathrm{sun}}$\fi} 
\def\nuzeros{\ifmmode \nu_{0,\mathrm{s}}\else $\nu_{0,\mathrm{s}}$\fi} 
\def\nuzeroff{\ifmmode \nu_{0,\mathrm{ff}}\else $\nu_{0,\mathrm{ff}}$\fi} 
\def\nuzerocold{\ifmmode \nu_{0,\mathrm{c}}\else $\nu_{0,\mathrm{c}}$\fi}
\def\nuzerohot{\ifmmode \nu_{0,\mathrm{h}}\else $\nu_{0,\mathrm{h}}$\fi}
\def\nuzeronear{\ifmmode \nu_{0,\mathrm{n}}\else $\nu_{0,\mathrm{n}}$\fi} 
\def\nuzeroame{\ifmmode \nu_{0,\mathrm{AME}}\else $\nu_{0,\mathrm{AME}}$\fi} 
\def\nuzerosd{\ifmmode \nu_{0,\mathrm{}}\else $\nu_{0,\mathrm{sd}}$\fi} 
\def\nuzerosrc{\ifmmode \nu_{0,\mathrm{src}}\else $\nu_{0,\mathrm{src}}$\fi} 
\def\nup{\ifmmode \nu_{\mathrm{p}}\else $\nu_{\mathrm{p}}$\fi} 
\def\alphasd{\ifmmode \alpha_{\mathrm{sd}}\else $\alpha_{\mathrm{sd}}$\fi} 
\def\Te{\ifmmode T_{\mathrm{e}}\else $T_{\mathrm{e}}$\fi} 
\def\kB{\ifmmode k_\mathrm{B}\else $k_{\mathrm{B}}$\fi}

\begin{document} 

   \title{\bfseries{\Cosmoglobe\ DR2. I. Global Bayesian analysis of \COBE-DIRBE }}

   \newcommand{\oslo}[0]{1}
\newcommand{\ubc}[0]{2}
\author{\small
D.~J.~Watts\inst{\oslo}\thanks{Corresponding author: D.~Watts; \url{duncan.watts@astro.uio.no}}
\and
M.~Galloway\inst{\oslo}
\and
E.~Gjerl\o w\inst{\oslo}
\and
M.~San\inst{\oslo}
\and
R.~Aurlien\inst{\oslo}
\and
A.~Basyrov\inst{\oslo}
\and
M.~Brilenkov\inst{\oslo}
\and
H.~K.~Eriksen\inst{\oslo}
\and
U.~Fuskeland\inst{\oslo}
\and
L.~T.~Hergt\inst{\ubc}
\and
D.~Herman\inst{\oslo}
\and
H.~T.~Ihle\inst{\oslo}
\and
J.~G.~S.~Lunde\inst{\oslo}
\and
S.~K.~Næss\inst{\oslo}
\and
N.-O.~Stutzer\inst{\oslo}
\and
H.~Thommesen\inst{\oslo}
\and
I.~K.~Wehus\inst{\oslo}
}
\institute{\small
Institute of Theoretical Astrophysics, University of Oslo, Blindern, Oslo, Norway\goodbreak
\and
Department of Physics and Astronomy, University of British Columbia, 6224 Agricultural Road, Vancouver BC, V6T1Z1, Canada\goodbreak
}

   \titlerunning{\Cosmoglobe: DIRBE reanalysis}
   \authorrunning{D.~Watts et al.}

   \date{\today} 
   
   \abstract{We present the first global Bayesian analysis of the time-ordered Diffuse Infrared Background Experiment (DIRBE) data within the \Cosmoglobe\ framework, building on the same methodology that has previously been successfully applied to \Planck\ LFI and \WMAP. These data are analyzed jointly with \COBE-FIRAS, \GAIA, \Planck\ HFI, and WISE observations, which allows for a more accurate instrumental and astrophysical characterization than possible through single-experiment analysis only. This paper provides an overview of the analysis pipeline and main results, and we present and characterize a new set of zodiacal light subtracted mission average (ZSMA) DIRBE maps spanning the wavelength range between 1.25 and 240\,$\mu$m. A key novel aspect of this processing is the characterization and removal of excess radiation between 4.9 and 60$\,\mu$m that appears static in solar-centric coordinates. The new DR2 ZSMA maps have several notable advantages with respect to the previously available maps, including 1) lower zodiacal light (and possibly straylight) residuals; 2) better determined zero-levels; 3) natively HEALPix tessellated maps with a $7\arcm$ pixel size; 4) nearly white noise at pixel scales; and 5) a more complete and accurate noise characterization established through the combination of Markov Chain Monte Carlo samples and half-mission maps. In addition, because the model has been simultaneously fitted with both DIRBE and HFI data, this is the first consistent unification of the infrared and CMB wavelength ranges into one global sky model covering 100\,GHz to 1\,$\mu$m. However, we do note that even though the new maps are improved with respect to the official maps, and should be preferred for most future analyses that require DIRBE sky maps, they still exhibit non-negligible zodiacal light residuals between 12 and 60$\,\mu$m. Further improvements should be made through joint analysis with complementary infrared experiments such \IRAS, \AKARI, WISE and SPHEREx, and thereby release the full combined potential of all these powerful infrared observatories. 
   }


   \maketitle

\setcounter{tocdepth}{2}
\tableofcontents
   
\section{Introduction}

The astrophysical sky contains a wealth of information about our own Solar System, the Milky Way, and the high-frequency universe in the infrared wavelength regime from roughly 1 to 1000\,$\mu$m \citep[e.g.,][]{johnson:1966,soifer:1987,gardner:2006}. These wavelengths have therefore been the target of many ground-breaking experiments during the last five decades, most of which have been satellite-based due to the high opacity of the Earth's atmosphere. The first transformational observations were made by the NASA-led Infrared Astronomical Satellite (\IRAS, \citealt{neugebauer:1984}), which observed the sky for ten months in 1983, covering four wavelength bands from 12 to 100$\,\mu$m. \IRAS\ revealed for the first time the intricate nature of thermal dust emission both in the Solar System and the Milky Way.

\IRAS\ was quickly followed by another NASA-led satellite experiment called Cosmic Background Explorer (\COBE, \citealt{boggess92}), which launched in 1989 and carried three instruments. One of these was the Diffuse Infrared Background Experiment (DIRBE; \citealp{hauser1998}), which observed the sky in ten wavelength bands between 1.25 to 240 microns, with the primary aim to characterize the statistical properties of the Cosmic Infrared Background (CIB; \citealp{partridge1967}). The CIB is thermal infrared radiation from both dust particles in distant galaxies and their redshifted starlight, and contains a large fraction of the total energy released in the Universe since the formation of galaxies. After an extended period of detailed analysis, clear CIB signatures were finally discovered in the DIRBE data, but confusion from both zodiacal light from the Solar system and thermal dust emission from the Milky Way made it difficult to fully reach DIRBE's original goal \citep{arendt1998,hauser1998,kelsall1998}. However, the fact that these emission processes are so bright also have ensured that the DIRBE data have had a far-reaching legacy value, and it remains one of the most important data sets for understanding zodiacal light emission to this date. The main goal of the work presented in this paper, and in its companion papers, is to resolve the most important and long-standing problems regarding the DIRBE data, and thereby finally release the full potential of these invaluable measurements.

Following DIRBE, almost a dozen other satellite experiments have targeted the same wavelengths with different angular resolution, sensitivity, and observation strategies, and today there exists a wealth of complementary and ancillary information that was not available between 1990 and 1994, when the official DIRBE analysis was completed. Two examples of such experiments are \AKARI\ \citep{murakami:2007}, which covered six bands from 9 to 180\,$\mu$m, and WISE \citep{wright:2010}, which covered four bands from 3.4 to 22\,$\mu$m. Another important example of a recent and highly complementary experiment is the optical \GAIA\ mission \citep{gaia:2016}, which recently completed a deep survey of stars in the Milky Way \citep{gaia:2018}.

Not only has great observational progress been made in terms of detailed measurements in the infrared regime during the last decades, but major breakthroughs have also been achieved both in terms of understanding the detailed structure of the Milky Way, and in how to analyse complex datasets optimally. One particularly striking example of this is provided by the cosmic microwave background (CMB) community, which through a long series of transformational experiments has revolutionized our understanding of the early universe; only a few examples include ACT \citep{fowler:2007}, BICEP/\textit{Keck} \citep{2014ApJ...792...62B}, \COBE\ \citep{mather:1994}, SPT \citep{carlstrom:2011}, and \WMAP\ \citep{bennett2012}. The current state-of-the-art in terms of full-sky CMB sensitivity is defined by ESA's \Planck\ satellite experiment \citep{planck2016-l01}. However, precisely because of its exquisite signal-to-noise ratio, a long series of key data analysis challenges had to be overcome before its full cosmological potential could be released. Indeed, \Planck\ was the first full-sky CMB experiment for which instrumental and astrophysical uncertainties dominated the total error budget, as opposed to white noise. As such, \Planck\ faced many of the same types of problems that DIRBE had experienced two decades earlier, and massive amounts of algorithm development efforts were spent by hundreds of scientists on resolving these.

One of the main lessons learned from \Planck\ was the importance of joint analysis of multiple complementary experiments, using information from one instrument to break the degeneracies in the others \citep[e.g.,][]{planck2014-a12}. Building on that early ground-breaking work in \Planck, a dedicated effort called \Cosmoglobe\footnote{\url{http://cosmoglobe.uio.no}} was started, with a very simple basic idea: All radio, microwave and infrared experiments measure fundamentally the same sky. However, due to technical limitations, each experiment only measures a relatively small part of the electromagnetic spectrum, and with limited angular resolution and sensitivity. At the same time, the field as a whole is currently at a stage where astrophysical uncertainties play a dominating role in understanding the systematic properties of each experiment. It is therefore natural to expect that better results may be obtained by analyzing multiple complementary experiments together, as opposed to each separately, and in effect use information from one experiment to break the basic degeneracies in another. The long-term goal of the \Cosmoglobe\ effort is therefore to establish one single state-of-the-art model of the astrophysical sky that covers the entire electromagnetic spectrum, using all available experiments at the same time. This is a monumental task, and it will require the combined effort of the entire astrophysical community in order to be successful \citep{bp05}. 

A second important lesson learned from \Planck\ was that, in order to properly mitigate all dominant systematic effects, it was no longer possible to consider each source of systematic uncertainty in isolation. Rather, it was necessary to perform a global integrated analysis in which all parameters are optimized simultaneously at the level of time-ordered data, whether they happen to be of instrumental or astrophysical origin. Two pioneering efforts in this direction were the \sroll\ \citep{sroll2} and \npipe\ \citep{npipe} data analysis pipelines, both of which were developed within the official \Planck\ consortium, and eventually formed the algorithmic basis for the \Planck\ PR3 \citep{planck2016-l01} and PR4 \citep{npipe} data releases, respectively. In particular, both \sroll\ and \npipe\ integrated knowledge about the astrophysical sky directly in their instrument calibration and mapmaking steps, even though neither actually fitted the corresponding astrophysical parameters themselves  during the low-level processing.

The first pipeline to perform true integrated global analysis of \Planck\ data was implemented in a computer code called \commanderthree\ \citep{bp03} by the \textsc{BeyondPlanck} collaboration \citep{bp01}. This extended earlier work on Bayesian component separation that was performed within the \Planck\ collaboration \citep{planck2014-a12}, and was implemented in terms of an end-to-end Bayesian Monte Carlo Gibbs sampler in which an explicit parametric data model was fitted to raw uncalibrated time-ordered data (TOD). As a result of this integrated analysis, a number of long-standing problems regarding the \Planck\ LFI data \citep{planck2016-l02} were resolved, in particular with respect to gain calibration, and the full LFI data set was now for the first time finally available for cosmological analysis \citep{bp10,bp11,bp12}.

This line of work was subsequently generalized by \citet{watts2023_dr1} to perform joint end-to-end Bayesian analysis of both the \WMAP\ and \Planck\ LFI data simultaneously. This turned out to be very effective, and the introduction of LFI measurements effectively resolved a number of long-standing calibration issues in the \WMAP\ data that never could be resolved with \WMAP\ data alone. The products from this analysis were released in March 2022 as ``\Cosmoglobe\ Data Release 1 (DR1)'', and defines today the state-of-the-art in terms of both \Planck\ LFI and \WMAP\ sky maps.

The current paper is the first of a series of papers in which we perform a similar analysis for the \COBE-DIRBE data, collectively referred to as \cosmoglobe\ Data Release~2 (DR2). This work is a major step forward in the \Cosmoglobe\ program by expanding the modelled frequency range by three orders of magnitude, and it is a first step towards merging the microwave and infrared fields into one joint effort. The reasons for considering \COBE-DIRBE in this first step, as opposed to \AKARI, \IRAS, or WISE, are two-fold. First and foremost, DIRBE has excellent systematics properties, both in terms of absolute calibration and zero-level determination, thermal stability, and in terms of a highly interconnected scanning strategy. At the same time, both its data volume and angular resolution are relatively modest, which  makes the computational load and debugging cycle very manageable. Overall, DIRBE is an ideal dataset for generalizing the previous CMB-oriented model and computer code into the infrared regime.

At the same time, the fundamental challenges faced by DIRBE are very similar to those faced by any other infrared experiment. In particular, the single most challenging aspect is the zodiacal light (ZL) emission. This is thermal emission and scattered sunlight from interplanetary dust (IPD) grains. The main difficulty when dealing with zodiacal emission contamination in infrared data is that the observed emission is highly dependent on the position of the observer, and as such, it cannot be modeled like a static foreground, as for instance Galactic foregrounds are treated in the CMB community. Rather, the state-of-the-art method to remove zodiacal emission from timestreams today is to use a three-dimensional parametric interplanetary dust model which describes the distribution of interplanetary dust within the solar system, and perform line-of-sight integration for every single time step. The IPD model most widely used today is the so-called K98 model \citep{kelsall1998} produced by the DIRBE team, or variants thereof \citep[e.g.,][]{planck2013-pip88}. In a companion paper, \citet{CG02_02} present a major step forward in terms of ZL modelling for the DIRBE experiment, as a key component of the current \cosmoglobe\ analysis. This progress is enabled by three main components. First, the usage of external data from \Planck, WISE, and \GAIA\ breaks key degeneracies between the ZL and the Galactic parameters. Second, fitting all parameters jointly with a modern Monte Carlo sampler allows the remaining degeneracies to be explored more efficiently than before. Third and finally, the current analysis characterizes and mitigates a source of excess radiation observed in the mid-infrared DIRBE channels that appears static in solar-centric coordinates. This radiation was noted already by \citet{leinert:1998}, but no corrections have until now been implemented and applied to the DIRBE data. The net result is a greatly improved ZL model that should be of great utility to the entire infrared community.

These improvements also lead to better cosmological and astrophysical interpretation with the DIRBE data. For example, as part of the current data release \citet{CG02_03} derive improved constraints on the CIB monopole spectrum with DIRBE data, while \citet{CG02_05} present a new three-component model of thermal dust emission in the Milky Way that will be of great interest for the CMB community in the search for primordial gravitational waves. Similarly, \citet{CG02_04} construct a new starlight model for DIRBE by combining WISE and \GAIA\ data, that allow for robust modeling of the wavelength channels between 1.25 and 25$\,\mu$m. Additionally, \citet{CG02_06} derive a full-sky map of ionized carbon (\CII) by combining spatial information in the DIRBE 140\,$\mu$m channel with spectral measurements from the \COBE-FIRAS instrument. While many issues still remain to be solved even after the current analysis, we argue that the products presented in the following redefines the standard for full infrared sky modelling, and the most of the methods described in the following are likely to be of direct use for the wide range of other infrared experiments, including \AKARI, \IRAS, and WISE. All products and computer codes are made publicly available\footnote{\url{https://github.com/Cosmoglobe/}} under an Open Source license.

The rest of the paper is organized as follows. In Sect.~\ref{sec:global_modelling} we review the \cosmoglobe\ data model and algorithms, and discuss the extensions needed for DIRBE analysis. In Sect.~\ref{sec:dirbe} we give an overview of the DIRBE instrument and data, as well as any preprocessing and data selection we apply to these, and in Sect.~\ref{sec:ancillary} we summarize the ancillary data sets used in the current processing. The actual results derived through this analysis are summarized in the next four sections. Section.~\ref{sec:chains} discusses the basic Markov chains produced by the algorithm in terms of burn-in and convergence, while Sect.~\ref{sec:noise_gof} focuses on instrumental noise estimation and overall goodness-of-fit. In Sect.~\ref{sec:excess} we provide the first systematic characterization of excess radiation for all DIRBE channels, and in Sect.~\ref{sec:maps} we present and characterize the new \cosmoglobe\ DR2 ZSMA maps. Finally, we conclude and discuss avenues for future work in Sect.~\ref{sec:conclusions}. 

\section{Global Bayesian modelling of the infrared sky}
\label{sec:global_modelling}

The use of Bayesian sampling methods have become widespread in the CMB
community \citep[e.g.,][]{cosmomc,dunkley2009b,handley:2015,planck2014-a12,millea:2019,planck2016-l06,Torrado:2020dgo,bp01,watts2023_dr1} during the last few decades for at least two important
reasons. First, for any analysis task that may be phrased in terms of
a classical parameter estimation problem with measured data $\dv$ and a
model with some set of unknown parameters $\omega$, the posterior
distribution $P(\omega\mid\dv)$ is a complete summary of the
information about $\omega$ contained in the current data, both in
terms of best-fit point estimates and corresponding
uncertainties. Second, both due to the innovation of a wide range of
efficient Monte Carlo sampling methods and the exponential growth of
computing power that took place until very recently, far more complex
models can be mapped out today than was possible only one or two
decades ago. As a particularly relevant case in point for the current
paper is \commander\ \citep{eriksen:2004,seljebotn:2019,bp03}, which is a Gibbs sampler designed to perform
end-to-end analysis with time-ordered data. While the primary
motivation for developing this machinery until today has been
CMB-oriented applications, we show in the following that the same
framework is also very well suited for analysis of observations in the
infrared regime, and, indeed, that it may be used to construct one
global model that includes both microwave and infrared wavelengths.

\subsection{Data model and posterior distribution}
\label{sec:datamodel}

The first step in any parametric Bayesian analysis is simply to write
down a model for the data in question. The quality of the
final results depends sensitively on the accuracy and completeness of
this model, which must be monitored through detailed goodness-of-fit
statistics, typically in the form of residual and $\chi^2$
measures. In practice, an initial model is typically established based
on a pre-existing knowledge about both the astrophysical sky and
instrument in question, and the model is then gradually refined until
the residuals are consistent with instrumental noise. The model
described in this section is the product of such a process
that has involved hundreds of trial runs, starting from a model very
similar to that described by the official DIRBE and \Planck\ teams,
but then gradually generalized with new parameters. In particular, the
current analysis follows closely in the footsteps of
\BP\ \citep{bp01} and \cosmoglobe\ DR1 \citep{watts2023_dr1}, which
implemented the first version of this algorithm, and applied it to
\Planck\ LFI and \WMAP, respectively. We refer the interested reader
to those papers (and references therein) for complete algorithmic
details.

As described in Sect.~\ref{sec:dirbe}, we will in the current analysis
focus on the so-called DIRBE Calibrated Individual Observations
(CIOs). Ideally, the optimal approach would in principle be to start from raw uncalibrated TOD, but those are not
publicly available. In addition, the CIO are easier to work with,
since they have been cleaned from low-level instrumental effects. On
the other hand, we note that this immediately implies that there are
important degrees of freedom, in particular with respect to gain and
zero-level determination, that rely directly on the official analysis,
and that may need to be revisited at a later stage. On the other hand,
the main residuals that emerge at the end of the current analysis
still appear to be dominated by astrophysical confusion rather than
gain errors, and moving on to uncalibrated TOD is therefore not yet a
top priority. We will in the following refer to the DIRBE CIOs simply
as ``TOD''.

\subsubsection{TOD model}

We adopt the following high-level parametric data model for the DIRBE
TOD,
\begin{align}
	\label{eq:model}
	\dv &=\G\P\B\sum_{c=1}^{n_{\mathrm{comp}}}\M_c\a_c+\s_{\mathrm{zodi}} +
          \s_{\mathrm{static}} + \n_\mathrm{corr} + \n_\mathrm w\\
        &\equiv \s_{\mathrm{tot}} + \n_{\mathrm{w}},
\end{align}
where $\dv$ denotes a stacked vector of all DIRBE TOD for all
frequency bands; $\G$ is an $n_{\mathrm{tod}}\times
n_{\mathrm{tod}}$ diagonal matrix with an overall constant gain
calibration factor per frequency channel; $\P$ denotes a satellite
pointing matrix, which we define in Galactic coordinates; $\B$ denotes
an instrumental beam (or point spread function) convolution operator;
the sum runs over $n_{\mathrm{comp}}$ astrophysical components, each
with a free amplitude $\a_c$ at some reference frequency and a mixing
matrix $\M_c$ which defines the effective scaling from the reference
frequency to an observed frequency for each component, taking into account the
bandpass of each detector; $\s_{\mathrm{zodi}}$ is a model of zodiacal
light emission from components that appear time-variable as seen from
Earth (e.g., the zodiacal cloud and asteroidal bands);
$\s_{\mathrm{static}}$ is an excess signal that appears stationary
with respect to the Earth-Sun system, discussed further in Sect.~\ref{sec:excess}; $\n_{\mathrm{corr}}$ is 
correlated instrumental noise (which for now is only fitted for the
lowest DIRBE frequency channel); and $\n_{\mathrm{w}}$ denotes white
instrumental noise. We also define $\s_{\mathrm{tot}}$ to be the sum
of all terms in the data model except for the white noise.

As we work with calibrated TOD, we set $\G=\tens{I}$ for now, but
note that this effective prior should be relaxed in future work, for
instance by using \COBE-FIRAS data as a calibration source in the
overlap frequency range between FIRAS and DIRBE. Similarly, both the
pointing $\P$ and the beam operator $\B$ are provided by the DIRBE
team, and we do not account for any uncertainties in these. However,
we do note that the DIRBE beams have an intrinsically square shape,
while our current beam convolution implementation only supports
azimuthally symmetric beams. This will necessarily lead to a residual
that should ideally be accounted for through full beam integration,
for instance using a \texttt{conviqt}-style algorithm
\citep{prezeau2010,keihanen2012}; this is left for future
work. Similar remarks apply to bandpass definitions as well; for now,
we neglect the uncertainty in the bandpass profiles provided by the
DIRBE team. For further details regarding the pointing, beam and
bandpasses, see Sect.~\ref{sec:dirbe}.

The sky model is described in detail by \citet{CG02_04} and \citet{CG02_05}, and
reviewed briefly in Sect.~\ref{sec:skymodel}. We define the set of all
linear sky component amplitude parameters as $\a_{\mathrm{sky}}$ and
the set of all spectral parameters as $\beta_{\mathrm{sky}}$ in the
following.

Our model for zodiacal light emission, $\s_{\mathrm{zodi}}$, is
described by \citet{CG02_02}, and the overall framework follows
closely that introduced by \citet{kelsall1998} (denoted ``K98'' in the
following) for the original DIRBE analysis. Specifically, we fit a
limited number of shape parameters per interplanetary dust component,
such as a smooth cloud and asteroidal bands, in addition to linear
emissivity and albedo parameters for each frequency channel. In total,
there are 64 free parameters in this model, and these are collectively
denoted $\zeta_{\mathrm{z}}$.

The term denoted $\s_{\mathrm{static}}$ has not been included in
previous DIRBE analyses, but is rather an important novel feature
presented in the current paper. We will return to its physical
interpretation in Sect.~\ref{sec:excess}, but note for now that it
models excess radiation in the DIRBE channels between 4.9 and
60$\,\mu$m not accounted for in the K98 model. In practice, this is
implemented in terms of a pixelized map, $\a_{\mathrm{static}}$, in
solar-centric coordinates, such that $\s_{\mathrm{static},\nu} =
\P_{\mathrm{sol},\nu}\a_{\mathrm{static},\nu}$, where
$\P_{\mathrm{sol},\nu}$ is the pointing matrix rotated into a
coordinate system where the Sun is always at coordinates
$(l,b)=(0^{\circ},0^{\circ})$, and the Ecliptic plane is aligned with
the equator. The amplitude map, $\a_{\mathrm{static},\nu}$, is fitted
independently for each frequency channel.

Next, we assume that the instrumental noise is piecewise stationary,
and we model it with an uncorrelated zero-mean Gaussian distribution
with a free standard deviation per sample, $\sigma_{\mathrm{n}}$ for
all channels except 240\,$\mu$m. The stationarity period is assumed to
be 24~hours, and the data are correspondingly processed in segments of
this length. For the 240\,$\mu$m channel we additionally include a
correlated noise term.  We assume that the time-domain noise power
spectrum of this component may be described by a standard $1/f$
profile of the form $P(f) = \sigma_{\mathrm{n}}^2
(1+(f/f_{\mathrm{knee}})^{\alpha})$, where the slope $\alpha$ and knee
frequency $f_{\mathrm{knee}}$ are fitted independently in each data
segment. Ideally, we would like to include this component in all
frequencies. However, we find that the current sky model is not yet
a sufficiently good fit at any of the other channels. In
total, we denote the sum of all noise parameters by
$\xi_{\mathrm{n}}$.

\subsubsection{Sky model}
\label{sec:skymodel}

The sky signal defined implicitly by the sum in Eq.~\eqref{eq:model}
is defined by \citet{CG02_04} and \citet{CG02_05}, and reads as follows in units
of brightness temperature and frequency,\footnote{Due to its
CMB-oriented origin, \commander\ uses brightness temperature and
frequency units for internal calculations, rather than flux density
and wavelength units which would be more natural for DIRBE. This has,
however, no actual effect on the final results, but only requires
appropriate unit conversions to be applied during input and output
operations. }
\begin{alignat}{4}
  \sum_{c=1}^{n_{\mathrm{comp}}} \M_c \a_c  = \,
  &\M_{\mathrm{mbb}}(\bcold,\Tcold, q_i;\nuzerocold,\{\Delta\nu_i\})\vec{a}_{\mathrm{cold}}
  && \textrm{(Cold dust)}\nonumber\\ %
  + &\M_{\mathrm{mbb}}(\bhot,\Thot, q_i;\nuzerohot,\{\Delta\nu_i\})
  \vec{a}_{\mathrm{hot}} && \textrm{(Hot dust)}\nonumber \\
  + &\M_{\mathrm{mbb}}(\bnear,\Tnear,q_i;\nuzeronear,\{\Delta\nu_i\}) \t_{\mathrm{near}}
  a_{\nu} && \textrm{(Nearby dust)} \nonumber \\
  + &\left(\frac{\nuzeroff}{\nu}\right)^2
  \frac{g_{\mathrm{ff}}(\nu;\Te) }{g_{\mathrm{ff}}(\nuzeroff;\Te)}
  \vec{t}_{\mathrm{ff}} && \textrm{(Free-free)} \nonumber\\
  + &\delta(\nu-\nu_{0,\mathrm{CO}}^i) \t_{\mathrm{CO}}
  h^{\mathrm{CO}}_{\nu,i} && \textrm{(CO)}\nonumber\\
	+ &\delta(\nu-\nu_{0,\CII}) \a_{\CII}
  h^{\CII}_{\nu} && (\CII) \nonumber\\
  + &U_{\mathrm{mJy}} \sum_{j=1}^{n_{\mathrm{s}}}
  f_{\GAIA,j} a_{\mathrm{s},j}, &\quad&
  \textrm{(Bright stars)} \nonumber\\
  + &U_{\mathrm{mJy}} \t_{\gaia,\mathrm{fs}}\, a_{\mathrm{fs},\nu}, &\quad&
  \textrm{(Faint stars)} \nonumber\\  
    + &U_{\mathrm{mJy}} \sum_{j=1}^{n_{\mathrm{e}}}
  M_{\mathrm{mbb}}(\beta_{\mathrm{e},j},
  T_{\mathrm{e},j})
  a_{\mathrm{e},j} && \textrm{(FIR sources)}\nonumber\\
  + &m_{\nu} && \textrm{(Monopole)}. \nonumber
\end{alignat}
In this expression, we have defined a function of the form
\begin{equation}
  M_{\mathrm{mbb}}(\beta, T, q_i;\nu_0, \{\Delta\nu_i\}) =
    \begin{cases}
      q_i & \nu \in \Delta\nu_i\\
      \left(\frac{\nu}{\nu_0}\right)^{\beta+1}
  \frac{\e^{h\nu_0/\kB T}-1}{\e^{h\nu/\kB T}-1} & \nu \notin \Delta\nu_i,
    \end{cases}       
\end{equation}
which represents a generalized modified blackbody function. However,
in addition to the usual emissivity index and temperature, $\beta$ and
$T$, this function takes a set of constant values, $q_i$, and
corresponding frequency ranges $\Delta\nu_i$. If the requested
frequency happens to lie in any one of $\Delta\nu_i$, then $q_i$ is
returned; otherwise the default is to return the standard modified
blackbody spectrum. Another point to note in the above equation is
that if a given amplitude is denoted by $\a$, then it is fitted freely
to the current data; if it is denoted by $\t$, it is fixed to an
external template.

As indicated by the above sky model, we fit a novel three-component
generalized modified blackbody model to account for thermal dust
emission across the combined DIRBE and \Planck\ HFI frequency range;
for full details, see \citet{CG02_05}. The three components correspond
to cold dust, hot dust, and nearby dust emission, respectively. All
three are modeled with spatially constant spectral parameters, and
only the cold and hot component amplitudes are fitted pixel-by-pixel;
the amplitude of the nearby component is fixed to the \GAIA-based dust
extinction template covering distances up to 1.25\,kpc produced by
\citet{edenhofer:2024}.  As such, this dust model has in fact only two
degrees of freedom per pixel, in addition to fewer than 30 spatially
constant SED parameters. This is an extremely economical model of
thermal dust emission, considering the fact that it describes the
entire combined frequency range covered by both \Planck\ HFI and
DIRBE, from 100\,GHz to 1\,$\mu$m.

To account for free-free emission, we adopt the model presented by
\citet{planck2014-a11}, both in terms of spatial distribution and
spectrum. The SED of this component is defined by the Gaunt factor,
$g_{\mathrm{ff}}(\nu; T_e)$ \citep{dickinson:2003,draine:2011}, which
corresponds to a shift in the spectral index of about $-0.14$ in the
CMB frequency range; however, at the very high DIRBE frequencies of up
to 300\,THz, it takes on significantly more extreme values, and this
should at least in principle provide greater sensitivity to the
electron temperature, $T_e$. For now, however, we adopt the $T_e$
distribution presented by \citet{planck2014-a11} as given, and will
rather attempt to actually fit $T_e$ in future work.

The next component corresponds CO line emission, produced by
transitions between two quantized angular momentum eigenstates in the
CO molecule. The resulting emission forms effectively a ladder in
frequency space in multiples of 115.27\,GHz, and \COBE-FIRAS
identified emission all the way up to 922\,GHz, albeit with low
sensitivity and angular resolution. In contrast, \Planck\ produced
high-resolution maps with high sensitivity of the $J$=1$\leftarrow$0,
2$\leftarrow$1, and 3$\leftarrow$2 transitions, which contributed to
the HFI\,100, 217, and 353\,GHz frequency maps, but was unable to
identify CO emission at higher frequencies due to strong thermal dust
emission. In the current work, we adopt the \citet{dame:2001} CO
$J$=1$\leftarrow$0 map as a fixed tracer for all variations of CO
emission, and we also adopt the line ratios, $h_{\nu,i}$, presented by
\citet{planck2014-a12} for the 100, 217, and 353\,GHz \Planck\
channels. As far as the current analysis is concerned, the CO
component is thus a fixed correction applied to the relevant
\Planck\ bands.

Similarly, the fourth component corresponds to \CII\ line emission,
which has a rest frequency of 1900\,GHz. As such, it only affects the
DIRBE 140\,$\mu$m map in our dataset, in addition to selected FIRAS
bands. In this case, we fit for a free amplitude per pixel with DIRBE
140\,$\mu$m, using the general sky model determined by near-by
channels to remove thermal dust emission, and then exploit the near-by
FIRAS channels to monitor the overall reconstruction
quality. The result is a novel full-sky \CII\ map with an angular resolution of
about $1^{\circ}$ FWHM.

The fifth and sixth components correspond to starlight emission, which
is relevant in the frequency range between 1 and 25\,$\mu$m; for full
details regarding this model, see \citet{CG02_04}. For these, we first
extract a baseline star catalog by thresholding the AllWISE
3.5\,$\mu$m catalog at magnitude 8, resulting in a set of about
783\,000 sources.\footnote{We have tried different thresholds, and
found that magnitude 8 provides a good compromise; magnitude 6 results
in obviously missing sources, while magnitude 10 leads to too many
unconstrained sources.} For each of these, we search the \GAIA\ DR2
catalog, and if this returns a positive star identification within a
radius of 20~arcsec, we record the object, and store the best-fit
temperature $T_\mathrm{s}$, surface gravity $g$, and metallicity
$[\mathrm{M}/\mathrm{H}]$ as determined by \GAIA. These are then used
to estimate the best-fit SED using the PHOENIX spectrum grid
\citep{husser:2013}, which is convolved with the bandpass and beam
profile of each DIRBE channel. The resulting bandpass- and
beam-convolved SED is denoted $f_{\GAIA,j}$, which is unique for each
star. We then fit one overall amplitude for each star to the four
highest DIRBE frequency bands between 1.25 and 4.9\,$\mu$m; we also
account for star emission in the 12 and 25\,$\mu$m bands, but these
bands are not used for the actual fit. A total of 717\,000 stars are
included in the fit of individual stars; these are denoted as ``bright
stars'' in the sky model.

The remaining 66\,000 bright WISE sources that do not have a
\GAIA\ counterpart are fitted with a standard modified blackbody
spectrum across the same frequency range, as described by the line
marked by ``FIR sources'' in the sky model. Algorithmically speaking,
this component is identical to the bright star component, except for
the parametric form of the SED.

The AllWISE catalog itself contains a total of about 747 million
sources, making it impossible to fit all of these with DIRBE without
introducing massive degeneracies. We therefore instead co-add the rest
of the AllWISE sources into a diffuse background map of faint sources
under the assumption that their mean SED is equal to the average of
the bright sources that actually are fitted as part of the
algorithm. Together, these three source components comprise an
unprecedented deep model for compact objects in DIRBE that has only
become possible due to WISE and \GAIA.

The tenth and final component is simply a monopole per frequency. For
DIRBE and FIRAS, this should ideally describe the CIB spectrum, but it
is also sensitive to zodiacal light and Galactic residuals. DIRBE was designed to have negligible straylight contamination, while internal thermal sources in FIRAS were explicitly measured by onboard thermistorse and subtracted before mapmaking. For
\Planck, the monopoles account for the arbitrary zero-levels present
in the \Planck\ PR4 maps. 

Finally, it is worth noting that there is no CMB component present in
the current sky model, even though it applies to the \Planck\ HFI as
well as DIRBE. Since the main focus in the current work is DIRBE, we
have chosen to pre-subtract any CMB component (including the solar CMB
dipole and relativistic quadrupole corrections) from each frequency
map. For this, we use the PR3 \commander\ CMB maps \citep{planck2016-l04}.
Similarly, we neglect the impact of synchrotron emission,
anomalous microwave emission, the Sunyaev-Zeldovich effect, and other
smaller contributions; these will instead be included in a future
analysis that also has HFI as a main science target.

\subsubsection{Posterior distribution}

In principle, all quantities on the right-hand side of
Eq.~\eqref{eq:model} are associated with free parameters and
uncertainties that should be estimated from the data, whether they are
of astrophysical or instrumental origin. We define the full set of
free parameters as $\omega = \{\G,\xi_{\mathrm{n}},
\beta_{\mathrm{sky}},\a_{\mathrm{sky}},\zeta_{\mathrm{z}},\a_{\mathrm{static}}\}$,
and our goal is now to derive an explicit expression for the global
posterior distribution, $P(\omega\mid\dv)$. This is most easily done
through Bayes' theorem,
\begin{equation}
P(\omega\mid\dv) = \frac{P(\dv\mid\omega) P(\omega)}{P(\dv)} \propto
\mathcal{L}(\omega) P(\omega).
\end{equation}
In this expression, $\mathcal{L}(\omega) \equiv  P(\dv\mid\omega)$ is
called the likelihood, $P(\omega)$ is called the prior; $P(\dv)$, called the evidence, is a
normalization constant that does not depend on $\omega$, which we
neglect in this work.

Under the common assumption that the white noise component is Gaussian
distributed with zero mean and some covariance matrix,
$\N_{\mathrm{w}}$, we can write the log-likelihood in the usual
explicit form,
\begin{equation}
-2\ln\mathcal{L}(\omega) = (\dv-\s^{\mathrm{tot}}(\omega))^t
	\N_{\mathrm{w}}^{-1}(\dv-\s^{\mathrm{tot}}(\omega))+\ln|\N_{\mathrm w}|,
\end{equation}
once again up to an irrelevant normalization constant, and we have for
notational compactness suppressed the fact that also $\N_{\mathrm{w}}$ has free
parameters.

Regarding $P(\omega)$, we will in this analysis operate primarily with
three types of priors. First, for zodiacal light parameters we adopt
uniform priors between pre-defined limits, to avoid the algorithms to
diverge into pathological solutions. Second, for
astrophysical spectral parameters, such as temperature and spectral
indices, we adopt products of uniform priors with broad limits and
Gaussian priors with spectral parameters informed by \Planck\ where
applicable. Finally, for a few select astrophysical components, for
instance free-free and carbon monoxide line emission, we adopt
existing spatial templates as delta function priors on the spatial
morphology, and only fit overall free amplitudes in the current
analysis. For full details regarding the use of priors for a given
component, we refer the interested reader to \citet{CG02_04} and \citet{CG02_05}.

\subsection{Gibbs sampling with \commanderthree}

As described in Sect.~\ref{sec:datamodel}, the current data model
contains millions of strongly correlated parameters, ranging from
affecting individual time samples (such as the correlated noise,
$\n_{\mathrm{corr}}$) to describing the astrophysical signal in the
form of a pixelized map (such as the cold dust amplitude
$\a_{\mathrm{c}}$) or a catalog (such as the bright star amplitude
$\a_{\mathrm{s}}$), to simultaneously affecting essentially every
single data point, such as the zodiacal light shape
parameters. Mapping out this distribution is therefore highly
non-trivial.

\begin{figure*}
	\centering
	\includegraphics[width=\linewidth]{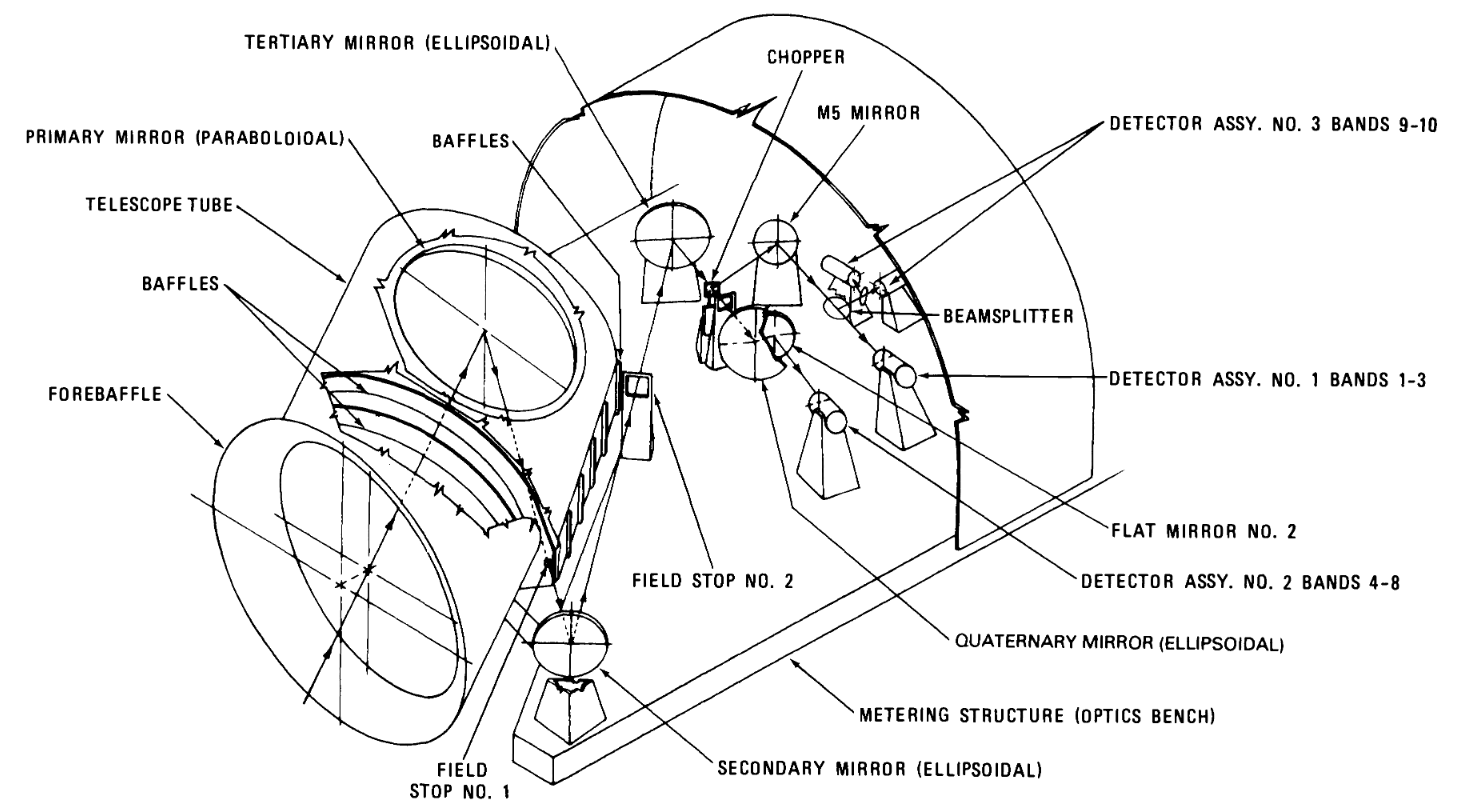}
	\caption{DIRBE optics module. The optical design includes two field stops, one of which is square, to reduce straylight contamination. Reproduced from \cite{magner87}.
	}
	\label{fig:optics_model}
\end{figure*}

So far, the only algorithm that has been demonstrated in practice to
work well on such complex end-to-end analysis problems
\citep{bp01,watts2023_dr1} is Gibbs sampling
\citep[e.g.,][]{geman:1984}, which is a special case of the
Metropolis-Hastings sampling algorithm. The defining
feature of this algorithm is that it loops over all free parameters
(which may be divided into groups), and draws a sample from each
conditional distribution. Returning to the defining data model in
Eq.~\eqref{eq:model}, and recalling that the set of
free parameters is $\omega = \{\xi_{\mathrm{n}},
\beta_{\mathrm{sky}},\a_{\mathrm{sky}},\zeta_{\mathrm{z}},\a_{\mathrm{static}}\}$,
we can immediately write down a corresponding Gibbs sampling chain of
the following form:
\begin{alignat}{10}
\xi_{\mathrm{n}} &\,\leftarrow P(\xi_{\mathrm{n}}&\,\mid &\,\dv,&\, &\,\phantom{\xi_n} &
\,\beta_{\mathrm{sky}}& \,\a_{\mathrm{sky}}, &\,\zeta_{\mathrm{z}},
&\,\a_{\mathrm{static}}&)\label{eq:gibbs_n}\\
\beta_{\mathrm{sky}} &\,\leftarrow P(\beta_{\mathrm{sky}}&\,\mid &\,\dv,&\, &\,\xi_n, &
\,\phantom{\beta_{\mathrm{sky}}}& \,\a_{\mathrm{sky}}, &\,\zeta_{\mathrm{z}}, &\,\a_{\mathrm{static}}&)\\
\a_{\mathrm{sky}} &\,\leftarrow P(\a_{\mathrm{sky}}&\,\mid &\,\dv,&\, &\,\xi_n, &
\,\beta_{\mathrm{sky}},& \,\phantom{\a_{\mathrm{sky}},}
&\,\zeta_{\mathrm{z}}, &\,\a_{\mathrm{static}}&)\\
\zeta_{\mathrm{z}} &\,\leftarrow P(\zeta_{\mathrm{z}}&\,\mid &\,\dv,&\, &\,\xi_n, &
\,\beta_{\mathrm{sky}},& \,\a_{\mathrm{sky}},
&\,\phantom{\zeta_{\mathrm{z}},} &\,\a_{\mathrm{static}}&)\label{eq:gibbs_zodi}\\
\a_{\mathrm{static}} &\,\leftarrow P(\a_{\mathrm{static}}&\,\mid &\,\dv,&\, &\,\xi_n, &
\,\beta_{\mathrm{sky}},& \,\a_{\mathrm{sky}}, &\,\zeta_{\mathrm{z}} &\,\phantom{\a_{\mathrm{static}}}&).\label{eq:gibbs_static}
\end{alignat}
Here, the symbol $\leftarrow$ indicates drawing a sample from the
conditional distribution on the right-hand side. However, we note that
our codes are also designed to perform maximum-posterior (or
likelihood) analysis, in which case we maximize the probability
distribution instead of drawing a sample from it.

The current state-of-the-art implementation in terms of CMB Gibbs sampling
is \commander\ \citep{Eriksen:2004ss}, which was used
extensively for the \Planck\ analysis. However, during the
\Planck\ analysis this code only supported high-level component
separation operations, and the low-level time-domain support was added
after the official end of \Planck. The first incarnation of this
end-to-end framework is called \commanderthree\ \citep{bp03}, which
was applied to the \Planck\ LFI data by the \BP\ collaboration
\citep{bp01}. Shortly after, a slightly extended version was applied
to the combination of \Planck\ LFI and \WMAP\ by
\citet{watts2023_dr1}, and the results from this analysis formed the
basis for \Cosmoglobe\ DR1. 

The existing \commanderthree\ implementation used for \BP\ and
\Cosmoglobe\ DR1 already provides sampling steps for most of the above
conditional distributions, and these can be reused with minimial
modifications. In particular, \citet{bp07} describe how to sample
instrumental gain; \citet{bp06} describe how to estimate instrumental
noise parameters, and \citet{bp02} discuss how to make optimal maps
with full noise propagation efficiently with Gibbs sampling; finally
\citet{bp13} describe how to sample from intensity foregrounds
posteriors. 

\begin{table}[t]
  \begingroup
  \newdimen\tblskip \tblskip=5pt
	\caption{List of celestial body flags.}
  \label{tab:planet_flags}
  \nointerlineskip
  \vskip -3mm
  \footnotesize
  \setbox\tablebox=\vbox{
    \newdimen\digitwidth
    \setbox0=\hbox{\rm 0}
    \digitwidth=\wd0
    \catcode`*=\active
    \def*{\kern\digitwidth}
    \newdimen\signwidth
    \setbox0=\hbox{-}
    \signwidth=\wd0
    \catcode`!=\active
    \def!{\kern\signwidth}
 \halign{
      \hbox to 1.5cm{#\leaderfil}\tabskip 1em&
      \hfil#\hfil\tabskip 1em&
      \hfil#\hfil\tabskip 1em&
      \hfil#\hfil\tabskip 1em&
      \hfil#\hfil\tabskip 1em&
      \hfil#\hfil\tabskip 1em&
      \hfil#\hfil\tabskip 1em&
      \hfil#\hfil\tabskip 1em&
      \hfil#\hfil\tabskip 1em&
      \hfil#\hfil\tabskip 1em&
      \hfil#\hfil\tabskip 1em&
      #\tabskip 0em\hfil\cr
    \noalign{\doubleline}
      \omit Object\hfil&
      \omit\hfil Radius ($^\circ$) \hfil\cr
      \noalign{\vskip 4pt\hrule\vskip 4pt}
      Moon****** & 10 \cr
      Mercury*** & *1 \cr
      Venus***** & *2 \cr
      Mars****** & *2 \cr
      Jupiter*** & *2 \cr
      Saturn**** & *1 \cr
      Uranus**** & *1 \cr
      Neptune*** & *1 \cr
      \noalign{\vskip 4pt\hrule\vskip 5pt} } }
  \endPlancktablewide \endgroup
\end{table}

While by far most of the code infrastructure required to process the
DIRBE TOD already exists, several of these steps and models discussed
above require slight modifications in order to work efficiently in a
production environment. In particular, efficient diffuse foreground
sampling for DIRBE is described by \citet{CG02_05} and the novel
starlight model and sampler are described by \citet{CG02_04}.

The ZL sampling step described by Eq.~\eqref{eq:gibbs_zodi}, however,
did not have support in the existing \commander\ implementation until
the current work, and had to be developed from scratch. An early step
towards this goal was described by \citet{san:2022}, who reimplemented
the default DIRBE zodiacal light model (K98; \citealp{kelsall1998}) in
Python. This served as the basis for the code developed here, which
now is a set of native \commander\ modules written in Fortran. The
full details of the new zodiacal light estimation framework, including
a significantly improved best-fit model with respect to K98, is
presented by \citet{CG02_02}.

Similarly, the sampling step for the static component amplitude,
$\a_{\mathrm{static}}$, also had to be developed from scratch for the
current work. However, in constrast to the ZL sampler, which required
a non-trivial amount of coding effort, the algorithm for
$\a_{\mathrm{static}}$ is very straight-forward. Based on the data
model in Eq.~\eqref{eq:model}, we first compute a residual that
removes all components except $\s_{\mathrm{static}}$,
\begin{equation}
  \r_{\mathrm{static}} = \dv - \left[\G\P\B\sum_{c=1}^{n_{\mathrm{comp}}}\M_c\a_c+\s_{\mathrm{zodi}} 
    + \n_\mathrm{corr}\right],
\end{equation}
and we then bin this time-ordered residual in solar-centric coordinates
according to the usual map-making equation for Gaussian noise
\citep[e.g.,][]{tegmark_mapmaking}. Explicitly, the appropriate
Gibbs sample is given by (see, e.g., Appendix~A in \citealp{bp01})
\begin{equation}
(\P_{\mathrm{sol}}^t\N_{\mathrm{wn}}^{-1}\P_{\mathrm{sol}})\a_{\mathrm{static}} 
  = \P_{\mathrm{sol}}^t\N_{\mathrm{wn}}^{-1}\r_{\mathrm{static}} + \P_{\mathrm{sol}}^t\N_{\mathrm{wn}}^{-\frac{1}{2}}\omega,
\end{equation}
where $\omega\sim N(0,1)$ is a vector of standard Gaussian
variates. Since the white noise component by definition is
uncorrelated, this equation may be solved pixel-by-pixel.

\section{Diffuse Infrared Background Experiment}
\label{sec:dirbe}

\begin{figure}
  \centering
  \includegraphics[width=\linewidth]{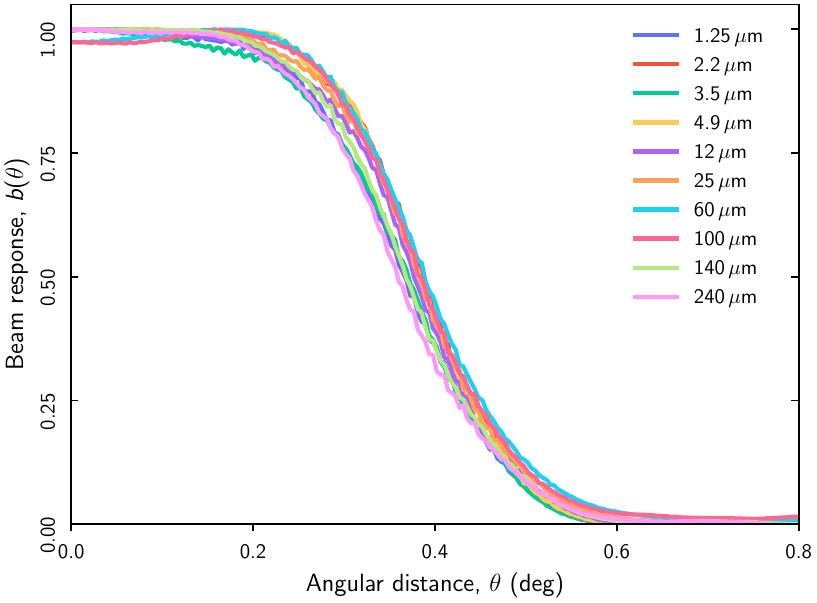}\\
  \includegraphics[width=\linewidth]{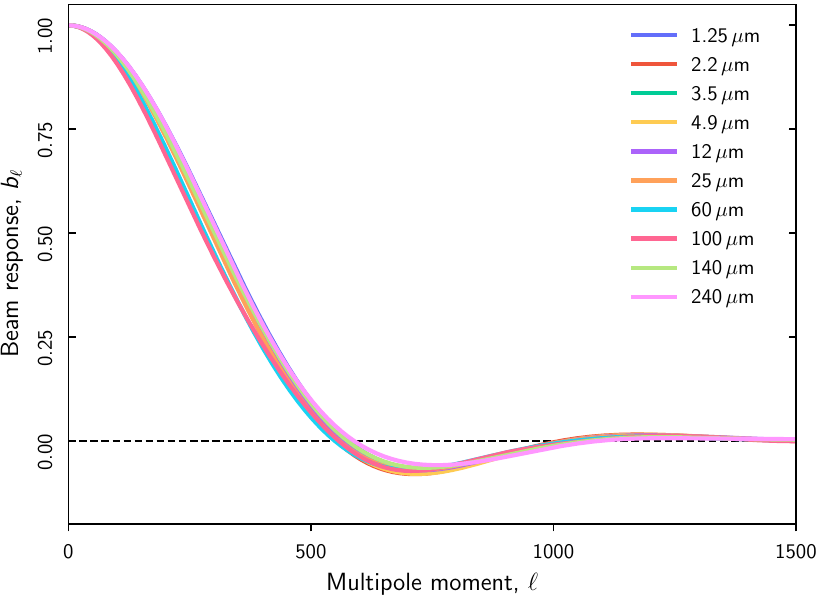}
  \caption{Symmetrized beam response functions for each DIRBE channel, both in real space (\emph{top}) and in harmonic space (\emph{bottom}).}
  \label{fig:beams}
\end{figure}

\subsection{The DIRBE instrument}

The Diffuse InfraRed Background Experiment (DIRBE) was one of three experiments on the Cosmic Background Explorer (\COBE) satellite \citep{boggess92}. DIRBE was designed to characterize the infrared sky from $1\,\mathrm{\mu m}$ to $240\,\mathrm{\mu m}$, with the sensitivity required to characterize thermal dust emission, zodiacal emission, and to detect the CIB \citep{silverberg93}. The DIRBE experiment was limited by its cryogenic requirement, using 600 L of superfluid \element[ ][4]{He}, cooling the instrument to 1.6 K.

\subsection{Pointing, beam and bandpass response}

The \COBE\ satellite followed a Sun-synchronous orbit at 900\,km, orbiting the Earth at a $99^\circ$ inclination every 103 minutes. The spacecraft rotated around its axis at a rate of 0.8 rpm, with the DIRBE optics pointed $30^\circ$ from the spin axis. Due to the orientation changes of the satellite throughout this orbit, DIRBE was able to observe approximately half the sky during the day at solar elongation angles of $64^\circ$--$124^\circ$.
The pointing was determined by interpolating on-board gyroscopic data with the positions of known stars in the short-wavelength bands.

The DIRBE optical design included several design solutions for calibration and stray light reduction. Straylight reduction was prioritized in the design, largely because of the difficulty in distinguishing this systematic effect from a true diffuse background. In particular, there are several straylight stops to reduce sidelobe contamination, mainly in the form of a square beam, as can be seen in Fig.~\ref{fig:optics_model}. In addition to the straylight reduction, DIRBE alternates observations between the sky and an internal calibration source that chops between the two light sources at a rate of 32 Hz.
All bands observe the same $0\fdg7\times0\fdg7$  field simultaneously, with small adjustments of the beam centroids depending on the location of the detectors. The light is divided using beam splitters to split the light into various detector assemblies. Detectors 1--3 were polarization-sensitive, with light parallel and perpendicular to the scan direction being detected. Because of the DIRBE scan strategy, the resulting maps show poor polarization angle coverage across the sky.

\begin{figure}
  \centering
  \includegraphics[width=\linewidth]{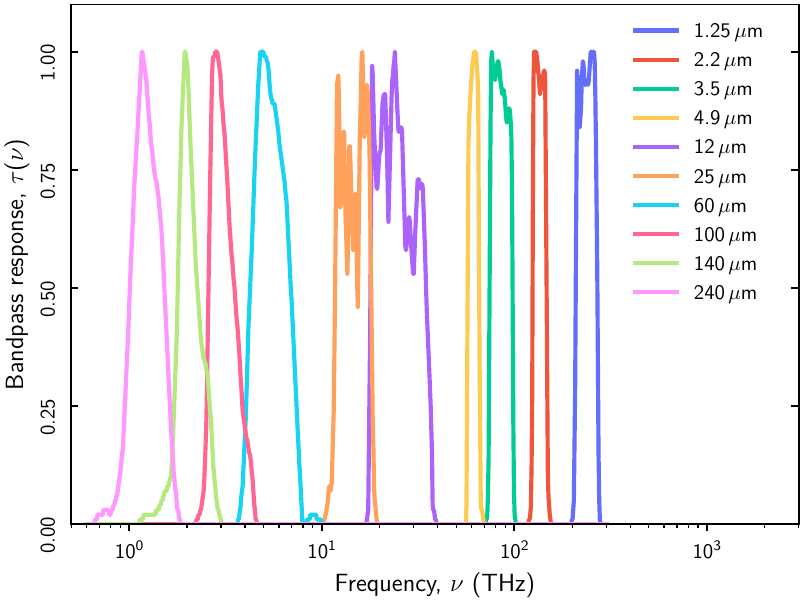}
  \caption{Bandpass response functions for each DIRBE channel, plotted as a function of frequency.}
  \label{fig:bandpass}
\end{figure}

The symmetrized beam, shown in the top panel of Fig.~\ref{fig:beams}, resembles a tophat with a slow falloff. The harmonic-space representation of the beam is therefore reminiscent of a sinc function, with oscillations about zero above $\ell\gtrsim500$. For a band-limited signal to be characterized fully in map space, the 42\arcm\ DIRBE beams must be represented with a 21\arcm\ or smaller pixelization. However, in order to be fully characterized in harmonic space, a requirement for the \commanderthree\ multi-resolution component separation, the pixelization scheme must have support up to $\ell\lesssim1500$. The original DIRBE maps have pixel size of 21\arcm, insufficient for harmonic space analysis.
Neither the original maps in Quadcube\footnote{
	Quadrilateralized Spherical Cube \url{https://lambda.gsfc.nasa.gov/product/cobe/skymap\_info\_new.html}
}
pixelization (resolution 9, pixel size 21\arcm) nor the CADE\footnote{Centre d'Analyse de Données Etendues, \url{http://cade.irap.omp.eu/dokuwiki/doku.php?id=dirbe}} reprojection into HEALPix ($N_\mathrm{side}=256$, pixel size 13\farcm7) have the required support over the full multipole range.

\begin{figure}
  \centering
  \includegraphics[width=\columnwidth]{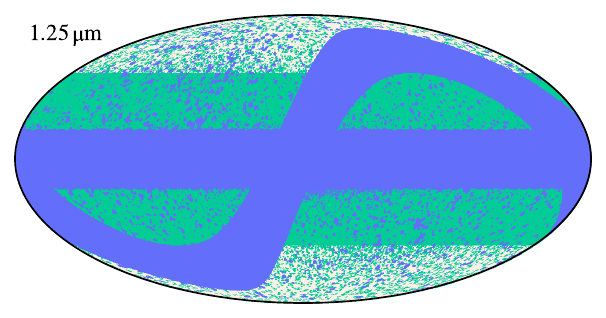}
  \includegraphics[width=\columnwidth]{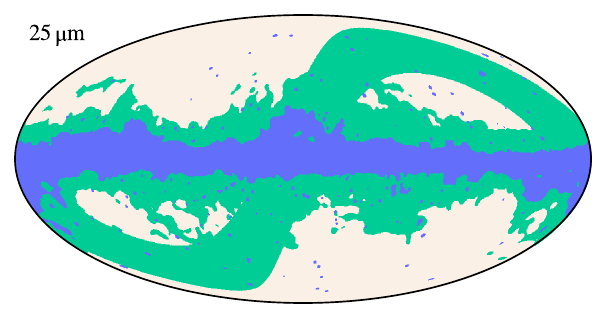}
  \includegraphics[width=\columnwidth]{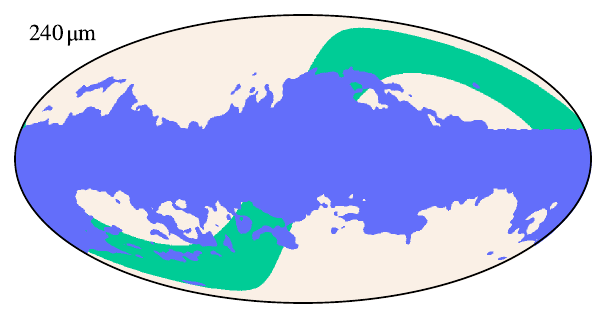}
	\caption{Processing masks used in the analysis. Green regions correspond to the general TOD processing masks, and the blue regions correspond to the zodiacal emission masks. In the $240\,\mathrm{\mu m}$ band, the zodiacal emission mask includes the entirety of the general TOD processing mask.}
  \label{fig:masks}
\end{figure}

The DIRBE central bandpasses, given in terms of central frequency and central wavelength, are given in Table \ref{tab:summary}, and the full bandpass responses as reported in \citet{cobe_exsupp} are shown in Fig.~\ref{fig:bandpass}.
The different detector technology for each band accounts for the different performances and systematics found in each of the bands. \citet{silverberg93} in particular highlight the Ge:Ga photoconductors' (bands 7 and 8) response to ionizing radiation in the South Atlantic Anomaly (SAA) being worse than other detectors, requiring long time for the detectors to return to normal. Similarly, the use of a composite Si bolometer for bands 9 and 10 partially explain the over an order of magnitude increase in noise when compared to adjacent bands.

\subsection{Data selection and masking}
\label{sec:data_selection}

In order to produce maps with full multipole support, we analyze the CIOs directly and convert the pointing into 7\arcm\  $N_\mathrm{side}=512$ pixels from the native resolution 15 Quadcube pixels with 20\arcs\ in the delivered CIOs.\footnote{\url{https://lambda.gsfc.nasa.gov/product/cobe/dirbe\_cio\_data\_get.html}} The conversion from CIO pixel indices to Galactic longitude and latitude is detailed in \citet{cobe_exsupp}, and is reimplemented in the Python preprocessing script \texttt{quadcube}.\footnote{\url{https://github.com/MetinSa/quadcube}}

The delivered CIOs are organized into 285 single-day files, with the datapoints ordered by Quadcube pixel index. The primary processing step was converting the pointing into $N_\mathrm{side}=512$ pixels using \texttt{quadcube}. The data are sampled at 8\,Hz and labeled by time index in seconds since January 1, 1981 00:00 UTC. Because the data are pre-calibrated and bad data are already removed, there are some gaps in the data, which we fill manually with an appropriate flag. Additional flags, such as excess noise, orbit and attitude errors, and presence of the SAA, are additionally extracted. In total, the data are placed in one hdf5 file per band, following the format enumerated in \citet{bp03}.
The planet flags are not present in the CIOs, and are regenerated beforehand. Using the radii as defined in Table~\ref{tab:planet_flags}, we mark data points within the pointing of each pixel. Note that this is not strictly optimal due to the non-circular beam shape, and can be optimized in future analyses.

\section{Ancillary data sets}
\label{sec:ancillary}

As demonstrated by the success of the \bp\ and \cosmoglobe\ projects, the use of complementary datasets with different angular resolution, frequency coverage, and observation strategies, can greatly improve the quality of low-level data processing. In this work, we use \Planck\ High Frequency Instrument, \WISE, \GAIA, and \COBE-FIRAS to better constrain our sky model and characterize the DIRBE data.

\subsection{\Planck\ HFI}

The \Planck\ High Frequency Instrument (HFI; \citealt{planck2016-l03}) observed the sky in six channels from 100\,GHz to 857\,GHz from May 2009--2013, with angular resolution of 10\arcm--\,4\arcm. While the primary purpose of the \Planck\ mission was to characterize fluctuations in the CMB, a large part of its scientific legacy comes from its observations of the far-infrared sky, with robust characterization of the Milky Way \citep{planck2013-XVII,planck2014-a12,planck2016-l03} and of CIB fluctuations \citep{planck2014-a12,planck2013-XVII,lenz2019,mccarthy:2024}.

In addition to its complementary observation strategy, \Planck's frequency coverage has a relatively lower expected amount of zodiacal emission, with a total expected amplitude of $\lesssim1\,\%$ before any subtraction \citep{maris2006c,planck2013-pip88}, as compared to almost 100\,\% of the signal in some DIRBE bands. All delivered \Planck\ maps have had an estimate of the zodiacal emission modeled using the 3D model derived by K98 with varying emissivities per component. While this technically is redundant information that could contaminate this joint analysis, the already low amplitude of zodiacal emission in the HFI maps limits the potential impact of using a technically incorrect zodiacal emission model. A full analysis fitting for zodiacal emission parameters using both HFI and DIRBE will be left for future work.

At the same time, CIB fluctuations with a similar SED to the Milky Way have been detected with high significance in the HFI data, and are directly visible in 353--857\,GHz maps at high Galactic latitudes. Incorrectly modeled, this could bias the Galactic thermal dust model and lead to an incorrect model of the sky in the DIRBE range. In order to avoid this, we remove the GNILC \citep{planck2016-XLVIII} estimate of the CIB from the HFI maps before including them in our analysis.

Since this work is primarily concerned with the DIRBE dataset, the modeling of CMB temperature fluctuations gives an unnecessary degree of freedom to be marginalized over. Therefore, we subtract the \commanderthree\ PR3 CMB temperature estimate from the \Planck\ HFI maps, effectively conditioning the entire Gibbs chain on this CMB estimate.
We use single \Planck\ detector maps to avoid the complication of subtle bandpass mismatches between nearby detectors. In total, we use the 100-1, 217-1, and 353-1 temperature maps and the total 545\,GHz and 857\,GHz maps, all from the PR4 release \citep{npipe}.

\subsection{\GAIA\ and \WISE}

In the near-infrared, most of the sky observed by DIRBE consists of stars and point sources. We therefore use catalogs derived from external datasets as fixed locations of each source, while fitting for the amplitudes in the Gibbs chain. As of this publication, the most complete catalog of stars in the  Milky Way comes from the \Gaia\ mission \citep{gaia:2016,gaia:2018}. In particular, physics models of 100\,million stars are provided in \Gaia\ DR2. Despite \Gaia\ operating between 330--1050\,nm, these models can be used as informative priors for the amplitude of stars at the DIRBE near-infrared bands. With a limiting magnitude of $G\sim20$ across the entire sky, this star catalog gives a complete survey of all stars that are detectable within the DIRBE bands.

Conversely, the \WISE\ satellite \citep{wright:2010} has mapped the sky at 3.4, 4.6, 12, and 22 $\mathrm{\mu m}$, with resolutions of 6\farcs11, 6\farcs4, 6\farcs5, and 12\farcs0. This gives a direct estimate of point source brightness and location, and allows for direct cross-matching with the \GAIA\ DR2 catalog. In order to leverage the \gaia\ data properly, we extract the SED for point sources in both \gaia\ and \WISE\ with $<8$ mag at $3.4\,\mathrm{\mu m}$. These SEDs are then scaled directly per star in the Gibbs chain to correct for absolute calibration differences between the \gaia+\WISE\ catalog and DIRBE. There are a total of 717\,454 stars in both catalogs. There are an additional 66\,217 extragalactic sources that exist in \WISE\ but not in \gaia. These are fit as modified blackbodies per source.

Within the Galactic plane, there are many stars per single DIRBE pixel. In order to avoid degeneracies between individual point sources, we create a map at $N_\mathrm{side}=512$ that includes all \WISE\ point sources that are $>8$ mag at $3.4\,\mathrm{\mu m}$ but have not been identified within the \gaia\ catalog. The SED's as derived by \gaia\ are then averaged over and used to scale the entire template. Within the Gibbs chain, this is sampled with a total scaling parameter, with a fixed map and relative amplitudes between different frequencies. For a full description of the star model, see the work in companion paper \citet{CG02_04}.

\subsection{\COBE-FIRAS}

The \COBE-FIRAS experiment was an absolutely calibrated differential Michelson Fourier transform interferometer that observed the full sky from 68\,GHz--2911\,GHz with 13.6\,GHz frequency resolution \citep{fixsen:1994,mather:1999}. 
FIRAS's primary goal was the characterization of the CMB blackbody spectrum \citep{mather:1994}, which motivated much of the experiment's design choices. In particular, its angular resolution of $7^\circ$ was chosen due to its focus on monopole characterization, while still allowing for foreground mitigation. Notably, the calibration for DIRBE and FIRAS were performed independently, although they have been compared explicitly by \citet{fixsen1997}. The possibility of improving the $140$ and $240\,\mathrm{\mu m}$ bands has already been noted to reduce the discrepancy between the two experiments. The possibility of joint determination of gain and zodiacal emission using both of these datasets will be explored in future work.

\begin{figure*}
    \centering
    \includegraphics[width=0.9\textwidth]{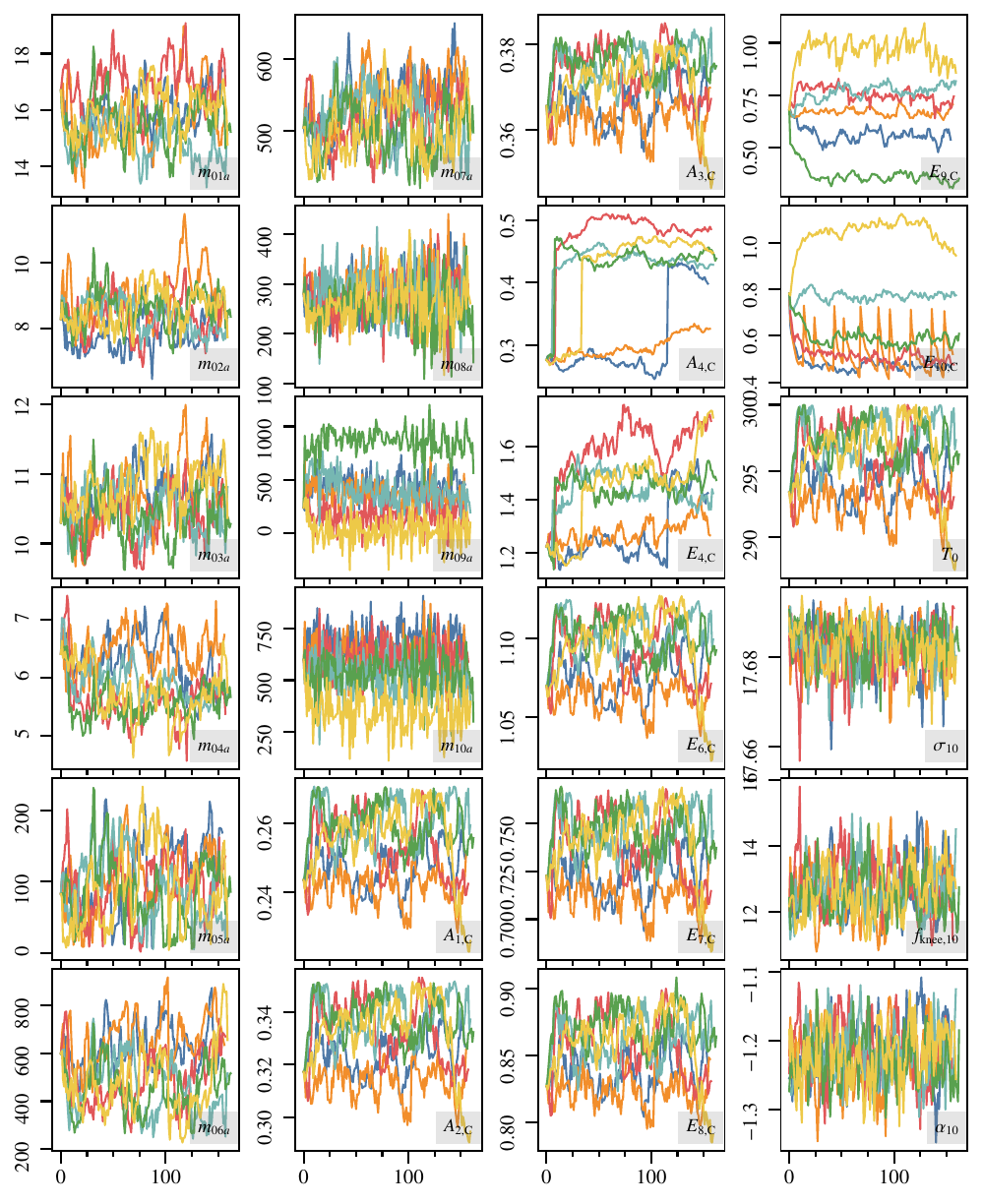}
	\caption{Trace plots of MCMC sampling chains. Each panel shows a series of sampled components for each Gibbs sample, including the ten DIRBE monopoles, the zodiacal dust cloud albedos $A$ and emissivities $E$, the ambient dust temperature at 1 AU $T_0$, and the instrumental parameters for band 10. Monopoles are plotted in units of $\mathrm{kJy\,sr^{-1}}$, $f_\mathrm{knee,10}$ in Hz, $\sigma_0$ in $\mathrm{MJy\,sr^{-1}}$, and $T_0$ in kelvin, while all other parameters are unitless.}
    \label{fig:trace}
\end{figure*}

\begin{figure*}
    \centering
    \includegraphics[width=1\textwidth]{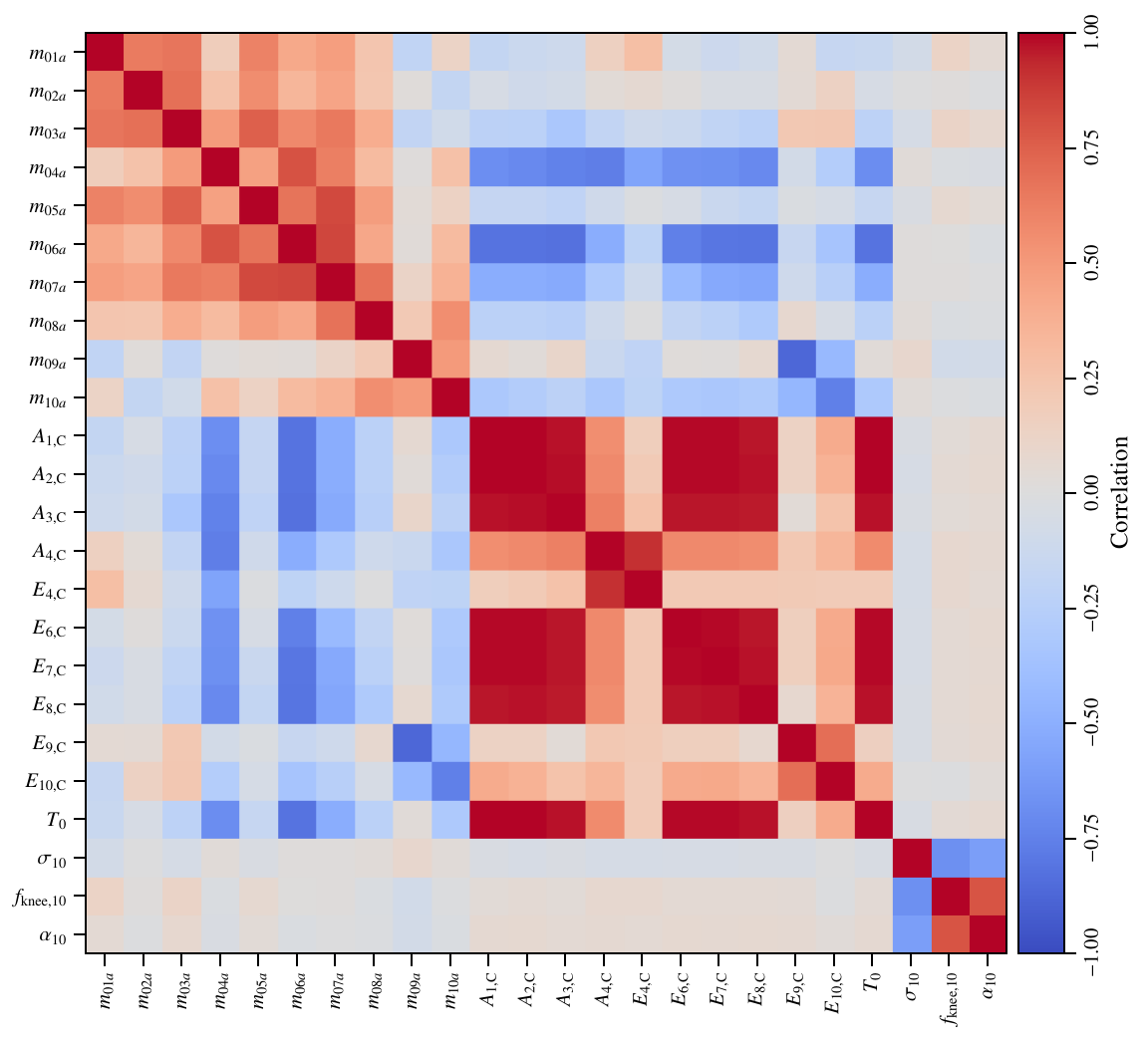}
	\caption{Correlations between a selection of all sampled parameters. The band monopoles are denoted $m$, while $A$ and $E$ are the albedos and emisivities of the zodiacal light cloud per band, $T_0$ is the dust temperature, and $\sigma_{10}$, $f_{\mathrm{knee}, 10}$, and $\alpha_{10}$ are the instrumental noise parameters in band 10, i.e., the 240\,$\mu$m channel.}
    \label{fig:correlations}
\end{figure*}

Due to the observation strategy, there are non-negligible correlations between nearby frequencies, and smearing along the scanning direction, corresponding to an effectively nonsymmetric beam. In order to take the beam into account, we explicitly smooth the sky model following the same prescription as in \citet{odegard:2019} when comparing with the FIRAS bands. 
The dense frequency spacing of the FIRAS data makes it ideal for determining the continuum behavior of sky emission and allows for identification of emission lines \citep{bennett:1994}. 
To avoid the correlation between nearby bands, we used a subset of the FIRAS bands, listed in Table \ref{tab:firas_bands}. A full comparison using the correlation between bands, similar to \citet{bianchini:2022}, will be performed in future analysis.

\begin{table}[t]
  \begingroup
  \newdimen\tblskip \tblskip=5pt
	\caption{List of FIRAS bands used in this analysis.}
	\label{tab:firas_bands}
  \nointerlineskip
  \vskip -3mm
  \footnotesize
  \setbox\tablebox=\vbox{
    \newdimen\digitwidth
    \setbox0=\hbox{\rm 0}
    \digitwidth=\wd0
    \catcode`*=\active
    \def*{\kern\digitwidth}
    \newdimen\signwidth
    \setbox0=\hbox{-}
    \signwidth=\wd0
    \catcode`!=\active
    \def!{\kern\signwidth}
 \halign{
      \hbox to 2.5cm{#\leaderfil}\tabskip 0em&
      #\tabskip 1em\hfil&
      #\tabskip 0em\hfil\cr
    \noalign{\doubleline}
	\omit\hfil Frequency  \hfil &  Wavelength \hfil & Purpose \cr
	\omit\hfil (GHz)      \hfil &  ($\mathrm{\mu m}$) \hfil &  \cr
      \noalign{\vskip 4pt\hrule\vskip 4pt}
       *108 & 2776 & \Planck\ 100\,GHz gain monitor \cr
       *149 & 2012 & \Planck\ 143\,GHz gain monitor \cr
       *217 & 1382 & \Planck\ 217\,GHz gain monitor \cr
       *353 & *849 & \Planck\ 353\,GHz gain monitor \cr
       *544 & *551 & \Planck\ 545\,GHz gain monitor \cr
       *857 & *350 & \Planck\ 857\,GHz gain monitor \cr
       1251 & *240 & DIRBE 240\,$\mu$m gain monitor \cr
       1809 & *166 & \CII\ SED constraints \cr
       1890 & *159 & \CII\ SED constraints \cr
       1904 & *157 & \CII\ SED constraints \cr
       1918 & *156 & \CII\ SED constraints \cr
       2081 & *144 & \CII\ SED constraints \cr
       2135 & *140 & DIRBE 140\,$\mu$m gain monitor \cr
       2802 & *107 & DIRBE 100\,$\mu$m gain monitor \cr
      \noalign{\vskip 4pt\hrule\vskip 5pt} } }
  \endPlancktablewide \endgroup
\end{table}

\subsection{Mask definitions}

In order to model the TODs accurately, we make use of processing masks depending on the band and component being modeled. In particular, estimating the instrumental noise properties and the zodiacal emission require a sufficiently accurate model, which can be easily biased in regions of excessively high emission.

For zodiacal emission, masking out the Ecliptic plane is necessary for all bands. As discussed in Sect.~\ref{sec:excess} and elaborated on in \citet{CG02_02}, the Ecliptic plane has a complex structure that is not well fit by current parametric models. In addition, incorrectly modeled Galactic emission in the form of thermal dust and stars can bias measurements as well. As shown in Fig.~\ref{fig:masks}, this mask, shown in blue, is wavelength-dependent, and is most aggressive in regions where zodiacal emission is brightest, and where point sources are brightest.
Conversely, at very long and very short wavelengths, a more aggressive Galactic mask is required. These are shown in green in Fig.~\ref{fig:masks}, and are defined again by regions in which the foregrounds are bright and not modeled sufficiently well.

\section{Markov chains, burn-in and convergence}
\label{sec:chains}

With the model in hand, we run \commanderthree\ on the data. We find that an average Gibbs iteration takes approximatly 500\,CPU-hrs, excluding initialization time. In order to ensure proper sampling of the full distribution, we run six independent \commanderthree\ runs initialized with their own random seeds. Each chain has 155, 157, 158, 160, 163, and 160 samples, giving a total of 953 samples. A total of one month walltime on 416 computing cores was required to produce these samples.

In order to assess burnin and convergence, we plot a range of parameters from the Gibbs chain in Fig.~\ref{fig:trace}. Included in this figure are the sampled monopoles, zodiacal dust albedos and emissivities, and instrumental parameters, including the estimated white noise in bands 1 and 10, and the correlated noise parameters $f_\mathrm{knee}$ and $\alpha$ for band 10. From this figure, some burnin can be seen in the monopoles, concluding at approximately sample 20.  At the same time, there is still drift within the emissivity and albedos, especially for parameters less strongly expressed in the model, such as $A_{4,\mathrm C}$ and $E_{9/10,\mathrm C}$.. However, since the monopoles are stable beyond this point, this is mainly due to degeneracies within the zodiacal model itself. This is explored in full detail in \citet{CG02_02}.

Discarding the burnin of 20 samples, we obtain a total of 833 independent Gibbs samples. In our analysis, we truncate all chains to be the length of the shortest chain, giving a total of 810 samples. Due to the extended burn-in of the zodiacal parameters, individual ZL parameters cannot be considered as fully converged. However, this does not affect the monopoles and frequency maps, which only depend on the sum of all ZL components, which are the main topic in this paper.

We can also compute the correlation between parameters in the Gibbs chain, as displayed in Fig.~\ref{fig:correlations}.  As expected, there are strong correlations between the zodiacal dust parameters. Likewise, there are strong correlations between monopoles of nearby frequency maps. There are small correlations, on the order of 10\,\%, between the monopoles and zodiacal dust parameters. While these correlations are expected, the relatively low correlation indicates that the zodiacal dust and the monopole signal are relatively decoupled.

The two instrumental noise parameters are chosen as representative of the behavior of instrumental noise parameters in general. The white noise level is correlated with the monopoles and local dust temperature $T_0$, demonstrating the dependence of some instrumental parameters on the final noise level. Conversely, band 10's noise parameters, while degenerate with each other as expected (see \citealt{bp04} and \citealt{bp06}), have negligible correlation with zodiacal dust and monopole parameters.

\section{Noise estimation and goodness of fit}
\label{sec:noise_gof}

We now turn our attention to aggregate posterior statistics, typically
in the form of posterior mean and rms estimates for each sampled
quantity, and we start with noise estimation and overall
goodness-of-fit statistics. The algorithms used in the current
analysis to estimate the instrumental noise parameters are identical
to those described by \citet{bp06} as applied to \Planck\ LFI, and we
refer the interested reader there for further details.

\subsection{Instrumental noise}

\begin{figure}
	\centering
	\includegraphics[width=\columnwidth]{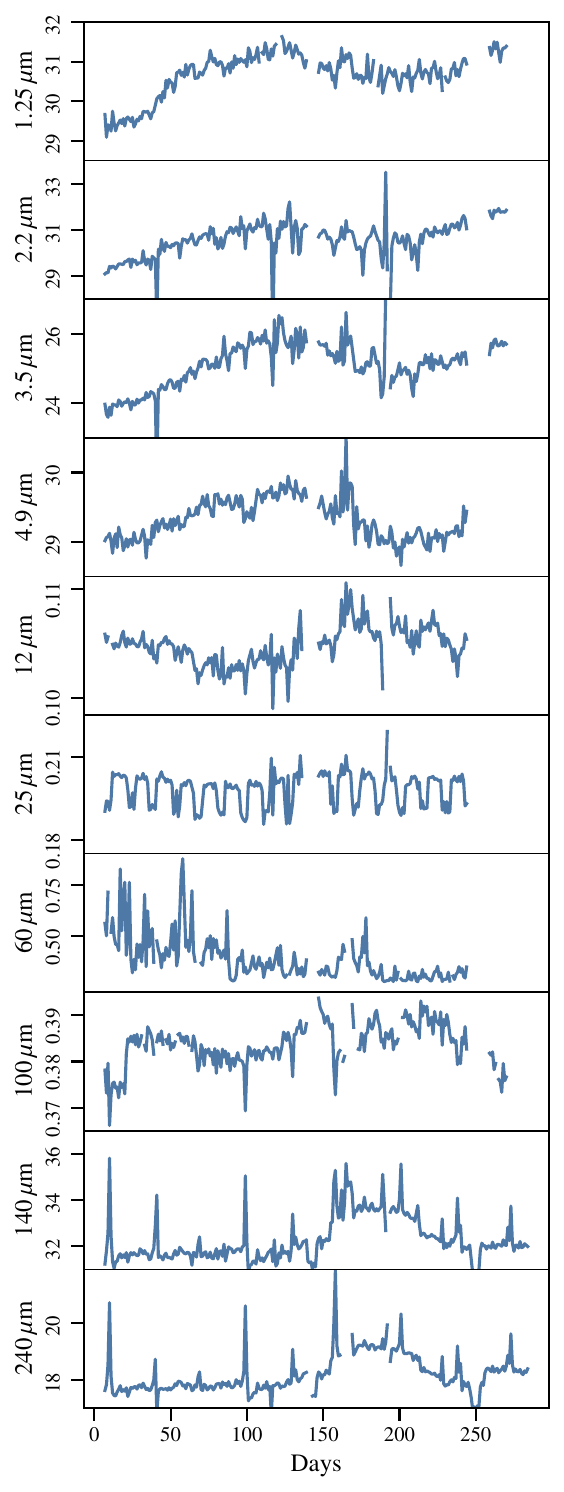}
	\caption{Instrumental noise $\sigma_0$ for each band, averaged over all Gibbs chains. Data not used in the Gibbs chain are not displayed. Bands $12$--$240\,\mathrm{\mu m}$ are plotted in units of $\mathrm{MJy\,sr^{-1}}$, while all others are in $\mathrm{kJy\,sr^{-1}}$.}
	\label{fig:sigma0}
\end{figure}

\begin{figure}
	\centering
	\includegraphics[width=\linewidth]{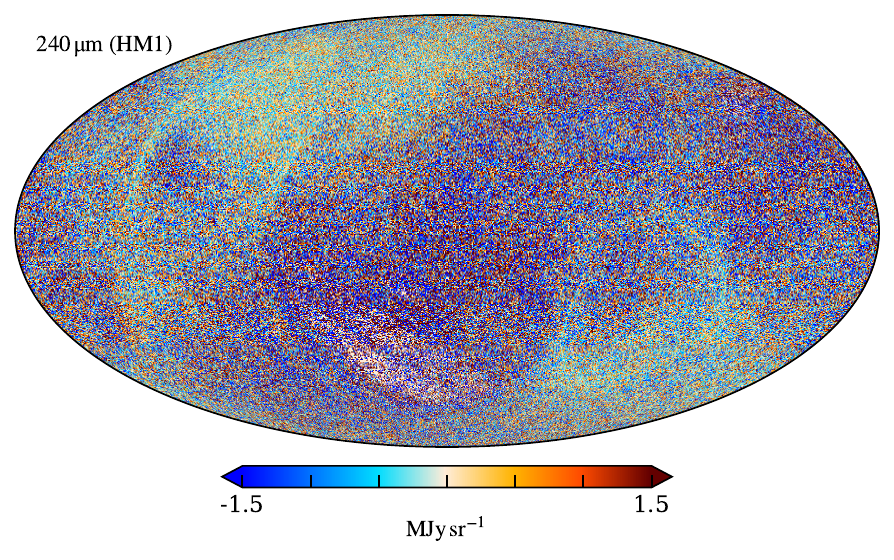}
	\caption{Realization of correlated noise for first half-mission of $240\,\mathrm{\mu m}$ band, from the 25th sample of the first chain.}
  \label{fig:ncorr_map}
\end{figure}

\begin{figure*}
  \centering
  \includegraphics[width=0.37\linewidth]{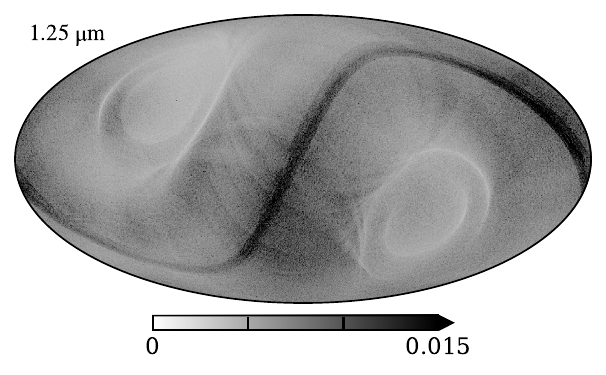}       
  \includegraphics[width=0.37\linewidth]{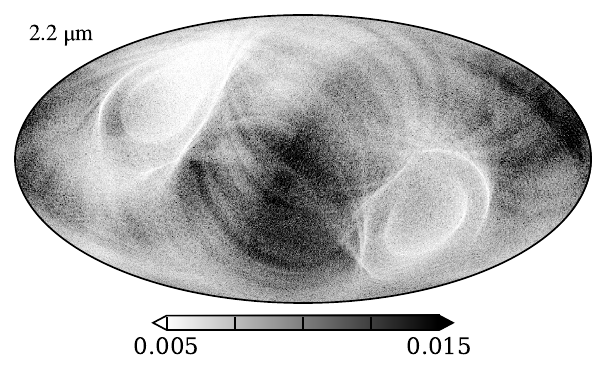}\\
  \includegraphics[width=0.37\linewidth]{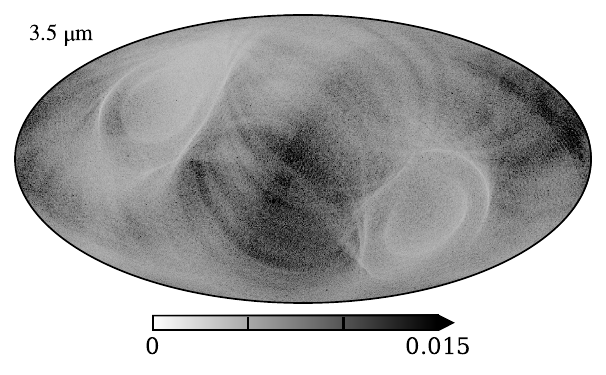}
  \includegraphics[width=0.37\linewidth]{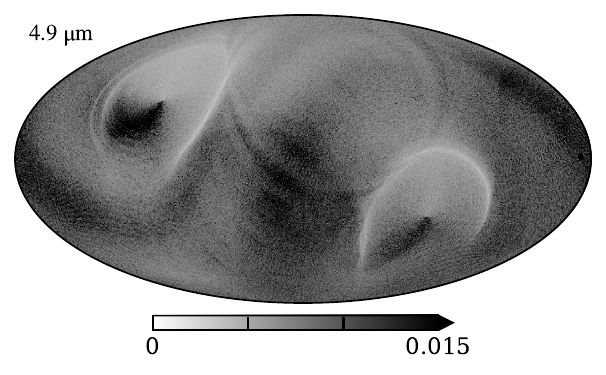}\\       
  \includegraphics[width=0.37\linewidth]{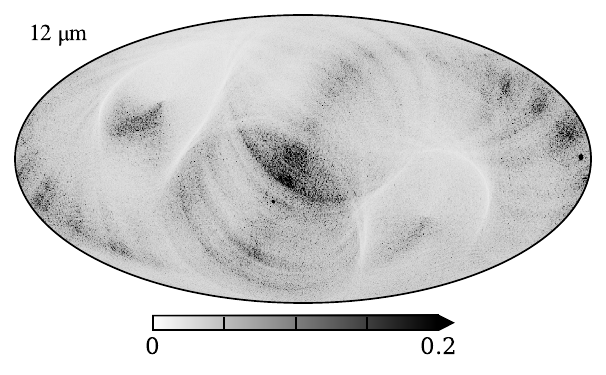}       
  \includegraphics[width=0.37\linewidth]{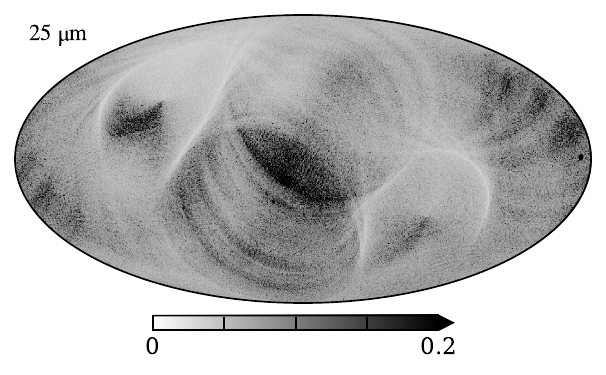}\\       
  \includegraphics[width=0.37\linewidth]{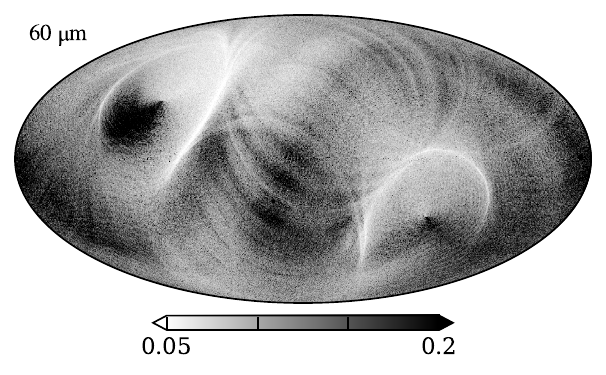}
  \includegraphics[width=0.37\linewidth]{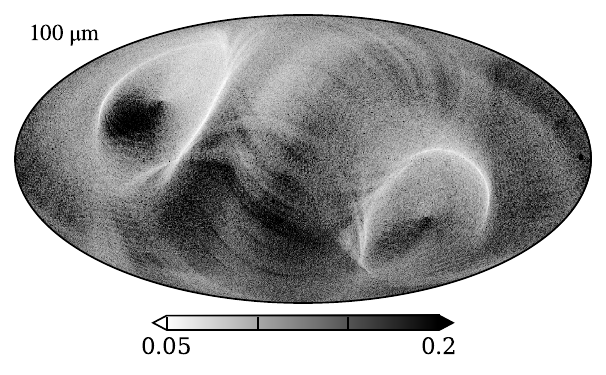}\\       
  \includegraphics[width=0.37\linewidth]{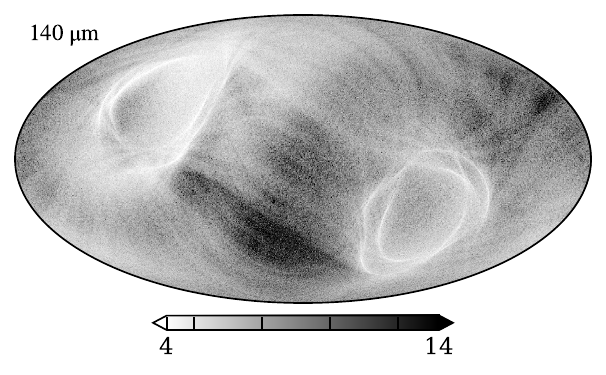}
  \includegraphics[width=0.37\linewidth]{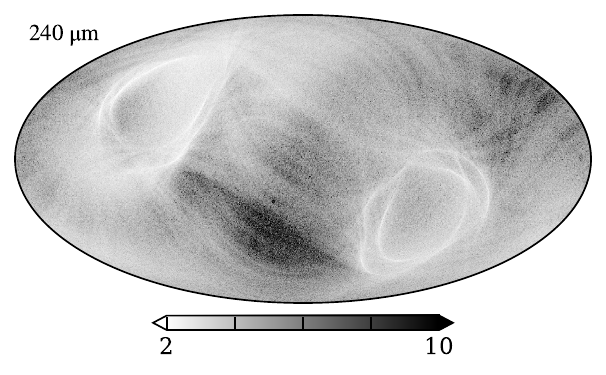}       
	\caption{White noise rms maps for each DIRBE channel. All maps are in units of $\mathrm{MJy\,sr^{-1}}$.}
  \label{fig:sigma0_map}
\end{figure*}

       \begin{figure*}
       	\centering
       	\includegraphics[width=0.35\linewidth]{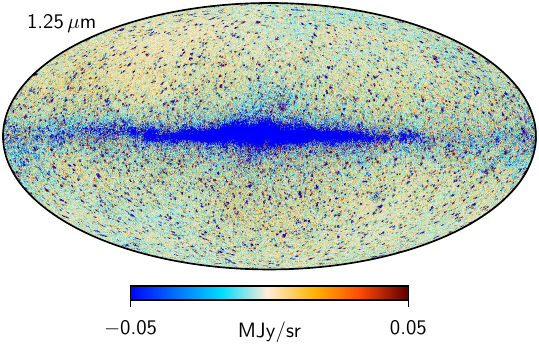}       
       	\includegraphics[width=0.35\linewidth]{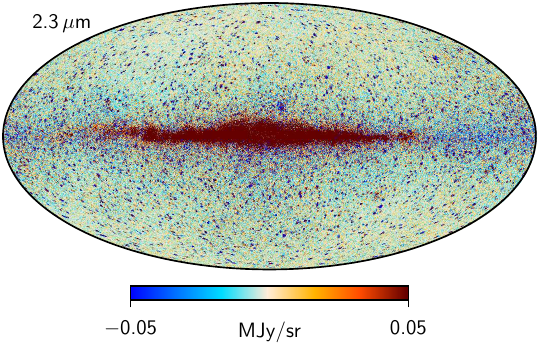}\\
        \includegraphics[width=0.35\linewidth]{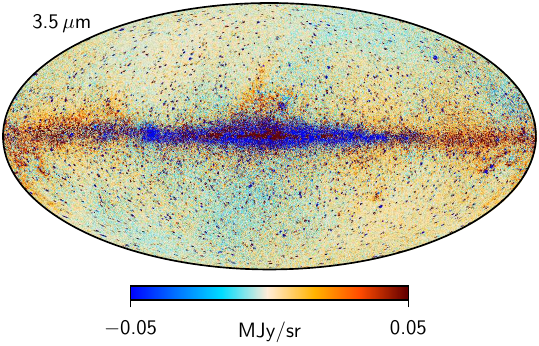}
        \includegraphics[width=0.35\linewidth]{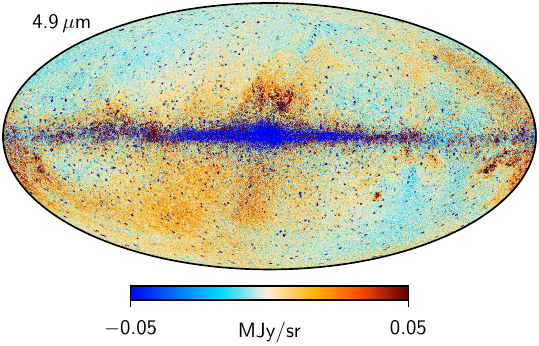}\\         
       	\includegraphics[width=0.35\linewidth]{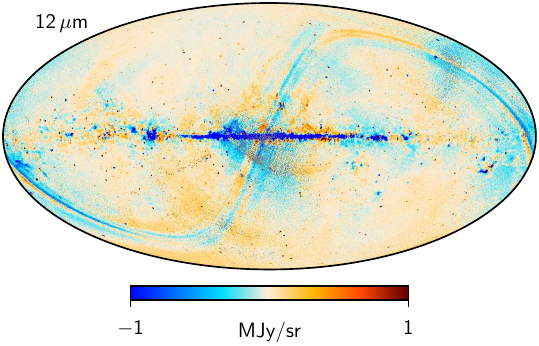}       
       	\includegraphics[width=0.35\linewidth]{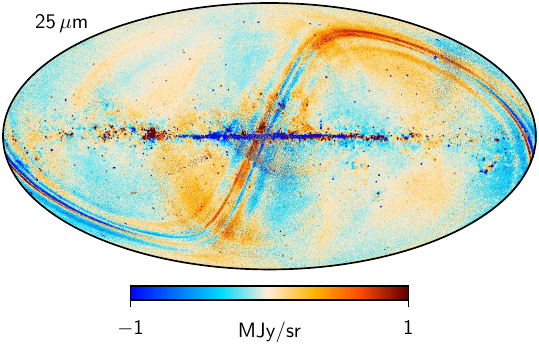}\\       
       	\includegraphics[width=0.35\linewidth]{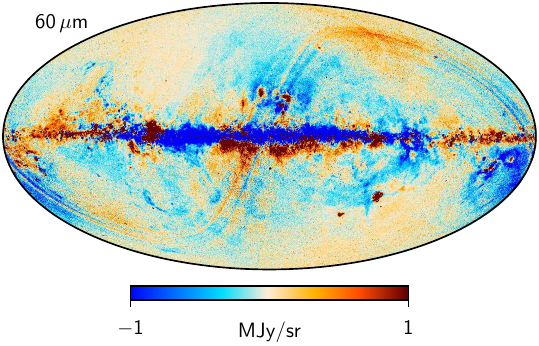}
       	\includegraphics[width=0.35\linewidth]{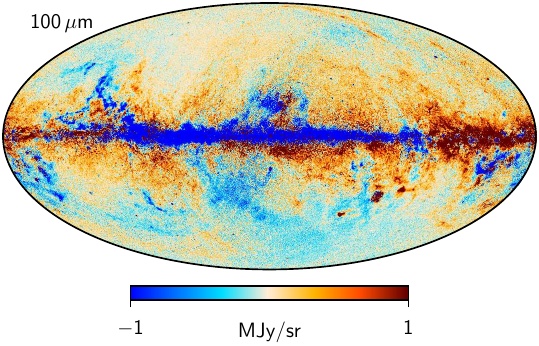}\\       
       	\includegraphics[width=0.35\linewidth]{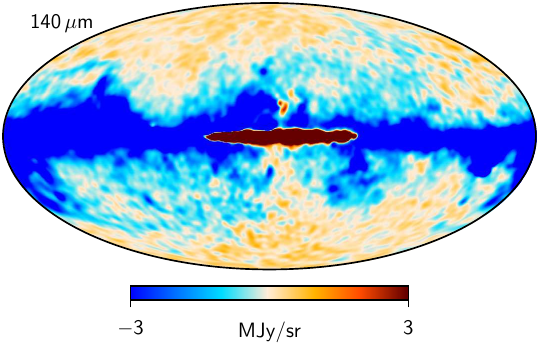}
       	\includegraphics[width=0.35\linewidth]{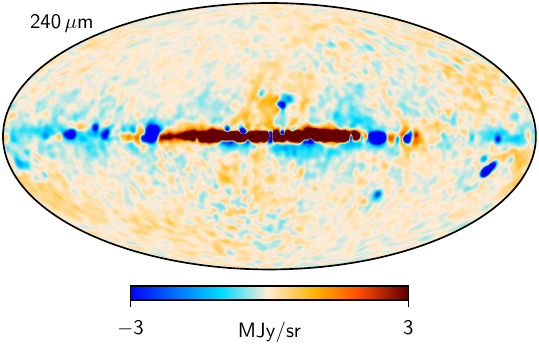}       
       	\caption{Data-minus-model residual maps for each band for one arbitrarily chosen Gibbs sample. The 140 and 240\,$\mu$m channels have been smoothed to an angular resolution of $3^{\circ}$, while all others are shown at their native resolution. }
       	\label{fig:res}
       \end{figure*}

       \begin{figure*}
       	\centering
       	\includegraphics[width=0.35\linewidth]{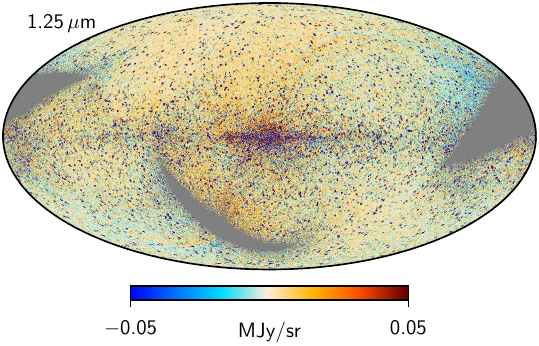}       
       	\includegraphics[width=0.35\linewidth]{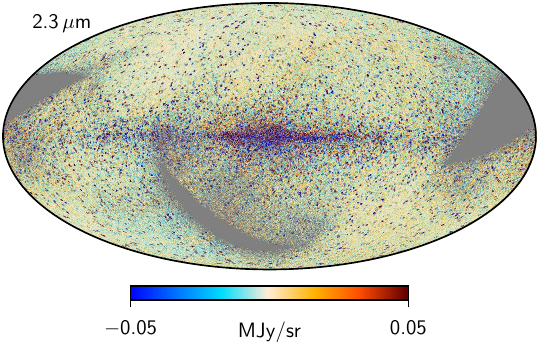}\\
        \includegraphics[width=0.35\linewidth]{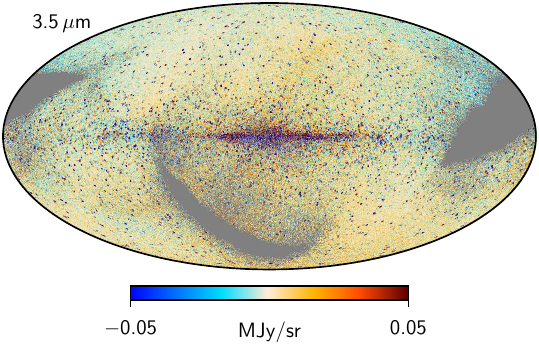}
        \includegraphics[width=0.35\linewidth]{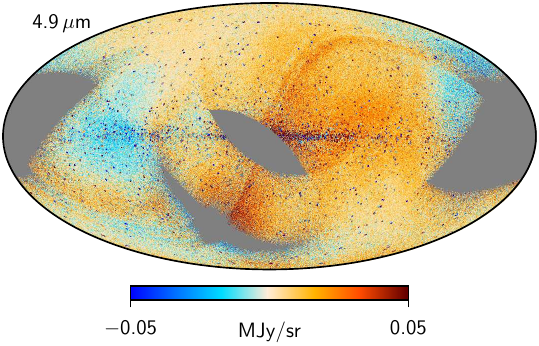}\\         
       	\includegraphics[width=0.35\linewidth]{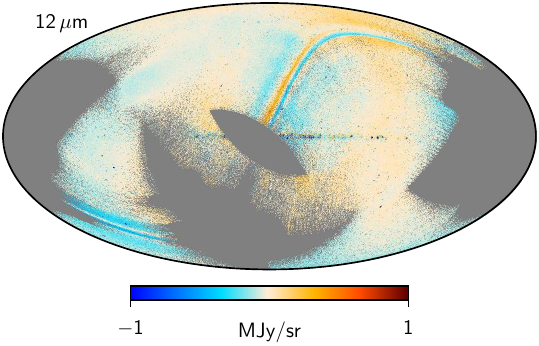}       
       	\includegraphics[width=0.35\linewidth]{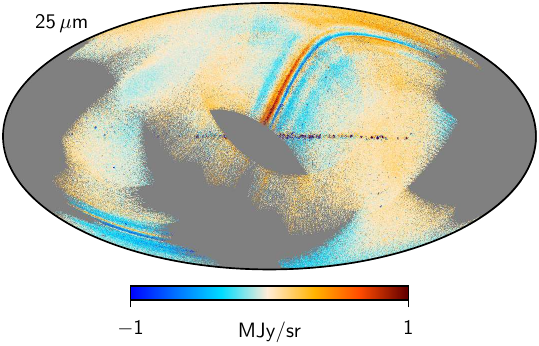}\\       
       	\includegraphics[width=0.35\linewidth]{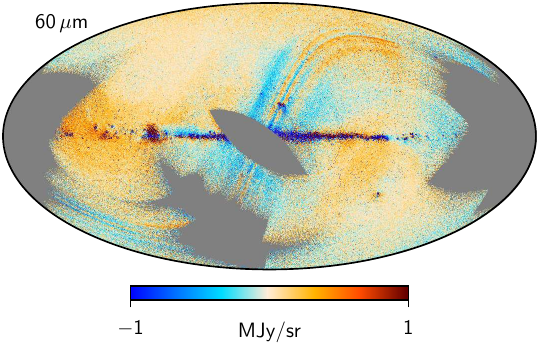}
       	\includegraphics[width=0.35\linewidth]{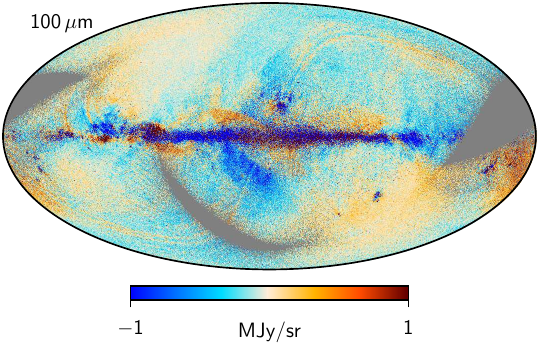}\\       
       	\includegraphics[width=0.35\linewidth]{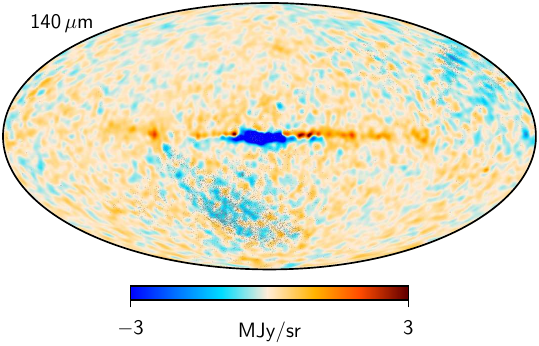}
       	\includegraphics[width=0.35\linewidth]{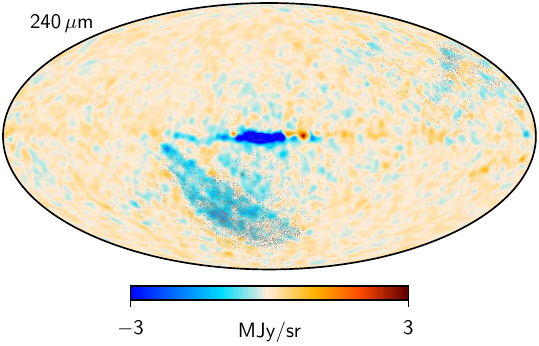}       
       	\caption{Zodi-subtracted half-mission half-difference maps for each channel for one arbitarily chosen Gibbs sample. The 140 and 240\,$\mu$m channels have been smoothed to an angular resolution of $3^{\circ}$, while all others are shown at their native resolution. }
       	\label{fig:hmhd}
       \end{figure*}

       \begin{figure}
	 \centering
	 \includegraphics[width=\columnwidth]{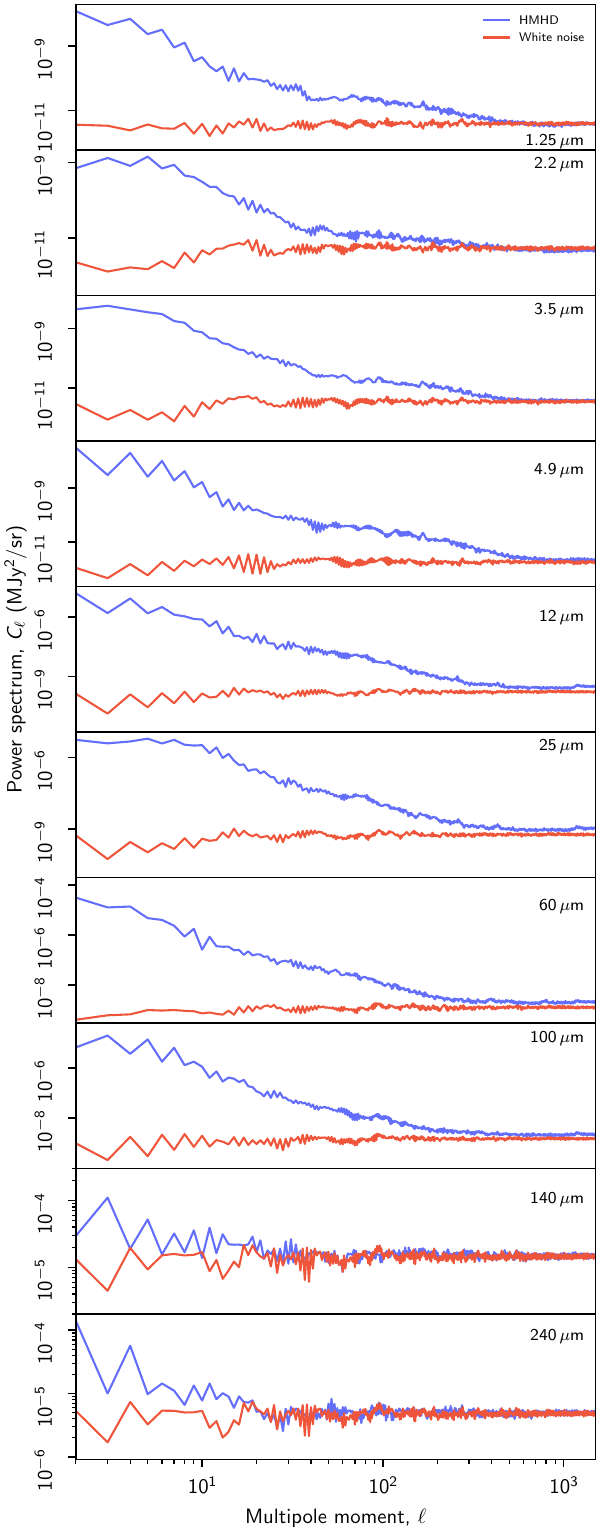}
	 \caption{Half-mission half-difference angular power spectra (blue curves) for each DIRBE channel computed with the CIB monopole masks defined by \citet{CG02_03}. The red curves show the power spectrum computed from one white noise realization with the same mask.}
	 \label{fig:hmhd_powspec}
       \end{figure}

       \begin{figure}
       	\centering
       	\includegraphics[width=\linewidth]{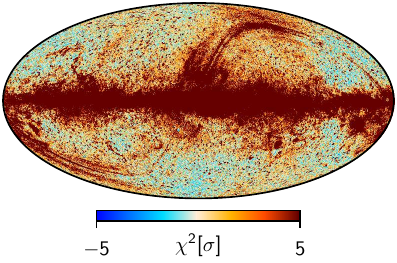}
       	\caption{Pixel-space reduced normalized $\chi^2$ in units of $\sigma$. The number of degrees-of-freedom per pixel is here assumed to be 103, which is equal to the sum of the number of observations per $N_{\mathrm{side}}=512$ pixel minus the number of freely fitted Galactic components. }
       	\label{fig:chisq}
       \end{figure}

As defined in Sect.~\ref{sec:datamodel}, $\xi_{\mathrm{n}}$ denotes
the set of all instrumental noise parameters in our data model, and
for all channels except 240$\,\mu$m this simply consist of a white
noise rms per TOD sample, $\sigma_0$, which is assumed to be constant
within each 24\,hr data segment. The corresponding posterior mean is
plotted as a function of observing day in Fig.~\ref{fig:sigma0}. Gaps
in each curve indicate observations that have been excluded from the
analysis, either due to the original DIRBE data quality flags or our
additional exclusion of the last two or four weeks of observations;
see Sect.~\ref{sec:data_selection}.

Several interesting features may be seen in this figure. Starting with
the 1.25$\,\mu$m channel as shown in the top panel, we notice an
average increase by about 3\,\% from the beginning to the end of the
survey. This increase is however not increasing uniformly, but rather
exhibits systematic variations as a function of time. Similar features
are observed for all the four shortest wavelength bands, both in terms
of absolute amplitude and general behaviour. Perhaps the most natural
explanation for such behaviour are changes in the thermal environment
of the DIRBE detectors, for instance due to varying levels of
radiation from the Sun or the Earth. As demonstrated by \citet{bp06}
within the \BP\ analysis by correlating $\sigma_0$ with house-keeping
focal plane thermometer measurements, this was the case for
\Planck\ LFI. We have not yet been able to locate
similar publicly available house-keeping information for DIRBE.

In contrast, the 25$\,\mu$m channel exhibits a qualitatively different
behaviour. In this case, $\sigma_0$ appears to jump between two
different stationary states that are separated by 5\,\% in
amplitude. At one level, this general behaviour appears structurally
similar to a phenomenon called ``popcorn'' or ``random telegraph''
noise, which was also observed in several \Planck\ detectors, in which
the noise level jumps between two discrete and well-defined
levels. However, the behaviour seen in Fig.~\ref{fig:sigma0} appears
more systematic than what is usually observed for popcorn noise, with
a very well-defined period of about 2 weeks. This time scale could
possibly suggest that the Moon plays a role in this behaviour, which
has an rotation period relative to the Earth of about 4~weeks.

Moving on, the 60$\,\mu$m channel exhibits much larger drifts than any
of the others, and changes by almost 50\,\% from the beginning to the
end of the survey. We also see a clear change in the level of
variations between scans as a function of the survey, with much
stronger variations in the first third of the mission. In contrast,
the 100\,$\mu$m channel appears much more stable, and is in fact
structurally quite similar to the short wavelength channels.

Finally, the 140 and 240$\,\mu$m channels behave yet again differently
from the other eight, with a very stable plateau during the first half
of the mission, but with a clear increase around Day 180. These two
channels, however, are internally very similar, and individual
features and spikes can be traced very accurately between the two. In
this respect, it is worth recalling the instrument layout shown in
Fig.~\ref{fig:optics_model}, where we see that these two channels are
co-located in the optical path, separated from the others. It is
therefore plausible that these two detectors experience a different
thermal environment than the others.

Even more notable than the time variations in the 140 and 240$\,\mu$m
channels are their much higher overall noise level, which is almost
two orders of magnitude higher than for the other channels. This is
due to the different detector technology used for these two channels
\citep{hauser1998}.  This also implies that these channels are where
the instrumental noise is best understood, allowing for a correlated
noise PSD as well as noise realizations to be fit and removed in the
$240\,\mathrm{\mu m}$ band. One realization of this correlated noise
map is shown for the first half-mission split in
Fig.~\ref{fig:ncorr_map}. Given that this map is derived directly from
the signal-subtracted frequency map, and therefore essentially acts as
a ``trash can'' for unmodelled effects, the abscence of large coherent
features provides strong evidence that the adopted signal model is
indeed able to account for all main effects. 

While Fig.~\ref{fig:sigma0} shows the white noise level in
time-domain, Fig.~\ref{fig:sigma0_map} shows the corresponding noise
rms as a function of position on the sky after accounting for the
number of observations per pixel. Starting once again with the
1.25$\,\mu$m channel, the smooth underlying variations that appear
nearly symmetric with respect to the Ecliptic poles are simply due to
the DIRBE scanning strategy, which effectively observe the Ecliptic
poles more often than the Ecliptic plane. The sharp band of higher
values along the Ecliptic plane is however not due to the scanning
strategy as such, but rather by the DIRBE quality flags which removes
near-planet observations. The 2.2 and 3.5$\,\mu$m channels show very
similar behaviour.

In general, the patterns seen in the 4.9--60$\,\mu$m channels also
appear broadly similar. However, in this case we can also see regions
with higher noise levels near the Ecliptic poles and south of the
Galactic center. These are primarily due to the excess radiation masks
defined in Sect.~\ref{sec:excess}, which removes a significant
fraction of the overall data, and affects some parts of the sky more
than others, depending on the specific orientation of the satellite at
any given time with respect to the Sun.

\subsection{Goodness of fit}

With basic noise statistics in hand, we are ready to consider the
overall goodness of fit of the model. The first such quality measure
we consider are simply the data-minus-model residual maps for each
wavelength band, and these are shown in Fig.~\ref{fig:res}. Starting
from the top, we first note that the color scale range spans
50\,$\mathrm{kJy\,sr^{-1}}$, while the natural plotting scale for the full sky signal
of this channel is typically 10\,M$\mathrm{Jy\,sr^{-1}}$. As such, the model accounts
for about 99\,\% of the total sky signal at high Galactic
latitudes. The same holds true for most other channels as well.

The dominant residual at short wavelengths is due to
residual starlight emission in the Galactic disk. At 1.25$\,\mu$m the starlight emission is
slightly over-subtracted, while at 2.2\,$\mu$m it is slightly
under-subtracted. In this respect it is important to note that the
starlight model presented by \citet{CG02_04} is based on the WISE
catalog, which includes about 747 million sources. While this is a
large number, it is still much lower than the total number of stars in
the Milky Way, which is about 100\,billion. The completeness of the
WISE catalog is, however, much higher at high Galactic latitudes than
in the central bulge, and it is therefore not surprising that the
model is not statistically adequate at low Galactic latitudes. 
       
In general, only very faint ZL residuals are seen in the two
shortest wavelength bands. In order to suppress these further, it is
worth considering fitting the albedo of the asteroidal bands
independently from the cloud; however, the signal-to-noise ratio of
the bands at these wavelengths is very low, and there is a significant
risk of introducing strong degeneracies with the starlight model by
doing so.

Significantly stronger ZL residuals are seen in the 3.5 to 60$\,\mu$m
channels, but still at the sub-percent level of the total
intensity. The asteroidal bands are particularly noteworthy at
25$\,\mu$m. In order to improve on these, higher angular resolution
would be extremely useful, and a future joint analysis with \IRAS\
and/or \AKARI\ should prove useful in reducing these residuals further.

Between 60 and 140$\,\mu$m, the dominant residuals are clearly due to
Galactic dust emission, and to improve on these, a more detailed
thermal dust model should be established. In this respect, it is
worth recalling that we currently only fit two degrees of freedom per
pixel for thermal dust emission from 100\,GHz to 1.25$\,\mu$m, and
there are therefore massive opportunities for refining the current
model without compromising the overall signal-to-noise ratio and
introducing uncontrollable degeneracies. Natural next steps are to
allow for spatial variations in the spectral parameters for each
thermal dust component, as well as to sub-divide the nearby dust
component into more local clouds \citep{CG02_05}.

Next, Fig.~\ref{fig:hmhd} shows zodi-subtracted half-mission half-difference (HMHD)
maps for each channel. These maps quantify seasonal variations in the
overall residuals, and put strong limits both on errors in the assumed
DIRBE-based calibration and in the overall ZL model. In particular, we
note that the Galactic plane is only barely visible in any of these
channels, and that indicates the DIRBE calibration is accurate to much
better than 1\,\% throughout the entire mission. Rather, the dominant
spatial structures in these maps appear to be zodiacal in nature, with
patterns matching those expected from convolving the ZL model with the
DIRBE scanning strategy.

Figure~\ref{fig:hmhd_powspec} shows the angular power spectra computed
from each of the HMHD maps as blue curves, compared with a single
white noise realization from the Gibbs chain, plotted as red curves. At low multipoles, we
see that the amplitude of the excess residuals typically are two
orders of magnitude higher than the white noise level, which indicates
that the current residuals are about one order of magnitude larger
than white noise in pixel space. Above $\ell\gtrsim100$ this
discrepancy falls smoothly to about unity due to the DIRBE
beam. Indeed, for the 1.25 to 3.5 and 14 to 240$\,\mu$m channels, the
agreement between the observed residual and the white noise model is
excellent. However, for the intermediate channels between
4.9 to 60$\,\mu$m  there is a discrepancy of about a factor of
two, and the origin of this is still under investigation.

Finally, Fig.~\ref{fig:chisq} shows the total reduced and normalized
$\chi^2$ as a function of pixel on the sky in units of $\sigma$; for
an exact definition, see \citet{bp10}. The number of degrees of
freedom per pixel is here assumed to be 103, which equals the total
number of individual frequency map pixels per $N_{\mathrm{side}}=512$
pixel (which is not equal to the number of data channels, because the
\Planck\ bands have higher resolution than DIRBE) minus the number of
diffuse components. In this figure, we clearly see both the Galactic
and ZL residuals, as discussed above. However, there are also large
extended regions for which the goodness of fit is within the expected
range of $\pm2\sigma$. This is a strong testament to the overall
quality of the data model defined by Eq.~\eqref{eq:model}. For
cosmological analyses of these data, the $\chi^2$ map in
Fig.~\ref{fig:chisq} serves as a useful starting point for mask
definitions. 

\begin{figure*}
  \centering
  \includegraphics[width=0.40\linewidth]{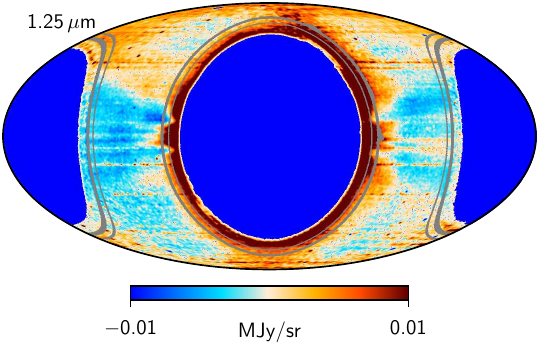}\hspace*{5mm}
  \includegraphics[width=0.40\linewidth]{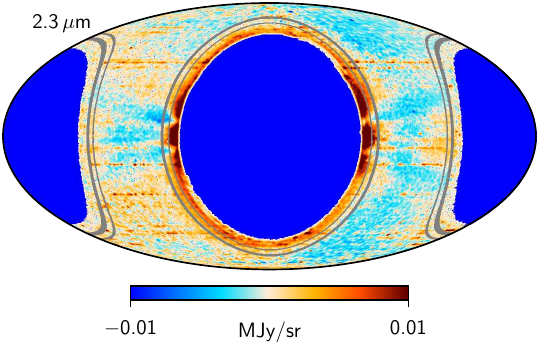}\\
  \includegraphics[width=0.40\linewidth]{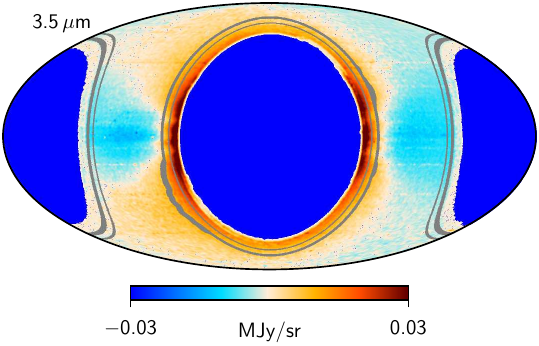}\hspace*{5mm}
  \includegraphics[width=0.40\linewidth]{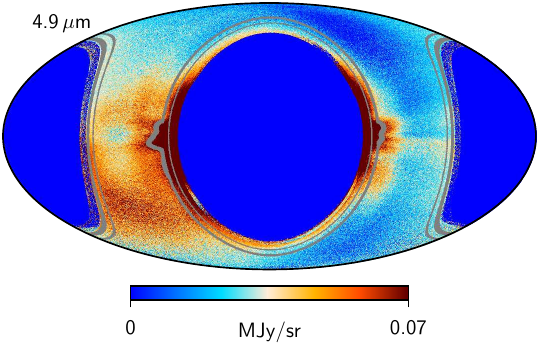}\\
  \includegraphics[width=0.40\linewidth]{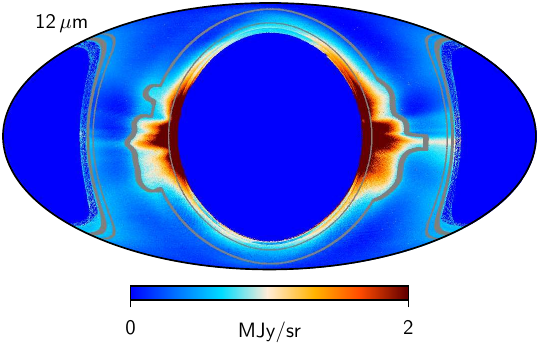}\hspace*{5mm}
  \includegraphics[width=0.40\linewidth]{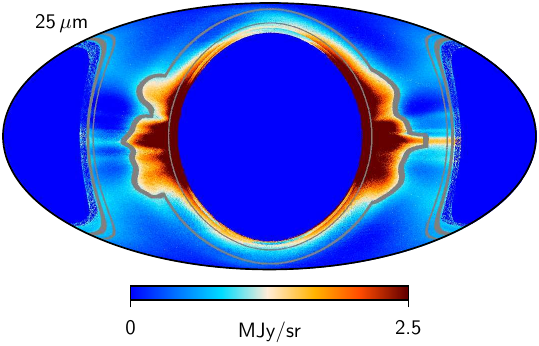}\\
  \includegraphics[width=0.40\linewidth]{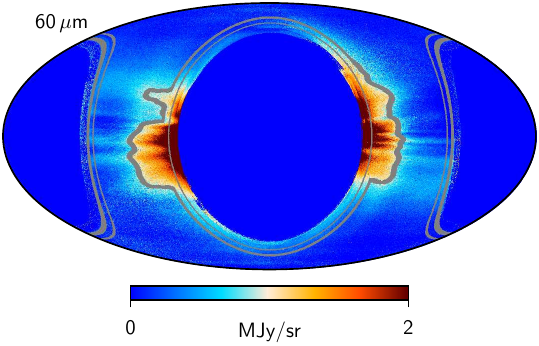}\hspace*{5mm}
  \includegraphics[width=0.40\linewidth]{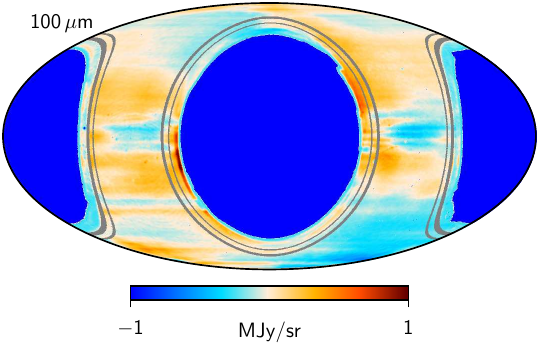}\\
  \includegraphics[width=0.40\linewidth]{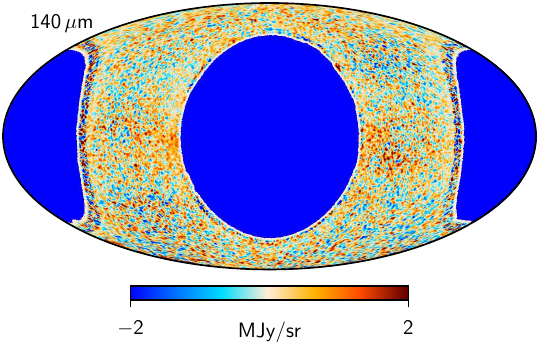}\hspace*{5mm}
  \includegraphics[width=0.40\linewidth]{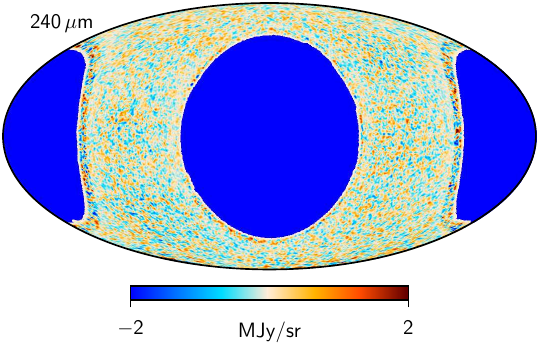}
  \caption{Solar-centric residual maps, derived by co-adding the residual TOD, $\r = \dv - \s_{\mathrm{fg}} - \s_{\mathrm{zodi}} - \n_{\mathrm{corr}}$, into solar-centric coordinates. The Sun is located in the center of each panel, and the equator is aligned with the Ecliptic plane. All maps have been smoothed with a $1^{\circ}$ FWHM Gaussian kernel. The gray boundaries indicate the solar-centric exclusion masks used for each channel; no masks are applied for 140 and 240\,$\mu$m.}
  \label{fig:solarmaps}
\end{figure*}

\section{Excess radiation model}
\label{sec:excess}

One of the key novel features of the current analysis is the inclusion
of $\s_{\mathrm{static}}$ in Eq.~\eqref{eq:model}. This component is
designed to account for excess radiation that appears static in
solar-centric coordinates. The existence of such radiation was already
noted by \citet{reach:1995} and \citet{leinert:1998}, but never
systematically characterized or corrected for in the final DIRBE data
processing.

As described in Sect.~\ref{sec:global_modelling}, we model any
potential excess radiation in the current analysis by subtracting all
other components from the raw TOD, and bin the residual TOD into
solar-centric coordinates. Because of this coordinate transformation,
the resulting component is not degenerate with most other components,
but only those that actually appear static in the Earth-Sun coordinate
system. In the current model, that applies only to two components in the
K98 ZL model, namely the so-called circumsolar ring and the
Earth-trailing feature \citep{kelsall1998}. These two physical
interplanetary dust (IPD) components are trapped in the Earth's
gravitational field, and follow the Earth's annual motion around the
Sun. As such, they appear to be static on the sky as seen from the
Earth, and they are therefore also fully degenerate with
$\a_{\mathrm{static}}$. For this reason, we make no attempt to refit
those two ZL components in the current analysis, but rather fix the
corresponding parameters at the respective K98 values. As a result,
$\a_{\mathrm{static}}$ captures any excess radiation beyond what is
described by the K98 model.

\subsection{Characterization}

In order to characterize the spatial morphology of
$\a_{\mathrm{static}}$ as a function of wavelength, we produced a
preliminary Gibbs chain as described by
Eqs.~\eqref{eq:gibbs_n}--\eqref{eq:gibbs_static} prior to the main
production run, while imposing no constraints on the effective sky
coverage of $\a_{\mathrm{static}}$. The main results from this
calculation are summarized in Fig.~\ref{fig:solarmaps} in terms of
posterior mean maps for each channel. Blue pixels corresponds to
directions on the sky that are never observed by the DIRBE instrument,
while the thin gray line corresponds to the edge of a set of
processing masks discussed further below. Each of these maps is thus
the full residual signal in the DIRBE data that is not
captured by the assumed ZL and astrophysical parametric model, binned
into solar-centric coordinates.

Browsing through the various panels, we can immediately make several
interesting observations. First, we see that the two channels in the
bottom row, i.e., the 140 and 240$\,\mu$m channels, appear for all
practical purposes consistent with instrumental Gaussian noise. There
are no signatures of any excess radiation in these channels, and that
is a strong testament to the efficacy of the current model at long
wavelengths.

The same also holds true to a large extent at the three shortest
wavelengths between 1.25 and 3.5$\,\mu$m. For these, most of the sky
does indeed appear to be dominated by noise, with three notable
exceptions. First, at low solar elongations $e$ -- that is, pixels
that lie close to the center -- one may see a slight positive and
circular excess along the boundary of the observed region. Second,
there are clear features with both positive and negative signs aligned
with equator, i.e., the Ecliptic plane. Finally, there is a faint
signature of large-scale features even at high Ecliptic latitudes, in
particular in the $3.5\,\mu$m channel.

However, this general picture appears quite different at the
intermediate wavelengths between 4.9 and 60$\,\mu$m. For each of
these, an excess radiation is measured with very high signal-to-noise
ratio. The most striking example is the 25\,$\mu$m channel, for which
the instrumental noise is fully sub-dominant.

When interpreting these maps physically, it is important to note that
any model error in the Galactic foreground model will turn into
horizontal stripes in these coordinates, due to the DIRBE's annual
motion around the Sun. In the current set of plots, this is most
clearly seen in the 100$\,\mu$m channel, for which several stripes are
clearly visible; these are most likely due to Galactic thermal dust
residuals. Conversely, the strong signal observed at $25\,\mu$m cannot
be explained in terms of Galactic residuals.

\begin{table*}[t]
  \begingroup
  \newdimen\tblskip \tblskip=5pt
  \caption{Key map-level characteristics of the \cosmoglobe\ DR2 ZSMA maps. The relative calibration, $\alpha$, is defined as the slope of a scatter plot between the old K98 and new DR2 maps, evaluated over either the default DR2 processing mask (``High latitudes'') or full sky. }
  \label{tab:summary}
  \nointerlineskip
  \vskip -3mm
  \footnotesize
  \setbox\tablebox=\vbox{
    \newdimen\digitwidth
    \setbox0=\hbox{\rm 0}
    \digitwidth=\wd0
    \catcode`*=\active
    \def*{\kern\digitwidth}
    \newdimen\signwidth
    \setbox0=\hbox{-}
    \signwidth=\wd0
    \catcode`!=\active
    \def!{\kern\signwidth}
 \halign{
      \hbox to 2.0cm{#\leaderfil}\tabskip 1em&
      \hfil#\hfil\tabskip 1em&
      \hfil#\hfil\tabskip 1em&
      \hfil#\hfil\tabskip 2em&
      \hfil#\hfil\tabskip 1em&
      \hfil#\hfil\tabskip 2em&
      \hfil#\hfil\tabskip 1em&
      \hfil#\hfil\tabskip 0em\cr
      \noalign{\doubleline}
	\omit&\multispan3\hfil\sc Bandpass$^{(\mathrm{a})}$ \hfil&\multispan2\hfil\sc Relative calibration, $\alpha$ \hfil&\multispan2\hfil\sc Noise rms (k$\mathrm{Jy\,sr^{-1}}$)$^{(\mathrm{b})}$ \hfil\cr
\noalign{\vskip -3pt}
\omit&\multispan3\hrulefill&\multispan2\hrulefill&\multispan2\hrulefill\cr
\noalign{\vskip 3pt} 
\omit\sc Channel ID\hfil& $\nu_{\mathrm{c}}$ (THz) & $\lambda_{\mathrm{c}}$ ($\mu$m) & $\Delta\nu/\nu\hfil$ & High lat & Full sky & K98 & DR2 \cr
\noalign{\vskip 3pt\hrule\vskip 5pt}
      *1 & 240  & 1.25 & 0.25 &  1.025 & 1.068 & ***1.0 & ***1.3 \cr
      *2 & 136  & 2.2  & 0.16 &  1.022 & 1.034 & ***1.2 & ***1.8 \cr
      *3 & 85.7 & 3.5  & 0.26 &  1.012 & 1.031 & ***1.1 & ***1.5 \cr
      *4 & 61.2 & 4.9  & 0.13 &  0.974 & 1.028 & ***1.3 & ***1.3 \cr
      *5 & 25.0 & *12  & 0.53 &  0.629 & 0.943 & ***3.6 & **8.8 \cr
      *6 & 12.0 & *25  & 0.34 &  0.206 & 1.019 & ***7.5 & **15 \cr
      *7 & 5.00 & *60  & 0.46 &  1.009 & 0.988 & **18 & **18 \cr
      *8 & 3.00 & 100  & 0.32 &  1.005 & 1.041 & **17 & **19 \cr
      *9 & 2.14 & 140  & 0.28 &  0.784 & 1.013 & 1530 & 1270 \cr
      10 & 1.25 & 240  & 0.40 &  0.841 & 1.012 & *860 & *740 \cr
      \noalign{\vskip 4pt\hrule\vskip 5pt} } }
  \endPlancktablewide 
  \tablenote {{\rm a}} Reproduced from the DIRBE Explanatory Supplement.\par
  \tablenote {{\rm b}} Noise sensitivity per $0.7^{\circ}\times 0.7^{\circ}$ pixel, averaged over the full sky.\par
\endgroup
\end{table*}

One possible hypothesis that could explain this signal is a yet
unknown, and unmodelled, IPD component, fully analogous to the
circumsolar ring and Earth-trailing feature. Indeed,
\citet{leinert:1998} refer to all the observed radiation simply as
``excess zodiacal light brightness due to the resonant dust ring
outside the Earth’s orbit''. However, that characterization was based
on a visualization that still included the circumsolar ring and
trailing feature contributions. In our maps, however, those
contributions are already subtracted, and the signals that we observe
in Fig.~\ref{fig:solarmaps} must therefore be due to unmodelled
components.

Considering the 25$\,\mu$m signal in greater detail, we both see a strong
excess that appears nearly symmetric with respect to the Sun at low
solar elongations, as well as a four-fold symmetric structure at
higher solar elongations. In contrast, the 4.9$\,\mu$m channel
exhibits a clear dipolar structure, in which the signal is clearly
brighter in the lower left quadrant than in the upper right
quadrant. At 60$\,\mu$m, there is a linear structure that extends from
the upper left to the bottom right quadrant. In general, it is not
trivial to envision a physical IPD structure that can explain all of
these structures simultaneously with reasonable assumptions regarding
its SED, and more work is certainly needed in terms of IPD modelling
to understand what the full set of constraints actually are.

However, there is another physical mechanism that is also worth
mentioning in this respect, namely straylight or sidelobe
contamination. This is a well-known problem for many, if not most,
high sensitivity experiments. As discussed in
Sect.~\ref{sec:dirbe}, the DIRBE optics were specifically designed to
minimize precisely such radiation. At the same time, there exists to
our knowledge no physical optical ray-tracing model for the DIRBE
instrument, similar to those produced with GRASP\footnote{\url{https://www.ticra.com/software/grasp/}} for \Planck\ and
\WMAP. Given the strong excesses seen in Fig.~\ref{fig:solarmaps}, it
seems well justified to establish such in the near future; either to
put strong numerical limits on the amplitude of such straylight if it
is indeed negligible, or to derive a model that can be used directly
to subtract the emission in the case that it is non-negligible.

\subsection{Mitigation through masking and subtraction}

As far as the current analysis is concerned, determining the true
physical origin of the excess signal is only of secondary importance,
and we choose for now to remain agnostic in this respect. The key
point at this stage, however, is to minimize its impact on the final
ZSMA maps, which serve as the inputs to any DIRBE-based cosmological
and astrophysical analysis. This can be done in two ways. First, one
may exclude any pixel in solar-centric coordinates with particularly
strong excess. This is similar to the approach taken by
\citet{kelsall1998}; while they did not produce detailed maps like
those in Fig.~\ref{fig:solarmaps}, they plotted residuals as a
function of solar elongations, and noted that particular strong
excesses were seen for $e<68^{\circ}$ and $e>120^{\circ}$, and all
those data were therefore excluded from the co-added ZSMA maps. Those
limits correspond to two concentric circles centered on the Sun in
Fig.~\ref{fig:solarmaps}. Secondly, for pixels that are only mildly
affected by the excess, we use $\a_{\mathrm{static}}$ as a template, and
subtract it from the TOD prior to mapmaking, as indicated in
Eq.~\eqref{eq:model}.

The solar-centric masks used in the current analysis are shown as
thick gray lines in Fig.~\ref{fig:solarmaps}. These were generated by
thresholding each excess maps after smoothing to $3^{\circ}$ FWHM. In
addition, a sharp cut in solar elongation with varying thresholds were
applied for each channel, similar to the K98 approach. For comparison,
the thin gray lines show the static solar elongation limts used in the
K98 analysis, and the excess signals seen between the thin and thick
lines are thus contamination that is entirely eliminated in the
current analysis, but still present in the official DIRBE maps.

\begin{figure*}
	\centering
	\includegraphics[width=0.96\linewidth]{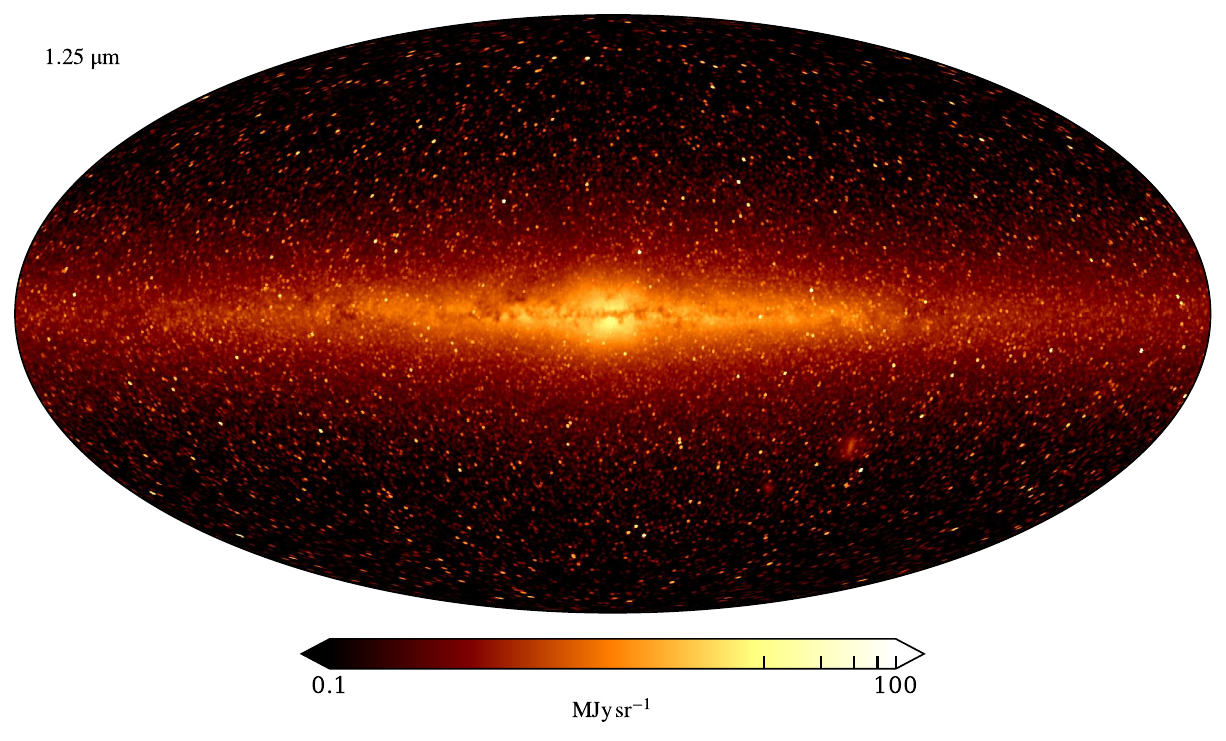}\\
	\includegraphics[width=0.96\linewidth]{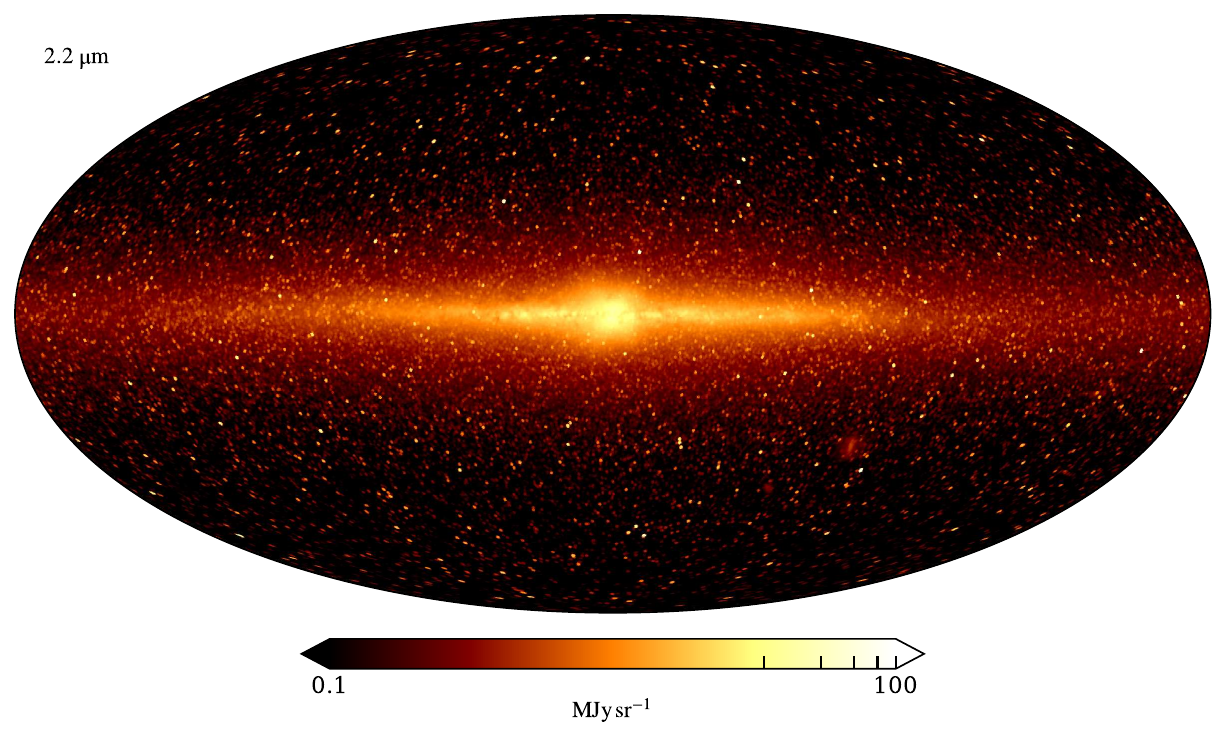}
	\caption{\cosmoglobe\ DR2 ZSMA maps at 1.25 (\emph{top}) and
          2.2$\,\mu$m (\emph{bottom}). Missing pixels have been replaced with
          the median of values within a $2^\circ$ radius.}
	\label{fig:freqmaps1_2}
\end{figure*}

\begin{figure*}
	\centering
	\includegraphics[width=0.96\linewidth]{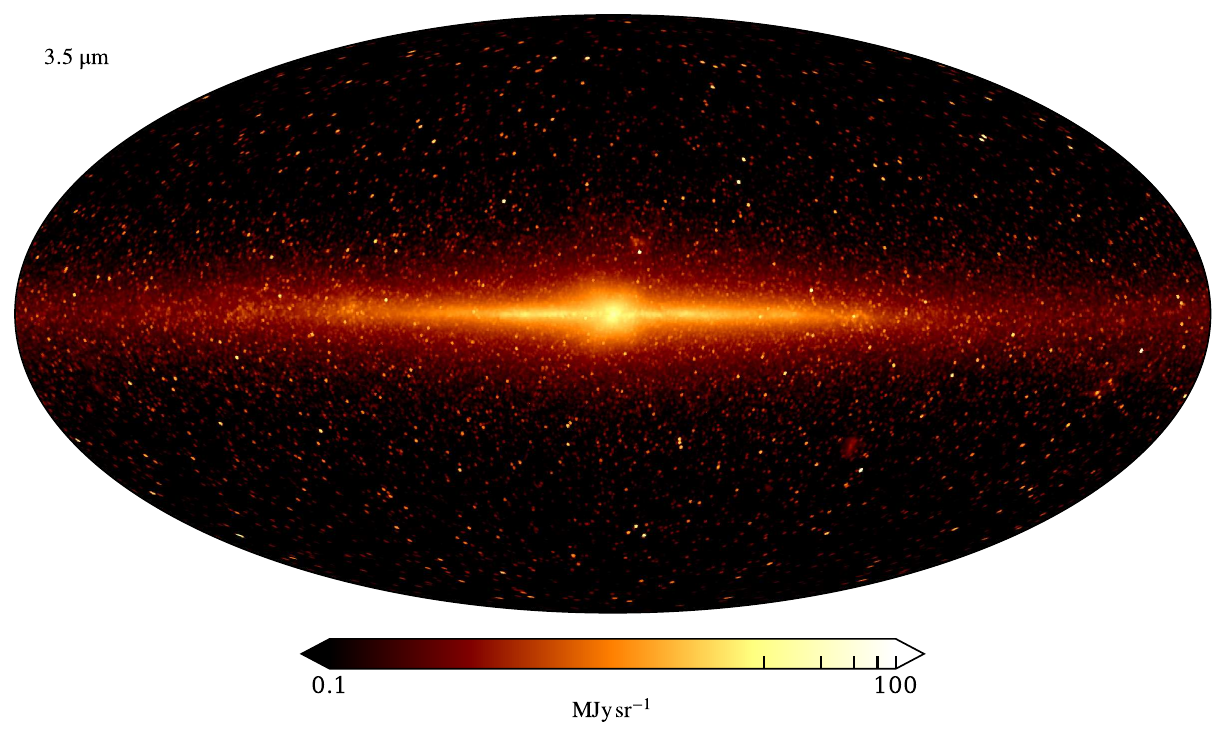}\\
	\includegraphics[width=0.96\linewidth]{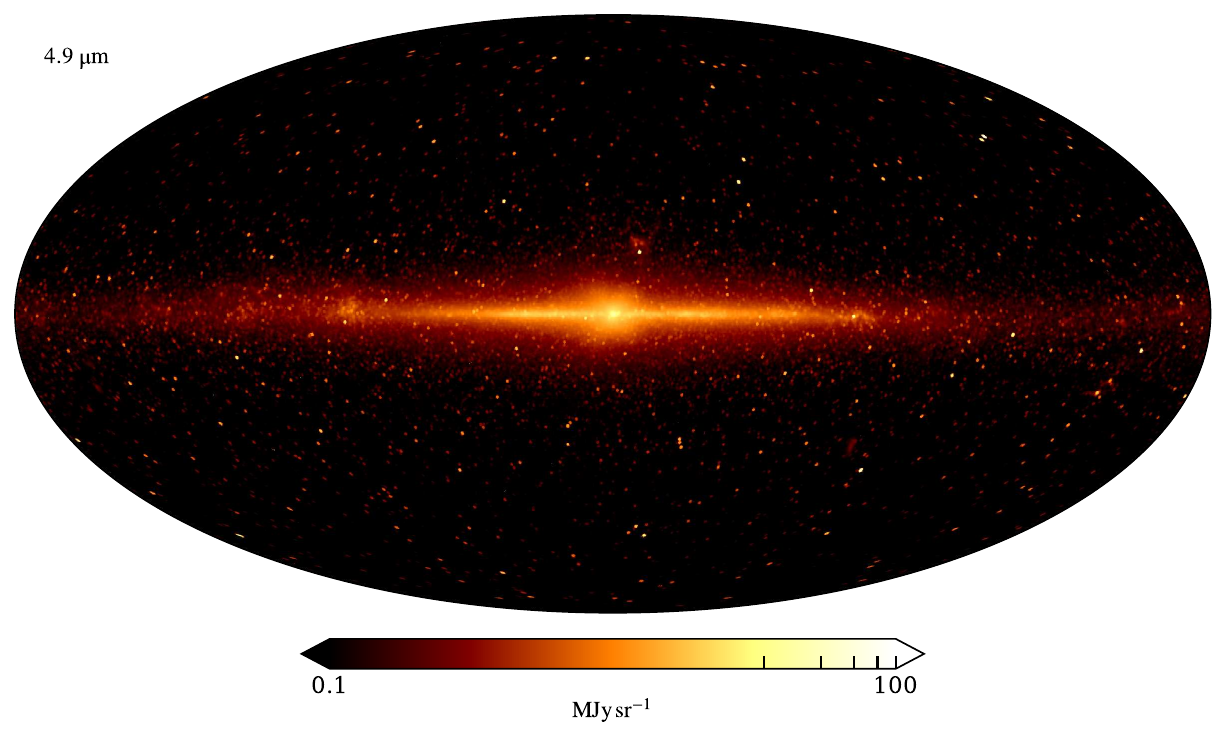}
	\caption{\cosmoglobe\ DR2 ZSMA maps at 3.5 (\emph{top}) and
          4.9$\,\mu$m (\emph{bottom}). Missing pixels have been replaced with
          the median of values within a $2^\circ$ radius.}
	\label{fig:freqmaps3_4}
\end{figure*}

\begin{figure*}
	\centering
	\includegraphics[width=0.96\linewidth]{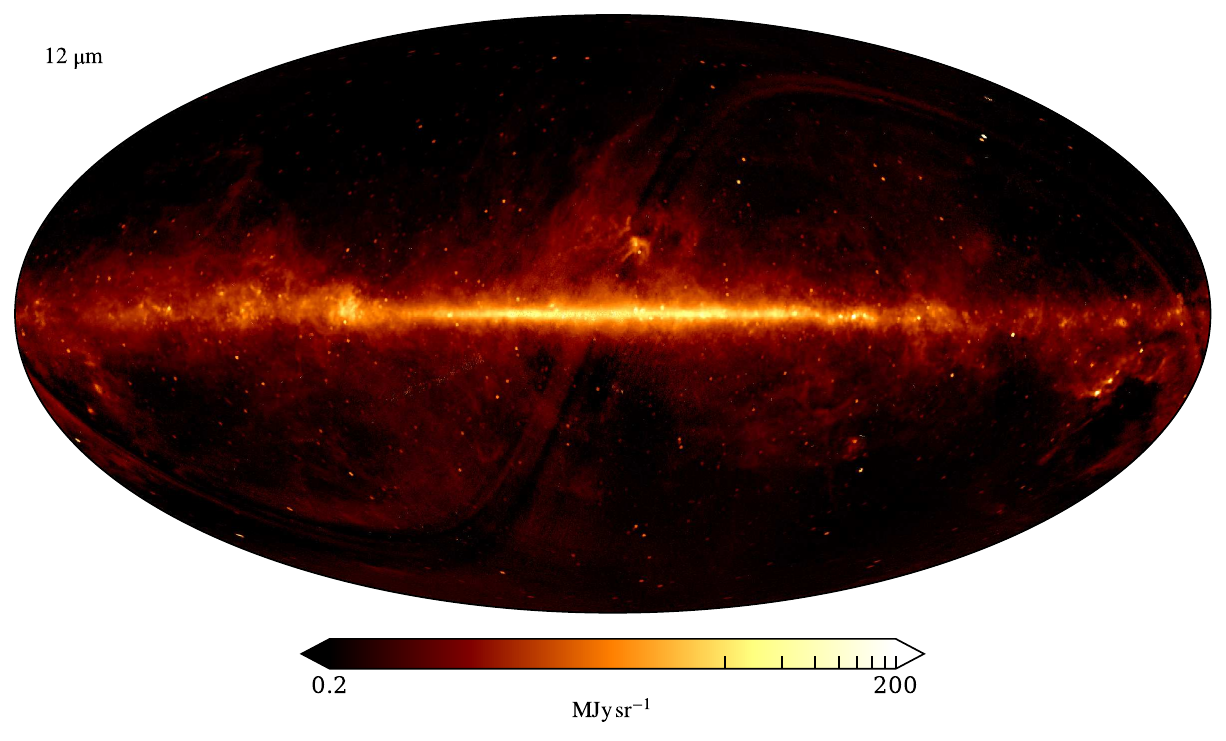}\\
	\includegraphics[width=0.96\linewidth]{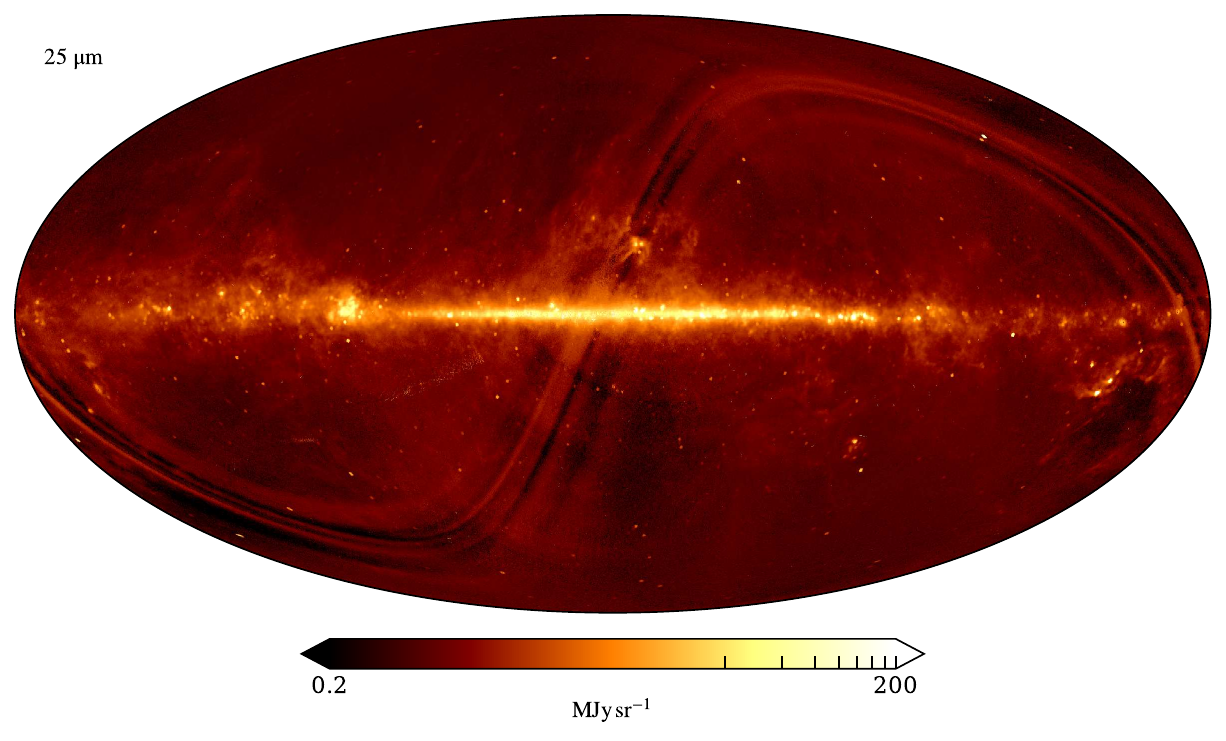}
	\caption{\cosmoglobe\ DR2 ZSMA maps at 12 (\emph{top}) and
          25$\,\mu$m (\emph{bottom}). Missing pixels have been replaced with
          the median of values within a $2^\circ$ radius.}
	\label{fig:freqmaps5_6}
\end{figure*}

\begin{figure*}
	\centering
	\includegraphics[width=0.96\linewidth]{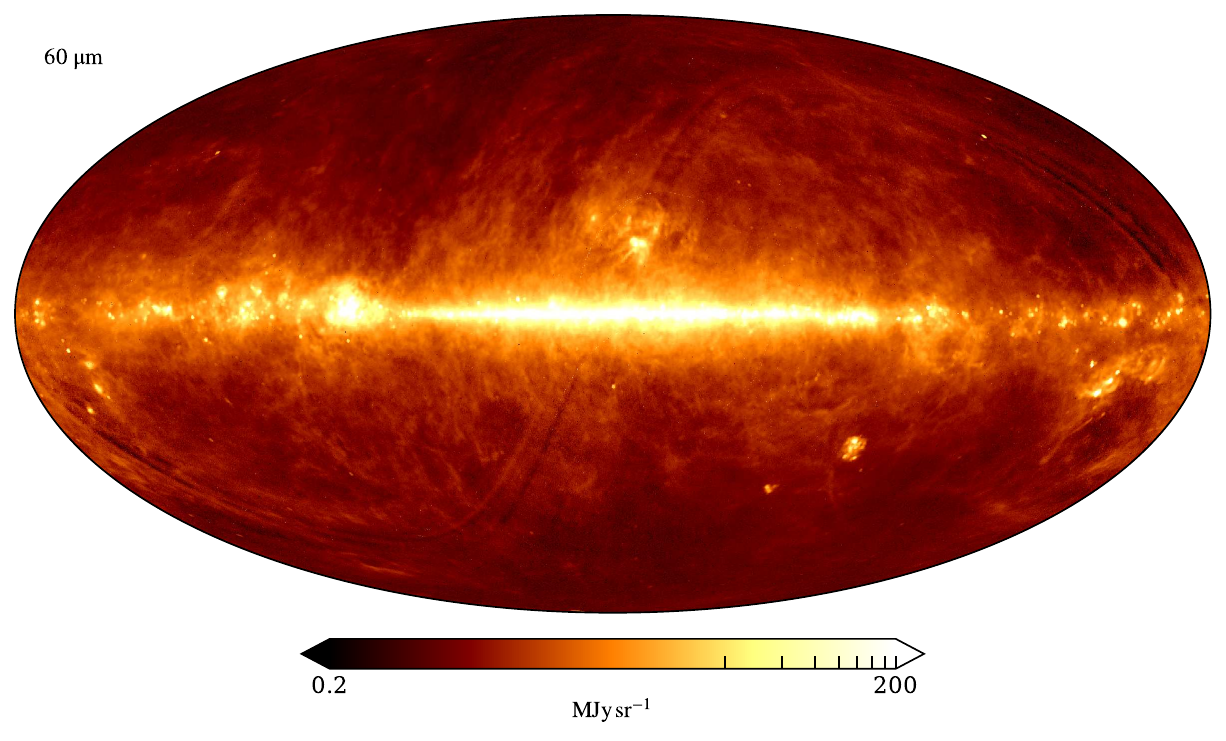}\\
	\includegraphics[width=0.96\linewidth]{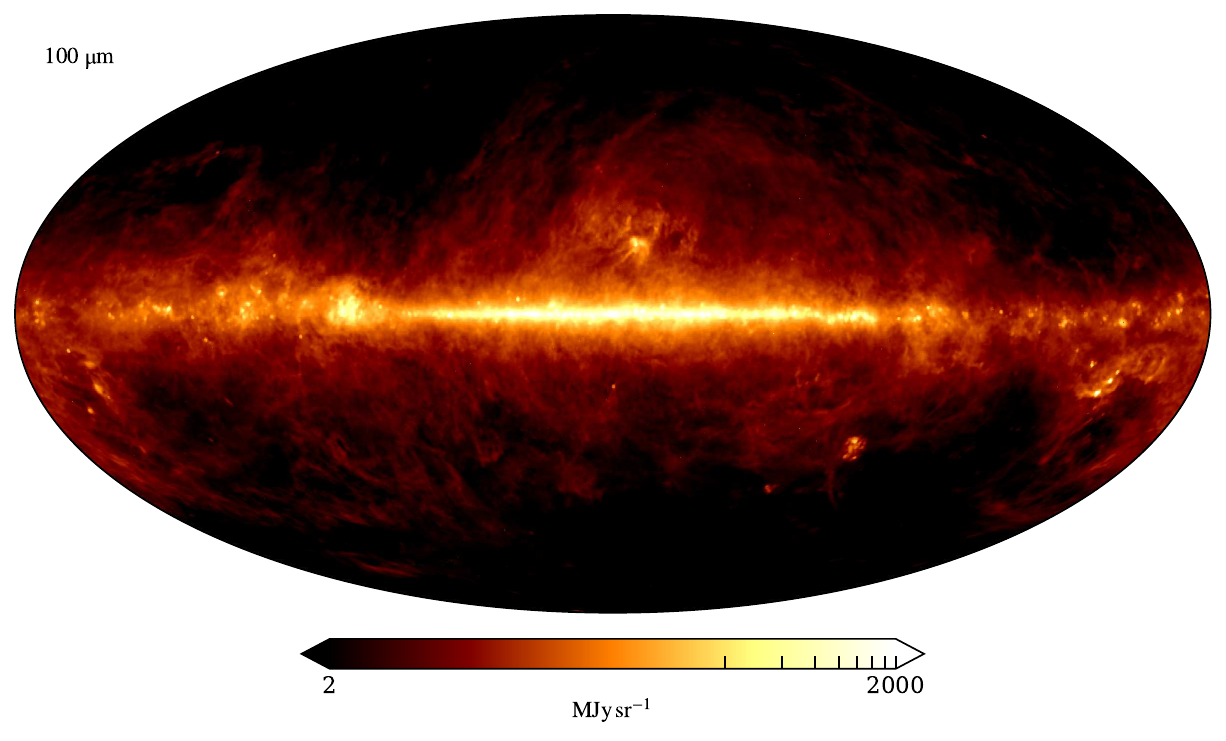}
	\caption{\cosmoglobe\ DR2 ZSMA maps at 60 (\emph{top}) and
          100$\,\mu$m (\emph{bottom}). Missing pixels have been replaced with
          the median of values within a $2^\circ$ radius.}
	\label{fig:freqmaps7_8}
\end{figure*}

\begin{figure*}
	\centering
	\includegraphics[width=0.96\linewidth]{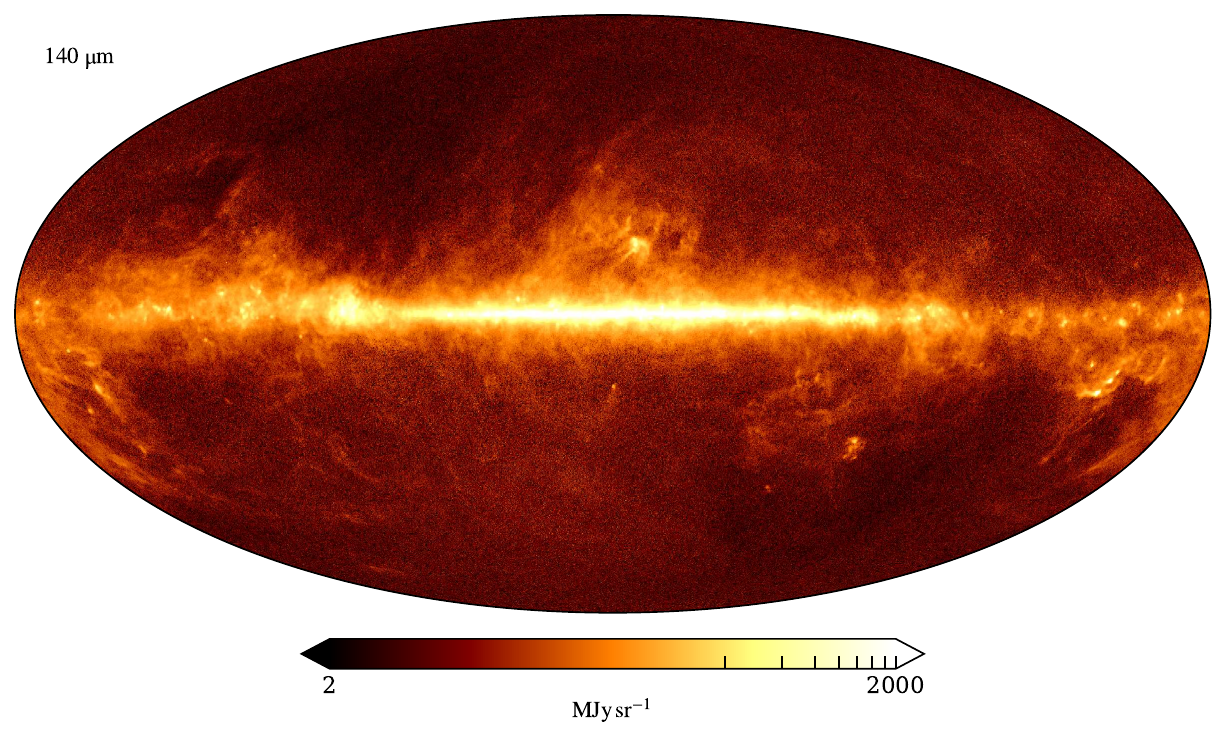}\\
	\includegraphics[width=0.96\linewidth]{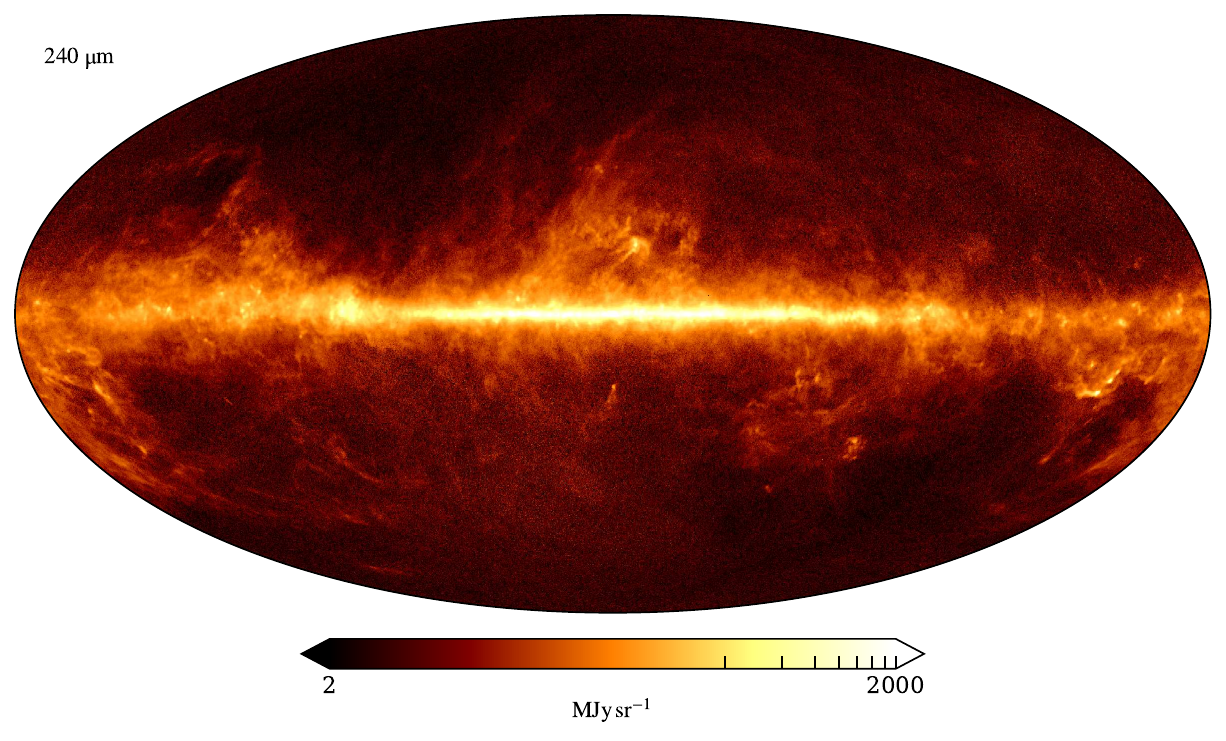}
	\caption{\cosmoglobe\ DR2 ZSMA maps at 140 (\emph{top}) and
          240$\,\mu$m (\emph{bottom}). Missing pixels have been replaced with
          the median of values within a $2^\circ$ radius.}
	\label{fig:freqmaps9_10}
\end{figure*}

\begin{figure*}
  \centering
  \includegraphics[width=0.37\linewidth]{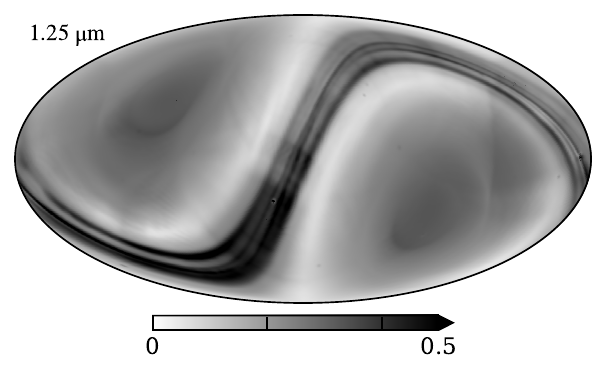}       
  \includegraphics[width=0.37\linewidth]{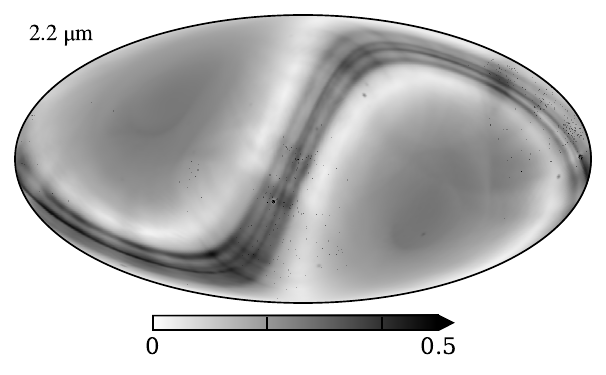}\\
  \includegraphics[width=0.37\linewidth]{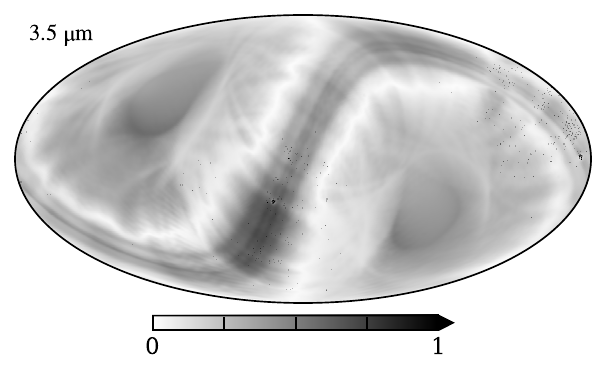}
  \includegraphics[width=0.37\linewidth]{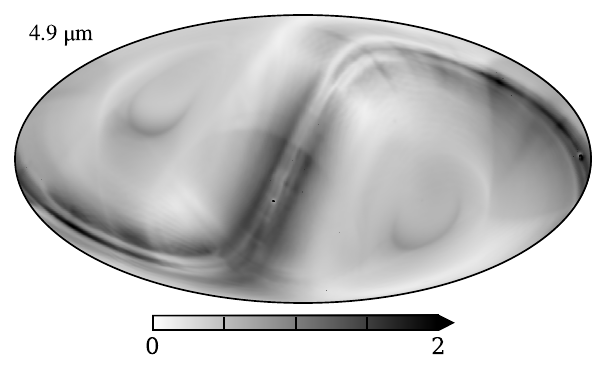}\\       
  \includegraphics[width=0.37\linewidth]{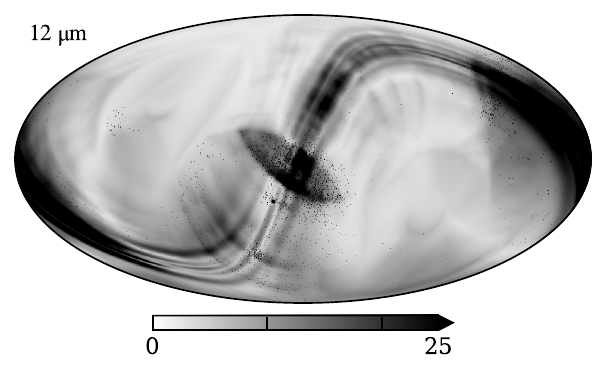}       
  \includegraphics[width=0.37\linewidth]{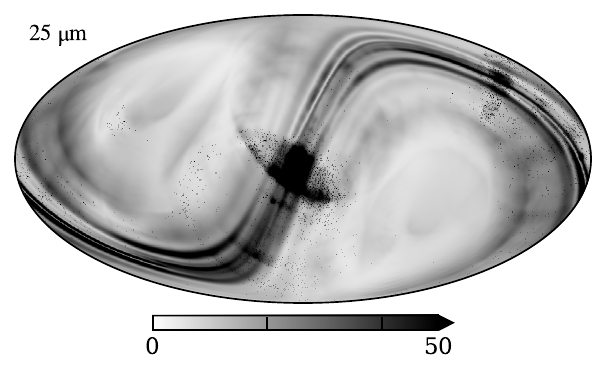}\\       
  \includegraphics[width=0.37\linewidth]{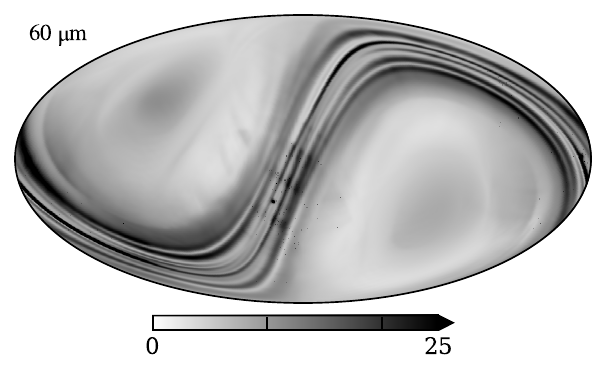}
  \includegraphics[width=0.37\linewidth]{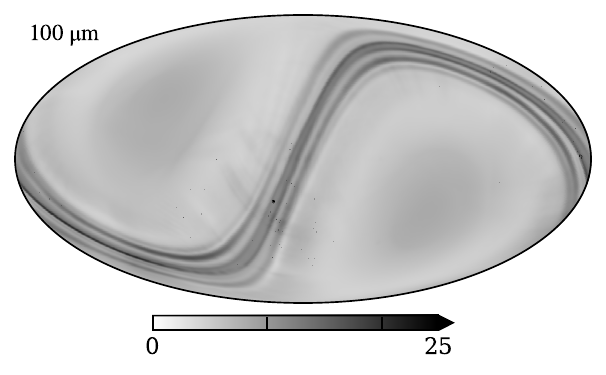}\\       
  \includegraphics[width=0.37\linewidth]{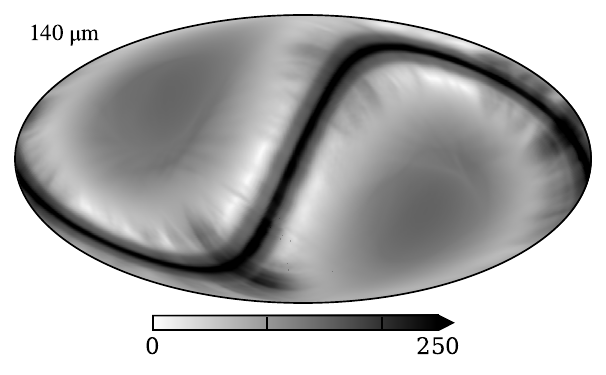}
  \includegraphics[width=0.37\linewidth]{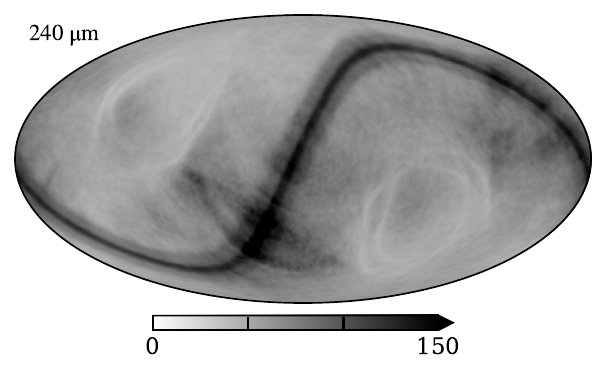}       
  \caption{Posterior rms maps for each DIRBE channel. These account
    for uncertainties due to model variations in each band, but not
    white noise. For white noise maps see Fig.~\ref{fig:sigma0_map}.
	All maps are in units of $\mathrm{kJy\,sr^{-1}}$.}
  \label{fig:rms}
\end{figure*}

While these masks eliminate the worst affected data,
highly significant excess radiation may still be seen in the unmasked
region for the 4.9--60$\,\mu$m channels. For these four channels we
let $\s_{\mathrm{static}}$ be non-zero in Eq.~\eqref{eq:model}, while
for the other six channels we set $\s_{\mathrm{static}}$ to zero, and
only apply the above masking procedure.

\subsection{Zero-level determination}

A side effect of applying a non-zero
$\s_{\mathrm{static}}$ correction is that the CIB monopole effectively
becomes unmeasurable at the corresponding frequencies; see
\citet{CG02_03}. The reason for this is simply that
$\a_{\mathrm{static}}$ is fitted freely pixel-by-pixel, and any
residual monopole that may remain in $\r_{\mathrm{static}}$ will
propagate directly into $\a_{\mathrm{static}}$. Conversely, any
monopole error in $\a_{\mathrm{static}}$ will propagate directly into
$m_{\nu}$ in Eq.~\eqref{eq:model} -- and those parameters are the CIB
monopole tracers in this analysis. This degeneracy is the main reason
for not applying the $\s_{\mathrm{static}}$ corrections to the
1.25--3.5 and 100--240$\,\mu$m maps.

\begin{figure*}
	\centering
	\includegraphics[width=0.33\linewidth]{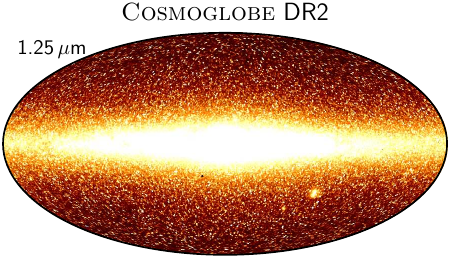}
        \includegraphics[width=0.33\linewidth]{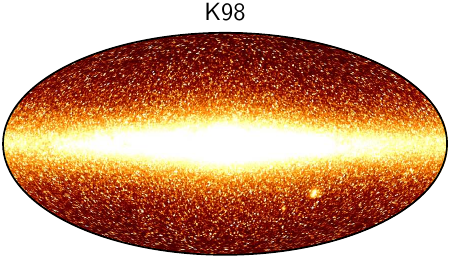}
        \includegraphics[width=0.33\linewidth]{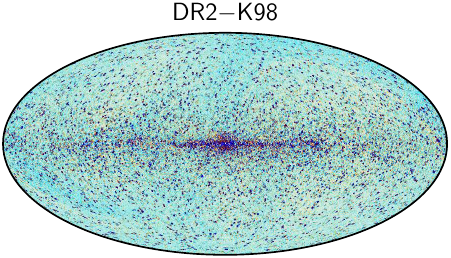}\\
	\includegraphics[width=0.33\linewidth]{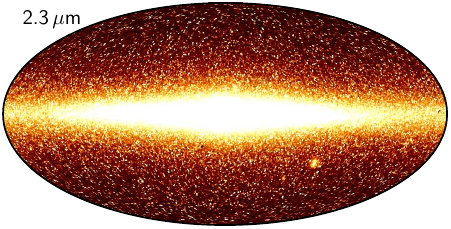}
        \includegraphics[width=0.33\linewidth]{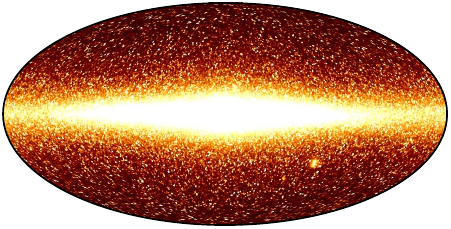}
        \includegraphics[width=0.33\linewidth]{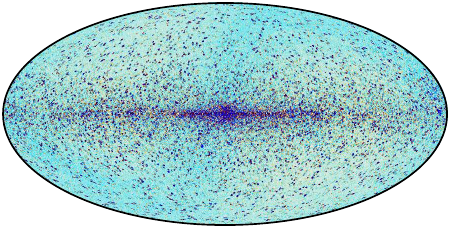}\\
	\includegraphics[width=0.33\linewidth]{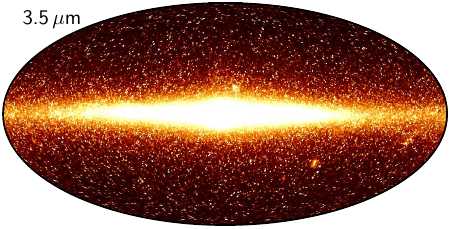}
        \includegraphics[width=0.33\linewidth]{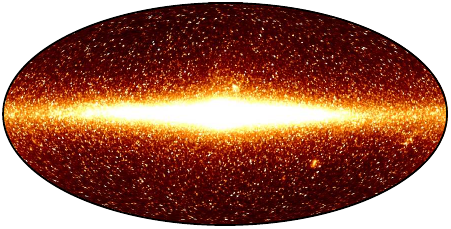}
        \includegraphics[width=0.33\linewidth]{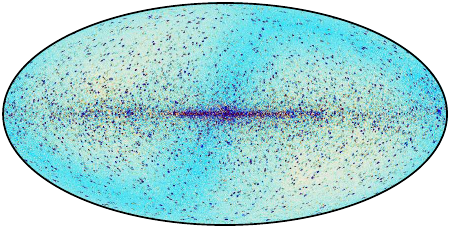}\\
	\includegraphics[width=0.33\linewidth]{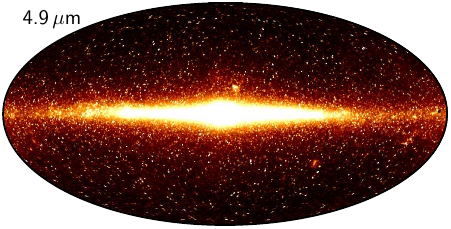}
        \includegraphics[width=0.33\linewidth]{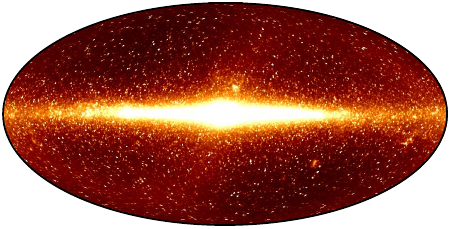}
        \includegraphics[width=0.33\linewidth]{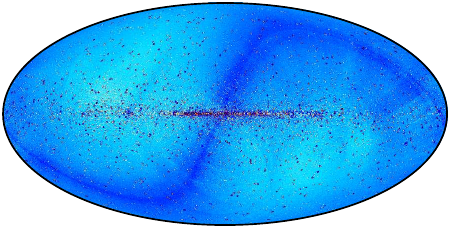}\\
	\includegraphics[width=0.33\linewidth]{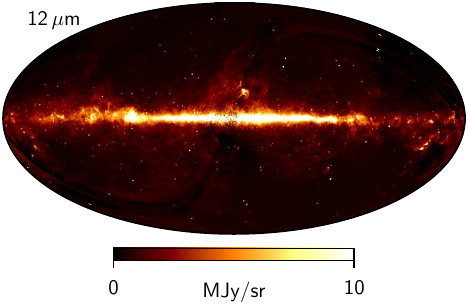}
        \includegraphics[width=0.33\linewidth]{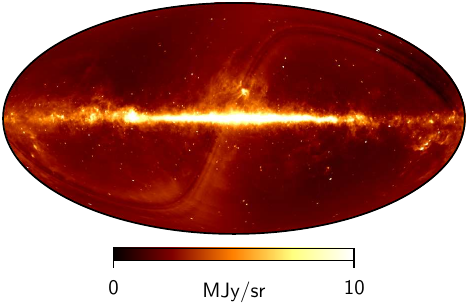}
        \includegraphics[width=0.33\linewidth]{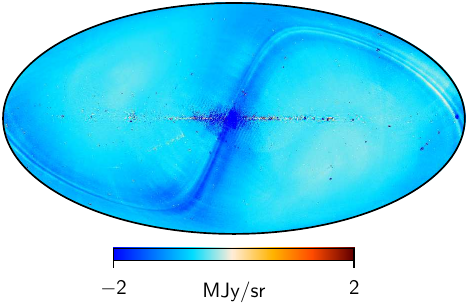}
	\caption{Comparison of \Cosmoglobe\ DR2 (\emph{left column}) and K98 (\emph{middle column}) zodiacal light subtracted mission average maps for the 1.25 to 12$\,\mu$m channels. Difference maps are shown in the rightmost column. Full maps are plotted with a non-linear color scale, while difference maps are plotted with a linear and symmetric color range.}
	\label{fig:freqmaps_cg_vs_dirbe1}
\end{figure*}

\begin{figure*}
	\centering
	\includegraphics[width=0.33\linewidth]{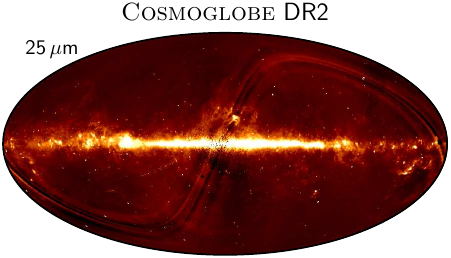}
        \includegraphics[width=0.33\linewidth]{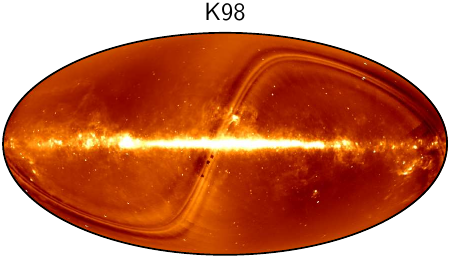}
        \includegraphics[width=0.33\linewidth]{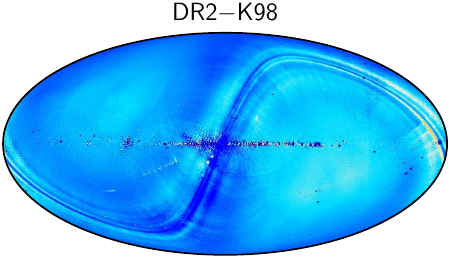}\\
	\includegraphics[width=0.33\linewidth]{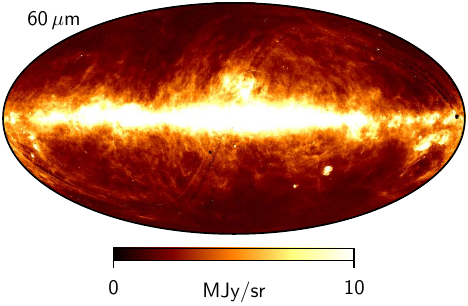}
        \includegraphics[width=0.33\linewidth]{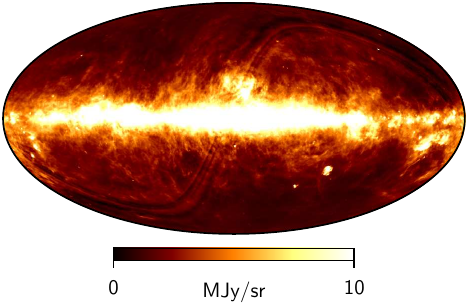}
        \includegraphics[width=0.33\linewidth]{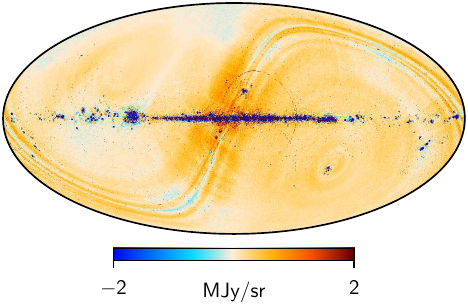}\\
	\includegraphics[width=0.33\linewidth]{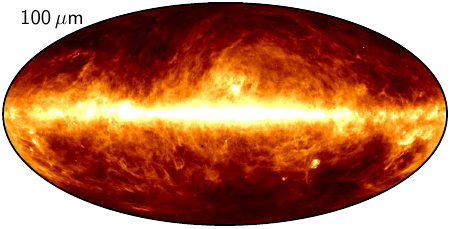}
        \includegraphics[width=0.33\linewidth]{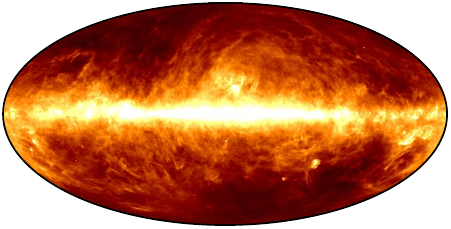}
        \includegraphics[width=0.33\linewidth]{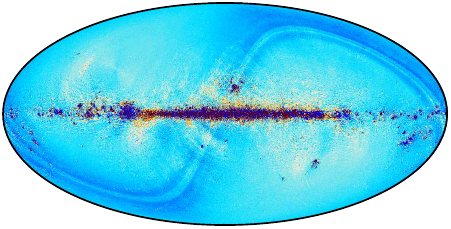}\\
        \includegraphics[width=0.33\linewidth]{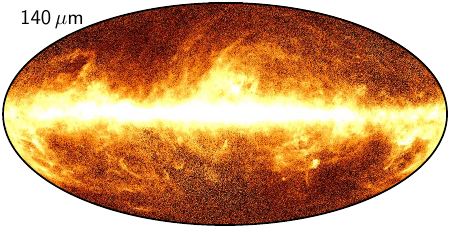}
        \includegraphics[width=0.33\linewidth]{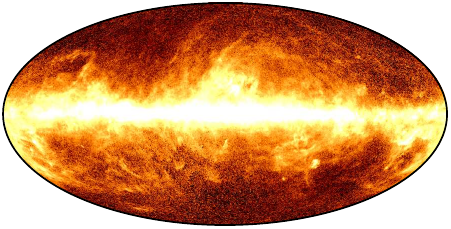}
        \includegraphics[width=0.33\linewidth]{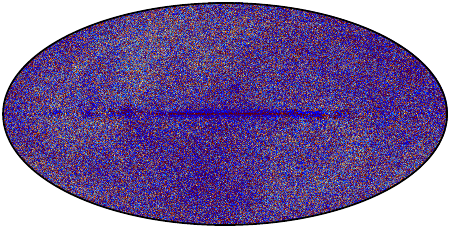}\\
        \includegraphics[width=0.33\linewidth]{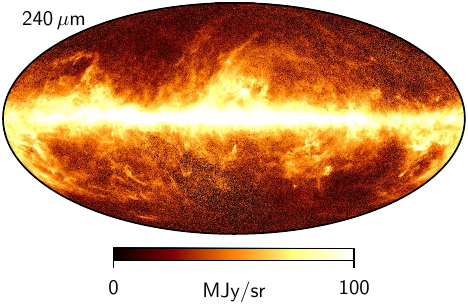}
        \includegraphics[width=0.33\linewidth]{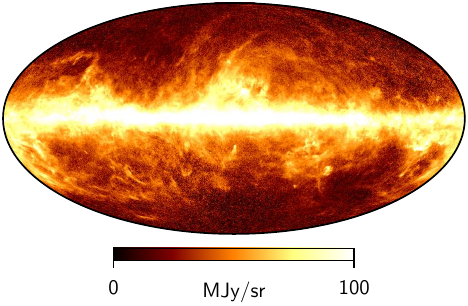}
        \includegraphics[width=0.33\linewidth]{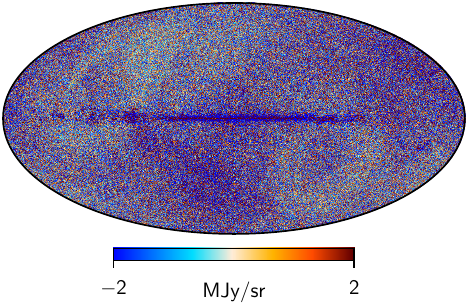}\\
	\caption{Same as Fig.~\ref{fig:freqmaps_cg_vs_dirbe1}, but for the 25--240$\,\mu$m channels.}
	\label{fig:freqmaps_cg_vs_dirbe2}
\end{figure*}

However, the ZSMA maps are useful for many other applications than CIB
monopole determination as well, and it is therefore important to
determine the zero-level of $\a_{\mathrm{static}}$ in a physically
well-motivated manner for the corrected channels. This is a priori
difficult, since we do not yet know the true physical origin of the
signal. However, whether it is due to an unmodelled ZL component or
instrumental straylight, it is reasonable to assume that the measured
intensity ideally should be strictly positive. This therefore defines
an absolute lower limit on the zero-level.

In practice, we set the zero-level of $\a_{\mathrm{static}}$ as
follows. We first smooth $\a_{\mathrm{static}}$ with a Gaussian kernel
of $2^{\circ}$ FWHM to reduce the impact of instrumental noise. We
then identify the lowest value in this smoothed map, and choose a
value that is slightly higher than this, to ensure that the value
propagated into the rest of the Gibbs chain represents a physically
meaningful value. We record the difference between the absolute lowest
value and the final selected value, and this difference may be
subtracted from the final ZSMA maps in case an absolute lower limit is
required. As discussed by \citet{CG02_03}, this procedure provides a
well-defined upper limit on $\a_{\mathrm{static}}$, and therefore also
a well-defined upper limit on $m_{\nu}$, but no lower limit on
$\a_{\mathrm{static}}$. 

To estimate an uncertainty for the zero-level of
$\a_{\mathrm{static}}$, we apply Gaussian smoothing kernels of 1, 2,
3, 4, and 5$^{\circ}$ FWHM's, and compute the standard deviation of
the resulting minima. The results from these calculations, both of the
corresponding uncertainties are tabulated in the seventh column of
Table~1 in \citet{CG02_03}.

\section{Frequency maps}
\label{sec:maps}

We now move on to the main products in the current paper, namely the
\cosmoglobe\ DR2 zodiacal light subtracted mission average maps.

\subsection{ZSMA frequency maps}

The individual DR2 ZSMA posterior mean maps are shown in
Figs.~\ref{fig:freqmaps1_2}--\ref{fig:freqmaps9_10}, plotted with
logarithmic color scales. Browsing through these in order, we find
that almost all of these appear visually consistent with Galactic
emission, and there are very few traces of residual ZL
emission. However, there is one striking exception to this, namely the
25$\,\mu$m channel. In this case, we see both a large residual
monopole and clear ZL band residuals. A minor exception is also the
12 and 60$\,\mu$m channels, which also shows slight evidence of residual ZL
band emission, although in this case the Galactic signal once again
clearly dominates. In the next section, we compare these maps with the
corresponding DIRBE K98 products.

\subsection{Angular power spectra}

\begin{figure}
	\centering
	\includegraphics[width=\columnwidth]{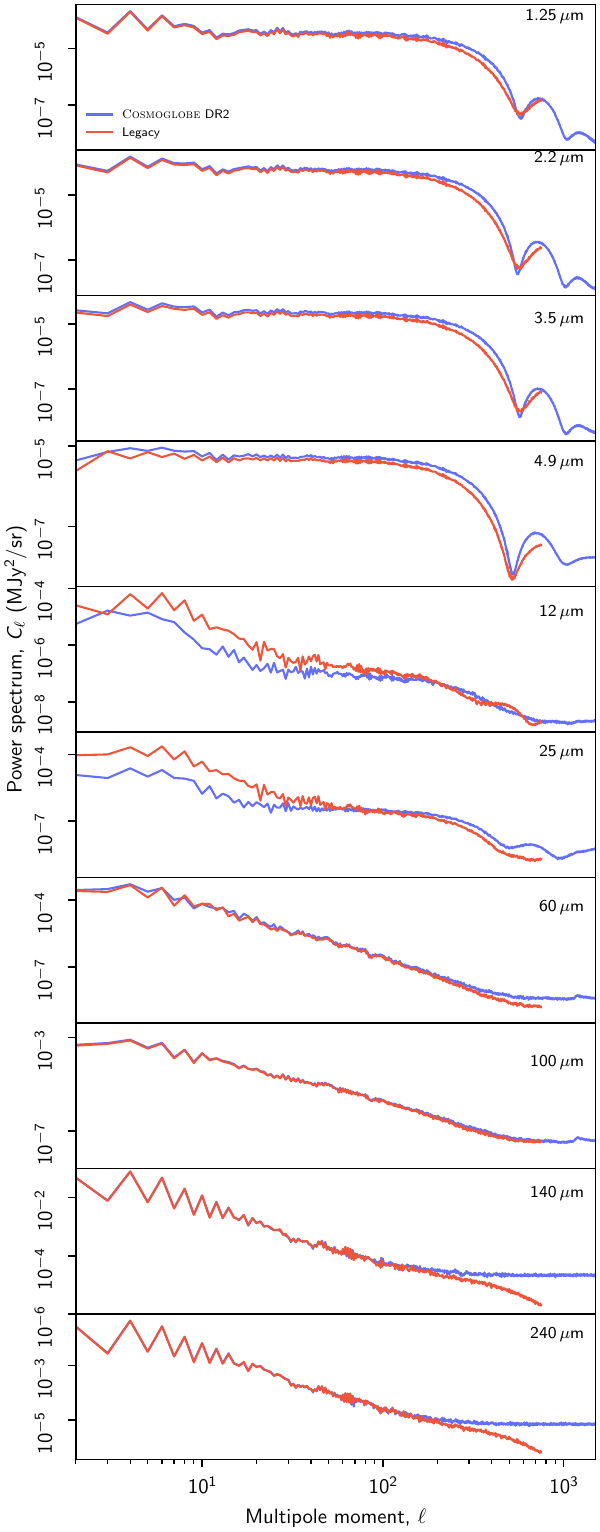}
	\caption{Comparison of angular power spectra computed from the DIRBE (red) and DR2 (blue) ZSMA maps.}
	\label{fig:powspec}
\end{figure}

Figure \ref{fig:rms} shows the posterior rms maps for each channel. It
is important to note that these do not account for the white noise of
each channel, but rather only summarize the systematic uncertainties
due to the other model components in Eq.~\eqref{eq:model}. To obtain
full uncertainties for each channel, the maps shown in
Figs.~\ref{fig:sigma0_map} and \ref{fig:rms} must be added in
quadrature. However, for most scientific analyses it is better yet to
analyze each individual Gibbs map sample separately, taking into
account only the white noise, and then build the desired summary
statistic from the ensemble of all available samples; for an example
of this procedure applied to the \Planck\ LFI and \WMAP\ data, see,
e.g., \citet{bp10}, \citet{bp11}, \citet{bp12}, \citet{bp13}, and
\citet{watts2023_dr1}.  For the 240\,$\mu$m channel only, one may
additionally see a low level of random noise, which is due to the
correlated noise component, a single realization of which is shown in
Fig.~\ref{fig:ncorr_map}. The overall amplitudes of these maps are
generally about one order of magnitude lower than the white noise
shown in Fig.~\ref{fig:sigma0_map}.

\subsection{Comparison with K98}

In Figs.~\ref{fig:freqmaps_cg_vs_dirbe1} and
\ref{fig:freqmaps_cg_vs_dirbe2} we compare the \cosmoglobe\ DR2 ZSMA
maps with the corresponding K98 maps as reprocessed by the CADE team;
see Appendix~A of \citet{paradis:2012} for algorithmic
details.\footnote{We refer to the reprocessed CADE maps also as
``K98'' in the following, but note that these maps may, at least in
principle, differ from the original Quadcube maps presented by
\citet{kelsall1998}. In particular, we expect that the pixel remapping
process used by CADE could have a non-trivial effect on the noise on
small angular scales. For a general discussion of noise in so-called
``drizzled'' maps, see \citet{Fruchter_2002}.} The right column shows
the difference between the two as defined by DR2$-$K98, so that a negative
difference indicates a higher intensity in K98. Inspecting these
maps channel-by-channel, we first note that the three shortest
wavelength bands appear visually very similar, and it is only through
the difference map it is possible to distinguish them. Here we see
that K98 has a brighter Ecliptic plane than DR2, but it is still very
difficult to determine by eye which of the two reveals a cleaner
Galactic signal. At 4.9$\,\mu$m, however, there is no longer any
visual ambiguity: Here we clearly see that the K98 map exhibits
ZL residual while the DR2 model is visually clean. The
same holds even more true at 12$\,\mu$m.

At 25$\,\mu$m, however, the relative improvements are less
striking. While the DR2 map does show lower residuals than K98 also in
this case, both of them  are contaminated to the extent that
the high-latitude regions are not useful for cosmological or
astrophysical analysis. Improving this channel is a
high-priority goal for future work. Similar considerations apply to
the 60\,$\mu$m channel, for which our map is clearly better than K98,
but faint signatures of the ZL bands are still visible.

Moving on, the 100\,$\mu$m channel is particularly interesting,
because the K98 map at this wavelength has served as a cornerstone for
Galactic thermal dust modeling for almost three decades, and it has
therefore had a massive impact in the community. By comparing the maps
shown in Fig.~\ref{fig:freqmaps_cg_vs_dirbe2}, we now see clearly that
the K98 version of that map contains significant ZL contamination. At
high latitudes, the correction is of order unity compared to the
actual Galactic signal, and it is therefore critically important to
revisit previously published thermal dust emission and extinction
parameters based on the K98 map. In contrast, the 140 and 240$\,\mu$m
channels appear very similar between the two analyses in terms of
large-scale structures.

In Fig.~\ref{fig:powspec} we plot the angular power spectra of both
the new DR2 (blue curves) and the old K98 (red curves) as evaluated
outside the DR2 processing masks. On large scales, we observe
behaviour that is consistent with the above visual impression; the two
map generations agree well on large angular scales for the
1.25--3.5$\,\mu$m and 60--240$\,\mu$m channels, while at the
intermediate channels the K98 maps exhibit clear excess from ZL
contamination.

However, these spectra additionally reveal notable differences on
smaller angular scales. First, we see that the K98 maps exhibit lower
power around the beam scale of $\ell\approx300$. We interpret this as
additional effective filtering from the coarse Quadcube
pixelization employed during the original mapmaking, coupled with the
subsequent CADE repixelization into HEALPix. Second, we note that the
DR2 spectra extend to twice the multipole range of those in the K98 maps, and
this is due to adopting a pixelization of $N_{\mathrm{side}}=512$. The
main motivation for this is clearly shown in the top four panels of
Fig.~\ref{fig:powspec}. The power spectrum shows clear structure all
the way up to the maximum limit of $\ell_{\mathrm{max}}=1500$, and
this is due to bright point sources coupled with the highly
non-Gaussian beam seen in Fig.~\ref{fig:beams}. Despite the fact that
the DIRBE resolution is only $42\arcm$ FWHM, its peculiar shape
requires a high pixel resolution to fully capture the full harmonic
bandwidth of the signal. Indeed, from these plots it appears as though
even $N_{\mathrm{side}}=512$ is formally sufficient, as the maps have
not yet reached their white noise limit at these
multipoles. The relatively coarse 8\,Hz sampling rate
of the DIRBE instrument implies that pixelization at
$N_{\mathrm{side}}=1024$ would result in a large number of missing
pixels, which is inconvenient to work with in practice; already with
the current $N_{\mathrm{side}}=512$ pixelization, our maps have some
missing pixels.

One final feature to note in Fig.~\ref{fig:powspec} is the fact that
the power spectra of the K98 maps at 140 and 240$\,\mu$m show signs of
additional smoothing at high multipoles. We have not been able to
conclusively identify the source of this effect, but users of the K98
maps should be aware of this additional smoothing.

We conclude this section by comparing the overall relative
calibration, $\alpha$, of the DR2 and K98 maps, as defined by the
slope of scatter plots between the two maps, evaluated either
outside the DR2 processing mask (Fig.~\ref{fig:masks}) or the full sky. The results from
these calculations are tabulated in the fifth and sixth columns of
Table~\ref{tab:summary}. As evaluated over the full sky, we see that
the overall calibration of the two maps agree to a few percent for
most channels, and our maps are generally slightly brighter. Again, we
interpret this difference as the effect of the coarser pixelization
used for the K98 maps. 

When considering high latitudes only, the best-fit slope differs
significantly from unity, and ranges between $\alpha=0.48$ and
1.11. This clearly illustrates the relative importance of the new maps
in terms of cosmological and astrophysical interpretation at high
latitudes; the improvements made in the current analysis are of order
unity in the low foreground regions of the sky. 

For completeness, the two rightmost columns in Table~\ref{tab:summary}
compare the average instrumental noise levels of the two analyses,
in which the K98 values are adopted from Table~1 of
\citet{hauser1998}. In general, the two estimates agree well, given
the large algorithmic differences adopted by the two analyses.

\section{Conclusions}
\label{sec:conclusions}

In this work, we have presented a joint analysis of DIRBE CIOs in a global Bayesian framework using external \Planck, \Gaia, WISE, and FIRAS data. In combination with an improved zodiacal dust model \citep{CG02_02}, an improved thermal dust model \citep{CG02_05}, and an infrared stellar model \citep{CG02_04}, we have produced maps using DIRBE data with unprecedented zodiacal dust removal and absolute monopole determination. As shown in \citet{CG02_03}, these have led to improved CIB monopole constraints across all existing DIRBE bands.

While part of this improvement comes from improved compute resources, the vast majority of the processing improvement comes from complementary datasets used in this analysis. The use of \Planck\ HFI and FIRAS data enabled the use of a thermal dust model that allowed for the unique features of the DIRBE data to shine through. Similar use of \Gaia\ and WISE data for near-infrared point source characterization allowed for robust determination of the monopole within the DIRBE bands as well. Together, this sky model allowed for DIRBE's unique zodiacal dust and monopole sensitivity to be fully utilized.

An especially prominent feature that was noted in the data was the discovery of excess radiation in solar-centric coordinates that cannot be modeled using existing models of zodiacal dust or Milky Way emission. While part of this emission could be attributed to poorly modeled zodiacal emission, the possibility of unmodeled straylight cannot be discounted. With the currently existing data, it is not possible to determine whether this excess radiation is from an astrophysical source or straylight, and investigating this should be a high priority for a future DIRBE analysis.

A major outcome of this work is the renewed possibility of using the DIRBE maps in the analysis of CMB experiments. As demonstrated here, it is possible to create a sky model that takes into account both high-resolution \Planck\ maps and the high-frequency DIRBE maps, provided that proper low-level TOD processing is included in the modeling. The fidelity of the DIRBE maps produced in this work will be indispensable for future analysis of \Planck\ HFI data, providing critical information about thermal dust and zodiacal light than cannot be constrained with Planck alone.

At the same time, this work demonstrates the need for more external data with complementary observing strategies. The existing \IRAS\ and \AKARI\ maps each have much higher resolution than DIRBE with bandpasses that overlap the brightest zodiacal emission bands. An analysis taking into account all of these data at the time-ordered level will both improve the instrument characterization for all of these experiments and improve the characterization of the infrared sky. In particular, the many lines of sight across several decades will allow for the most precise model of zodiacal dust possible, while giving an as-of-yet unattainable view into the CIB.

\begin{acknowledgements}
  We thank Tony Banday, Johannes Eskilt, Dale Fixsen, Ken Ganga, Paul
  Goldsmith, Shuji Matsuura, Sven Wedemeyer, and Janet Weiland for useful suggestions
  and guidance.  The current work has received funding from the
  European Union’s Horizon 2020 research and innovation programme
  under grant agreement numbers 819478 (ERC; \textsc{Cosmoglobe}),
  772253 (ERC; \textsc{bits2cosmology}), and 101007633 (MSCA;
  \textsc{CMBInflate}).  Some of the results in this paper have been
  derived using healpy \citep{Zonca2019} and the HEALPix
  \citep{healpix} packages.  We acknowledge the use of the Legacy
  Archive for Microwave Background Data Analysis (LAMBDA), part of the
  High Energy Astrophysics Science Archive Center
  (HEASARC). HEASARC/LAMBDA is a service of the Astrophysics Science
  Division at the NASA Goddard Space Flight Center. This publication
  makes use of data products from the Wide-field Infrared Survey
  Explorer, which is a joint project of the University of California,
  Los Angeles, and the Jet Propulsion Laboratory/California Institute
  of Technology, funded by the National Aeronautics and Space
  Administration. This work has made use of data from the European
  Space Agency (ESA) mission {\it Gaia}
  (\url{https://www.cosmos.esa.int/gaia}), processed by the {\it Gaia}
  Data Processing and Analysis Consortium (DPAC,
  \url{https://www.cosmos.esa.int/web/gaia/dpac/consortium}). Funding
  for the DPAC has been provided by national institutions, in
  particular the institutions participating in the {\it Gaia}
  Multilateral Agreement.
  We acknowledge the use of data provided by the Centre d'Analyse de Données Etendues (CADE), a service of IRAP-UPS/CNRS (http://cade.irap.omp.eu, \citealt{paradis:2012}). 
\end{acknowledgements}

\bibliographystyle{aa}
\bibliography{CG_bibliography,Planck_bib}
\end{document}